\documentclass[twocolumn,showpacs,preprintnumbers,amsmath,amssymb, nofootinbib]{revtex4}

\usepackage{graphicx}
\usepackage{dcolumn}
\usepackage{bm}
\usepackage{url}
\usepackage{tabularx}

\newcommand{\be}{\begin{equation}}
\newcommand{\ee}{\end{equation}}

\newcommand{\bea}{\begin{eqnarray}}
\newcommand{\eea}{\end{eqnarray}}
\newcommand{\Dxx}{D_{xx}}
\newcommand{\Dpp}{D_{pp}}
\newcommand{\ddp}{{\partial\over\partial p}}
\newcommand{\gray}{$\gamma$-ray~}
\newcommand{\grays}{$\gamma$-rays~}
\newcommand{\epm}{$e^{\pm}$\,}

\newcommand{\bebe}{$^{10}$Be/$^9$Be}

\def\araa{{Ann.\ Rev.\ Astron.\ Ap.}}

\def\apj{ApJ}
\def\apjl{{ApJ\ (Lett.)}}
\def\apjs{{ApJ\ Suppl.}}

\def\aa{{A\&A}}
\def\aap{{A\&A}}
\def\mnras{{MNRAS}}

\def\pre{{Phys. Rev. E}}
\def\physrep{{Phys.~Rep.}}   

\begin{document}


\title{The impact of propagation uncertainties on the potential dark
  matter contribution\\ to the Fermi LAT mid-latitude \gray data} 

\author{Daniel T. Cumberbatch}
 \email{D.Cumberbatch@sheffield.ac.uk}
 \author{Yue-Lin Sming Tsai}
 \email{php06yt@sheffield.ac.uk}
\affiliation{Astroparticle Theory \& Cosmology Group, Department of
  Physics \& Astronomy, The University of Sheffield, Hicks Building,
  Hounsfield Road, Sheffield, S3 7RH, U.K.} 

\author{Leszek Roszkowski}
 \email{L.Roszkowski@sheffield.ac.uk}
\affiliation{Astroparticle Theory \& Cosmology Group, Department of
  Physics \& Astronomy, The University of Sheffield, Hicks Building,
  Hounsfield Road, Sheffield, S3 7RH, U.K. and
  The Andrzej Soltan Institute for Nuclear Studies, Warsaw, Poland.} 
   
\begin{abstract}

\noindent We investigate the extent to which the uncertainties associated with the propagation 
of Galactic cosmic rays impact upon estimates for the \gray flux from the mid-latitude region. 
We consider contributions from both standard astrophysical background (SAB) processes as 
well as resolved point sources. We have found that the uncertainties in the total \gray flux from 
the mid-latitude region relating to propagation parameter values consistent with local B/C and 
\bebe~data dominate by 1-2 orders of magnitude. These uncertainties are reduced to 
less than an order of magnitude when the normalisations of the SAB spectral components are 
fitted to the corresponding Fermi LAT data. We have found that for many propagation parameter 
configurations (PPCs) our fits improve when an extragalactic background (EGB) component is 
simultaneously fitted to the data. We also investigate the improvement in our fits when a flux 
contribution from neutralino dark matter (DM), described by the Minimal Supersymmetric Standard 
Model, was simultaneously fitted to the data. We consider three representative cases of neutralino DM 
for both Burkert and Einasto DM density profiles, in each case simultaneously fitting a boost factor 
of the DM contribution together with the SAB and EGB components. We have found that for several 
PPCs there are significant improvements in our fits, yielding both substantial EGB and DM components, 
where for a few of these PPCs the best-fit EGB component is consistent with recent estimates by the 
Fermi Collaboration.
\end{abstract}

\maketitle

\section{Introduction}
\label{sec:intro}
\noindent A plethora of astrophysical data ranging from Galactic to cosmological scales indicate the (gravitational) influence of otherwise non-interacting particles in our Universe. This highly abundant, yet elusive, dark matter (DM), undoubtedly occupies one of the hot seats in current astroparticle physics research. 

This is especially true in light of the recent analyses of data from
direct detection experiments, which have yielded new improved bounds but also intriguing
hints of a possible signal (see, e.g., \cite{Ahmed:2009zw, Aalseth:2010vx}).  In
addition, the Large Hadron Collider (LHC) \cite{lhc}, set to
probe new areas of physics, in particular, supersymmetry (SUSY), which has spawned some of the most compelling DM
candidates, has also been recently activated.  (For a review of SUSY DM see, e.g.,~\cite{Jungman:1995df,Bertone:2004pz}.)

An alternative strategy to identifying DM is to reconcile discrepancies between current astrophysical predictions for various radiation (e.g., positrons, antiprotons and $\gamma$-rays) and current observations, then subsequently constrain the properties of DM particles whose annihilation (or decay) products may result in such discrepancies. To do this, one crucially has to determine how the particle and electromagnetic radiation emerging from DM annihilations, {\it as well as} the radiation emerging from the interactions of cosmic rays (CRs) with background nuclei and interstellar radiation fields, propagate through the interstellar medium (ISM), whilst specifying the distribution of contributing DM and astrophysical sources. Unfortunately, the number of Galactic properties relevant to CR propagation are vast, their effect on these {\it indirect signals} are largely ambiguous, and the data constraining them (e.g., local nuclear abundance ratios) is limited. 

However, the Large Area Telescope (LAT) of the recently launched Fermi Gamma-ray Space Telescope \cite{Fermi} (Fermi LAT henceforth) is designed to observe \grays with energies 20\,MeV$\lesssim E\lesssim\,$300\,GeV, which are typical of the energies expected from the annihilation products of the popular group of DM candidates known as weakly interacting massive particles (WIMPs) \cite{Baltz:2006sv}.
Fermi LAT has acquired data from both the mid-latitude region (MLR), $10^{\circ}<|b|<20^{\circ}, 0^{\circ}<l<360^{\circ}$ \cite{porter_ICRC_2009, Abdo:2009mr, collaboration:2010nz}, and the Galactic centre (GC) \cite{Goodenough:2009gk, fermidata}. However, here we only utilise data from the MLR since it is likely to be dominated by \grays originating from local sources and hence is limited in its contamination from astrophysical sources concentrated near the GC \cite{porter_ICRC_2009}.

In this paper we utilise the publicly available \texttt{GALPROP} package \cite{galprop} to investigate how different propagation parameter configurations (PPCs) affect corresponding predictions for (i) the local abundance of CR nuclei and (ii) the \gray flux from the MLR. 
For the former we deduce which PPCs are consistent with current measurements of the abundance ratios B/C and $^{10}$Be/$^9$Be. We then utilise these PPCs in the latter analysis, calculating the corresponding \gray spectra when including a successively increasing number of components, including the contributions from DM.  We assume that the DM solely consists of neutralinos described by the Minimal Supersymmetric Standard Model (MSSM) (see, e.g., \cite{Bertone:2004pz}), and select several benchmark points and density profiles to illustrate the spread of contributions that can be obtained.\\

\noindent This paper is organised as follows.
In Sec.\,\ref{sec:DGB} we briefly describe the processes responsible for the main contributions to the background diffuse \gray emission in the context of Fermi LAT. 
In Sec.\,\ref{sec:CRprop} we describe the model of CR propagation utilised in this study and the relevant aspects of the \texttt{GALPROP} code used to conduct our simulations. 
In Sec.\,\ref{sec:results} we present our results for the \gray flux emerging from the MLR when considering purely astrophysical sources.
In Sec.\,\ref{sec:DM} we briefly describe the MSSM and the criteria invoked to select our DM candidate points. We then display results for the \gray flux from the MLR when including contributions from these candidate points.
Finally in Sec.\,\ref{sec:summary} we summarise our results.

\section{The background diffuse \gray emission}
\label{sec:DGB}
\subsection{Signal components}
\label{subsec:signal_components}

\noindent The Galactic diffuse \gray emission primarily originates from
the interaction of high energy charged particles (known as cosmic rays) with interstellar nuclei and
radiation fields through a variety of physical mechanisms,
including: bremsstrahlung; inverse Compton scattering (ICS); $\pi^0$-decay; and synchrotron
emission, each of which we now briefly describe.

Bremsstrahlung occurs in the Galaxy primarily when high energy $e^{\pm}$ are deflected by the Coulomb field of interstellar atoms/nuclei, primarily H and He. Here, we utilise the original \texttt{GALPROP} code v50.1p to calculate the spectra of \grays resulting from such interactions.

Inverse Compton scattering primarily occurs in the Galaxy when high energy $e^{\pm}$ interact with photons of the interstellar radiation field (ISRF), up-scattering them to higher energies, resulting in energy loss for the traversing $e^{\pm}$. 
In this paper, we use a version of \texttt{GALPROP}, that we have modified, to calculate the Galactic \gray flux contribution from the MLR from ICS, taking into account three target photon distributions constituting part of the ISRF \cite{Porter:2005qx, Moskalenko:2005ng}: optical (i.e., starlight), the far infra-red (FIR) background, and the cosmic microwave background (CMB). 

Fluxes of \grays from $\pi^0$-decay primarily occur in the ISM by $pp$\,-chain reactions, generally resulting in the production of two $ \gamma$-rays. Whilst ICS and bremsstrahlung emission are, as we shall see, strongly dependent on the propagation parameters describing the diffusion of CRs and interstellar nuclei, \grays from $\pi^0$-decay, which are almost immediately produced where the parent $\pi^0$ is created, are particularly dependent on the initial distributions of nuclei within the Galaxy. Here, we adopt the so-called  {\it conventional}  \texttt{GALPROP} model for the initial nuclear abundances \cite{footnote:conv_model}.

Finally, synchrotron emission results from the deflection of charged particles by the Galactic magnetic field, resulting in electro-magnetic radiation emission. Unlike bremsstrahlung, ICS and $\pi^0$-decay, the flux contribution from synchrotron emission is negligible at the energies sensitive to Fermi LAT (see, e.g.,~\cite{Regis:2008ij}, \cite{Orlando:2009}). Hence, we ignore such contributions throughout this study.

In addition to the above, we must account for the contributions from the hundreds of high energy \gray point sources identified by Fermi LAT \cite{Abdo:2009a}. To account for these, we utilise results from the analysis presented in \cite{collaboration:2010nz} (see Sec.\,\ref{sec:results} for further details).

Further, we expect an additional, isotropic, extragalactic background (EGB) component, which includes the sum of contributions from unresolved point sources, diffuse emission from signatures of large scale structure formation, diffuse emission resulting from the interaction of ultra-high energy CRs and relic photons, etc. (see, e.g., \cite{Venters:2009gv, Venters:2010bq, collaboration:2010nz}). 
Despite this, if one adopts the estimate of the EGB, as recently determined by the Fermi collaboration \cite{collaboration:2010nz}, one finds that the diffuse \gray flux in the direction of the GC still dominates the EGB. However, the Galactic component decreases rapidly with increasing latitude to the extent where in the MLR the estimated EGB becomes comparable at energies $E\lesssim1\,$GeV. Therefore, for Fermi LAT \gray flux measurements from the MLR, observed at energies 20\,MeV$\lesssim E\lesssim$300\,GeV, if one adopts the Fermi estimate the EGB it should be acknowledged in any corresponding analysis. 
In Sec.\,\ref{sec:results} and Sec.\,\ref{sec:DM} we account for the EGB by simultaneously fitting a \gray flux component, described by the power-law:
\begin{equation}
E_{\gamma}^2\frac{{\rm d}\Phi}{{\rm d}E_{\gamma}}=AE^{\gamma},
\label{eq:egbflux}
\end{equation}
\noindent where $A$ and $\gamma$ are fitted parameters,  in addition to contributions from other relevant sources, to the total \gray flux $E_{\gamma}^2\frac{{\rm d}\Phi}{{\rm d}E_{\gamma}}$ observed from the MLR by Fermi LAT.


\subsection{Fermi LAT mid-latitude observations}
\label{subsec:Fermi}
\noindent The Large Area Telescope on board NASA's Fermi Gamma-ray Space Telescope \cite{Fermi} has measured the diffuse \gray emission with unprecedented sensitivity and resolution, being over an order of magnitude more sensitive than its predecessor, the Energetic Gamma-Ray Experiment Telescope (EGRET) \cite{Morselli:2002nw}.
In this study, we utilise results from Fermi LAT's observations of the diffuse \gray emission from the MLR during Fermi LAT's initial 10 months of scientific observation \cite{porter_ICRC_2009, Abdo:2009mr, collaboration:2010nz}.

The mid-latitude sky region was selected for the initial studies of Fermi LAT since it maximises the contribution to the diffuse \gray background produced within several kiloparsecs of the Sun, hence minimising uncertainties associated with CR diffusion. At smaller latitudes, contributions from sources near to the GC become dominant, significantly increasing astrophysical ambiguities, while at larger Galactic latitudes the emission is increasingly affected by contamination from charged particles misclassified as photons as well as increased uncertainties in the model used to estimate the background diffuse \cite{porter_ICRC_2009, Abdo:2009mr}. 
These results are in strong contention with those from EGRET, especially at energies $\gtrsim1$\,GeV, where it was proclaimed that EGRET observed an excess in diffuse \gray emission when compared to theoretical models that correctly reproduce fluxes of directly measured CR nucleon and electron spectra \cite{s&m_apj_2000}.

\section{Propagation of CR nucleons in the galaxy}
\label{sec:CRprop}
\subsection{Description of propagation model}

\noindent The composition and energy spectra of the vast
majority of CRs currently traversing the Galaxy originate from the
nuclear interactions (also known as {\it spallations}) of an initial
distribution of energetic particles, mainly protons and electrons,
with nuclei residing in the ISM, and, if electrically charged, their
electromagnetic interactions with the Galactic magnetic field and
ISRF.  
The astrophysical sources giving rise to these
initial distributions of CRs are believed to include mainly supernovae remnants
\cite{Aharonian:2006ws, Ahlers:2009ae, blandford:1987}, and, to a lesser extent,
pulsars \cite{Aharonian:1995, Hooper:2008kg, Profumo:2008ms, Buesching:2008hr,
Gendelev:2010fd}, compact objects in nearby binary systems \cite{Khangulyan:2007si} and stellar winds \cite{Benaglia:2002wa, Pittard:2006us, Reimer:2005hy}. (For a review of potential CR sources see, e.g.,~\cite{Hinton:2009}.)  The energetic particles observed by X-ray
experiments to be accelerating away from these sources (see, e.g.,~\cite{Leyder:2007gn, Fender:2004gg} and references therein), propagate several
kiloparsecs through the ISM where they lose energy via
bremsstrahlung and ICS, as discussed in
Sec.\,\ref{subsec:signal_components}, as well as ionization and Coulomb
interactions, and also synchrotron emission as they interact with the
Galactic magnetic field. These processes modify the energy
spectra and composition of CRs, whilst producing secondary
particles and \gray radiation.

Throughout the rest of this paper, we refer to the diffuse \gray flux
contributions resulting from the interaction of high energy CRs,
originating from the astrophysical sources mentioned above, with both CR
nuclei residing in the ISM and the ISRFs 
as the {\it standard astrophysical background} (SAB) component.

In order to simulate the propagation of CRs within the Galaxy, we
utilise the 2D model of the publicly available numerical propagation
code \texttt{GALPROP}. The intricacies
of the \texttt{GALPROP} propagation models are described in detail
elsewhere (see, e.g.,~\cite{Strong:1998pw, Strong:2007nh}), however in the following we briefly summarise
the basic features relevant to the present study.

The \texttt{GALPROP} code attempts to numerically solve the
propagation equation Eq.\,(\ref{eq:diff_eq}), for a given source
distribution of all CR species, and boundary conditions defining the
region of propagation, known as the {\it diffusion zone}, beyond which
free-particle escape is assumed. The \texttt{GALPROP} propagation
equation is written as
\begin{eqnarray}
\label{eq:diff_eq}
{\partial \psi \over \partial t} 
&=&
q(\vec r, p) 
+ \vec\nabla \cdot ( \Dxx\vec\nabla\psi - \vec V\psi )\nonumber\\
&&+ \ddp\, p^2 \Dpp \ddp\, {1\over p^2}\, \psi
- {\partial\over\partial p} \left[\psi\,\frac{{\rm d}p}{{\rm d}t}
- {p\over 3} \, (\vec\nabla \cdot \vec V )\psi\right]\nonumber\\
&&- {1\over\tau_f}\psi - {1\over\tau_r}\psi,\nonumber\\
\end{eqnarray}
where $\psi=\psi (\vec r,p,t)$ is the particle density per unit of
particle momentum, $q(\vec r, p)$ is the source term, $\Dxx$ is the
spatial diffusion coefficient, $\vec V$ is the convection velocity
associated with Galactic winds, re-acceleration is described as
diffusion in momentum space and is determined by the coefficient
$\Dpp$, $\tau_f$ is the time-scale for nuclear fragmentation, and
$\tau_r$ is the time-scale for the radioactive decay of nuclei.

The Galactic magnetic field is generally believed to follow the spiral pattern of its stellar population \cite{Borriello:2008gy, Tinyakov:2001ir}, and there are several studies that utilise models reflecting this when investigating CR diffusion (see, e.g., \cite{Borriello:2008gy}). 
However, as mentioned in Sec.\,\ref{subsec:Fermi}, a study of the MLR limits the relevant sources to local phenomena and hence minimises the effect of the large scale structure of the Galaxy. Here we adopt the following model for the Galactic magnetic field that is default to \texttt{GALPROP} and independent on azimuthal angle:
\be
B(r,z)=B_0\,{\rm exp}\left(-r/r_0-|z|/z_0\right),
\ee
where the scale lengths $r_0$ and $z_0$ indicate the approximate extent of the field. 
Consequently, it is also appropriate that the diffusion zone be cylindrically symmetrical, and is defined here to be a slab of half-thickness $L$ and radius $r_{\rm max.}=r_0=20$\,kpc \cite{footnote:rmax}.

The spatial diffusion coefficient $\Dxx = \beta D_0(R/R_0)^{\alpha}$ is taken here to be a function of particle rigidity $R$,
where the relativistic factor $\beta=v/c$ is a consequence of a random-walk (diffusion) process, and $R_0$ is a reference rigidity set equal to $4\,$GV. Where diffusive re-acceleration is considered, the momentum-space diffusion coefficient $\Dpp$ is taken here to be related to $\Dxx$ by the following relation, derived in \cite{Seo1994},
\be
\Dpp\,\Dxx=\frac{4\,p^2\,v_{\rm A}^2}{3\left(2-\alpha\right)\left(4-\alpha\right)\left(2+\alpha\right)\alpha}, 
\ee
where $v_{\rm A}$ is the Alfv\'en velocity, describing the propagation of low-frequency electromagnetic waves, known as Alf\'ven waves, generated by the motion of ions in the ISM relative to the Galactic magnetic field \cite{gedalin1993, footnote:reac}.

The convection velocity $V(z)$ incorporated into the original \texttt{GALPROP} code to describe the effect of Galactic winds on CR particle abundances at the periphery of the diffusion zone is assumed to increase linearly with distance from the Galactic plane (i.e., $\frac{{\rm d}V}{{\rm d}x}=\frac{{\rm d}V}{{\rm d}y}=0$, $\frac{{\rm d}V}{{\rm d}z}>0$), which we adopt here for a range of values of $\frac{{\rm d}V}{{\rm d}z}$. This linear convection model implies a constant adiabatic energy loss, which is consistent with CR-driven MHD wind models \cite{Zirakashvili96, Ptuskin:1997, Strong:1998pw}.

The \texttt{GALPROP} code incorporates the energy injection spectrum for nucleons, which is assumed to be a power-law in momentum, d$q(p)/$d$p\propto p^{-n_g}$, with multiple breaks: $n_g=n_{g_1},\,n_{g_2}$ below/above a reference rigidity with default value $R=R_n=9$\,GV. This form is shared by the injection spectrum for \epm, d$q(p)/$d$p\propto p^{-e_g}$, again with multiple breaks: $e_g=e_{g_0}$ for rigidities $R<R_1$, $e_g=e_{g_1}$ for $R_1<R<R_2$ and $e_g=e_{g_2}$ for $R>R_2$, where $R_1=4$\,GV and $R_1=10^6$\,GV are the adopted default values. In both cases, the spectra of these particles are modified as they lose energy via ionisation and Coulomb interactions (nuclei) as well as bremsstrahlung, ICS and synchrotron processes (\epm).

Lastly, we mention that, when calculating the local abundance of CR nuclei, we incorporate the estimated effects of the solar cycle on their energies and fluxes following the prescriptions of the {\it force field} approximation \cite{perko:1987}, as invoked by Maurin {\it et al.} \cite{maurin:2001}. Here, the 
interstellar energy $E^{\rm IS}$ of a nucleus of charge $Ze$ and atomic number $A$ is shifted according to the relation
\be
\frac{E^{\rm TOA}}{A} = \frac{E^{\rm IS}}{A} - \frac{|Ze|\phi}{A},
\label{eq:sma}
\ee
\noindent where $E^{\rm TOA}$ is the {\it modulated} energy of the nucleus at the top of the atmosphere (TOA), and $\phi$ is the solar modulation potential which varies over the 11 year solar cycle. In this study, whilst we assume a time-independent value of $\phi$, we utilise a large range of values to  illustrate the importance of such variations.
Given Eq.\,(\ref{eq:sma}), the relationship between the interstellar flux $\Phi^{\rm IS}$ of interstellar nuclei to that at the Earth $\Phi^{\rm TOA}$ is related by
\be
\frac{\Phi^{\rm TOA}(E^{\rm TOA})}{\Phi^{IS}(E^{\rm IS})}\simeq\left(\frac{(E^{\rm TOA})^2-(Am_p)^2}{(E^{\rm IS})^2-(Am_p)^2}\right),
\label{eq:smb}
\ee
\noindent where $m_p$ is the proton rest mass.

\subsection{Evaluation of models: B/C and \bebe~ratios}
\label{subsec:b/c}
\noindent In the previous section we described and highlighted the substantial flexibility in the current modelling of the propagation of CRs within the Galaxy. In this section we illustrate the result of imposing constraints on the values of these parameters when invoking the criterion that PPCs must generate results that are consistent with current measurements of the local interstellar nuclear abundance ratios B/C and \bebe. These ratios are not the only two which can be used to potentially constrain propagation parameters, but are arguably the most stringent, which explains their common appearance in the relevant literature (see, e.g.,~\cite{Strong:1998pw, Delahaye:2007fr, Delahaye:2010ji, Putze:2008ps, Maurin:2010zp, Putze:2010zn, Maurin:2001sj, Regis:2008ij, Strong:2001fu, Pato:2010ih}). Other possible ratios include (Sc+Ti+V)/Fe, $^{26}$Al/$^{27}$Al, $^{36}$Cl/Cl and $^{54}$Mn/Mn (see, e.g.,~\cite{Strong:2001fu, Putze:2010zn}), however we neglect the use of such data within this study. 

For each of the PPCs that we considered, we utilised the \texttt{GALPROP} package to calculate the local abundance ratios B/C and \bebe~as a function of nuclear kinetic energy, and subsequently compute their corresponding $\chi_{\rm LAR}^2$ value (expressed as $\chi_{\rm LAR}^2$ per data point), using current measurements of the local abundance ratios (LAR) B/C and \bebe~ measurements, defined as
\be
\chi_{\rm LAR}^2=\frac{\sum_{j}\sum_i^{N_j}\frac{\left(D_{ij}-T_{ij}\right)^2}{\sigma_{ij}^2}}{\sum_jNj}
\label{eq:chisq}
\ee
\noindent where $D_{ij}$ and $\sigma_{ij}$ are the respective central values and 1$\,\sigma$ errors
of the $i$th data point of the $j$th experimental data set (either B/C or \bebe), $T_{ij}$ is our predicted abundance ratio at the energy  corresponding to the data point $D_{ij}\pm\sigma_{ij}$. Hence, the inner sum of the numerator of Eq.\,(\ref{eq:chisq}) is over all data points associated with each experiment, and the outer sum is over all experimental data sets. We then normalise this expression by dividing by the total number of data points from all experiments considered. This quantity is equivalent to a maximum likelihood method, assuming complete ignorance with respect to B/C and \bebe~ measurements, and that the individual measurements are statistically independent from one another with Gaussian likelihoods.

To calculate our results we firstly conducted a broad scan of propagation parameter space involving 12 independent parameters. For convenience, we list these parameters in Table\,\ref{tab:paramdef}.
\begin{table}[t]
\begin{tabular}{|c|c|}
\hline\hline
Parameter & Definition \\
\hline
\hline
$D_{0}$ & Normalisation of spatial diffusion coefficient  \\
\hline
$B_0$ & Galactic magnetic field normalisation \\
\hline
$\alpha$ & Spatial diffusion spectral index \\
\hline
 $z_0$& Scale length of Galactic magnetic field ($z$-dir.) \\
\hline
 $e_{g_0}$& Index for $e^{\pm}$ injection spectrum ($R<R_1$) \\
\hline
$e_{g_1}$& Index for $e^{\pm}$ injection spectrum ($R_1<R<R_2$) \\
\hline
$n_{g_1}$& Index for nucleon injection spectrum ($R<R_n$) \\
\hline
$n_{g_2}$& Index for nucleon injection spectrum ($R>R_n$) \\
\hline
$v_{\rm A}$& Alf\'ven velocity\\
\hline
$L$& Half-depth of Galactic diffusion zone\\
\hline
$\frac{{\rm d}V}{{\rm d}z}$& Convection velocity gradient ($z$-direction) \\
\hline
$\phi_{\odot}$& Solar modulation potential\\
\hline\hline
\end{tabular}
\caption{A reference table of the 12 propagation parameters involved in our investigation, where rigidities $R_1=4$\,GV, $R_2=10^6$\,GV and $R_n=9$\,GV. }
\label{tab:paramdef}
\end{table}
Our broad scan consisted of approximately 3000 different PPCs, using different values of the 8
parameters: $D_{0}$, $\alpha$, $n_{g_{1}}$, $ n_{g_{2}}$, $v_{\rm A}$,
$L$, $\frac{{\rm d}V}{{\rm d}z}$ and $\phi_{\odot}$, displayed in
Table\,\ref{tab:param}.
\begin{table}[t]
\begin{tabular}{|c|c|c|c|}
\hline\hline
$D_{0}/10^{28}$ & $\alpha$ & $n_{g_{1}}$ & $ n_{g_{2}}$  \\
\hline
5, 6.5, 8.5 & 0.2, 0.4, 0.6 & 1.7, 2.15, 2.6 & 2.0, 2.3, 2.6\\
\hline
\hline
$v_{\rm A}$ & $L$ & $\frac{{\rm d}V}{{\rm d}z}$ & $\phi_{\odot}$\\
\hline
20, 35, 50 & 4, 7, 11 & 2, 6, 10 & 100, 250, 450\\
\hline\hline
\end{tabular}
\caption{Ranges of the 8 parameters varied in our initial broad scan
  consisting of approximately 3000 PPCs, expressed as:
  $D_{0}/10^{28}$\,(cm$^2$\,s$^{-1}$), $\alpha$,
  $n_{g_{1}}$, $ n_{g_{2}}$,
  $v_{\rm A}$\,(km\,s$^{-1}$), $L$\,(kpc), $\frac{{\rm d}V}{{\rm
      d}z}$\,(km\,s$^{-1}$\,kpc$^{-1}$) and $\phi_{\odot}$\,(MV).}
\label{tab:param}
\end{table}
The values of the 4 remaining parameters: $B_0$, $e_{g_0}$, $e_{g_1}$ and $z_0$, were taken to be constants equal to the default values specified in the ``conventional'' \texttt{GALPROP} model 
(5\,$\mu$G, 1.60, 2.50 and 2\,kpc, respectively), since we found that changing their values from these had an insignificant impact on the local nuclear abundance ratios. 
(However, these parameters have a significant effect on the \gray flux from the MLR, and in Sec.\,\ref{sec:results}, where we present our results for the \gray flux, we discuss how we utilised PPCs with different values of these parameters.) 

Following our initial scan we performed a second higher resolution scan, involving a further 3000 points, located closely in propagation parameter space to those that possessed $\chi_{\rm LAR}^2<30$ following our initial scan.
(We note that our high resolution scan on occasion utilised values of propagation parameters slightly outside of those ranges specified in Table\,\ref{tab:param} in order to determine our best-fit results. This is illustrated in the Appendix to this paper, where we display some of the PPCs utilised in our investigation.)
\begin{figure}[t]
	\begin{center}
	\includegraphics[width=6.6cm, keepaspectratio]{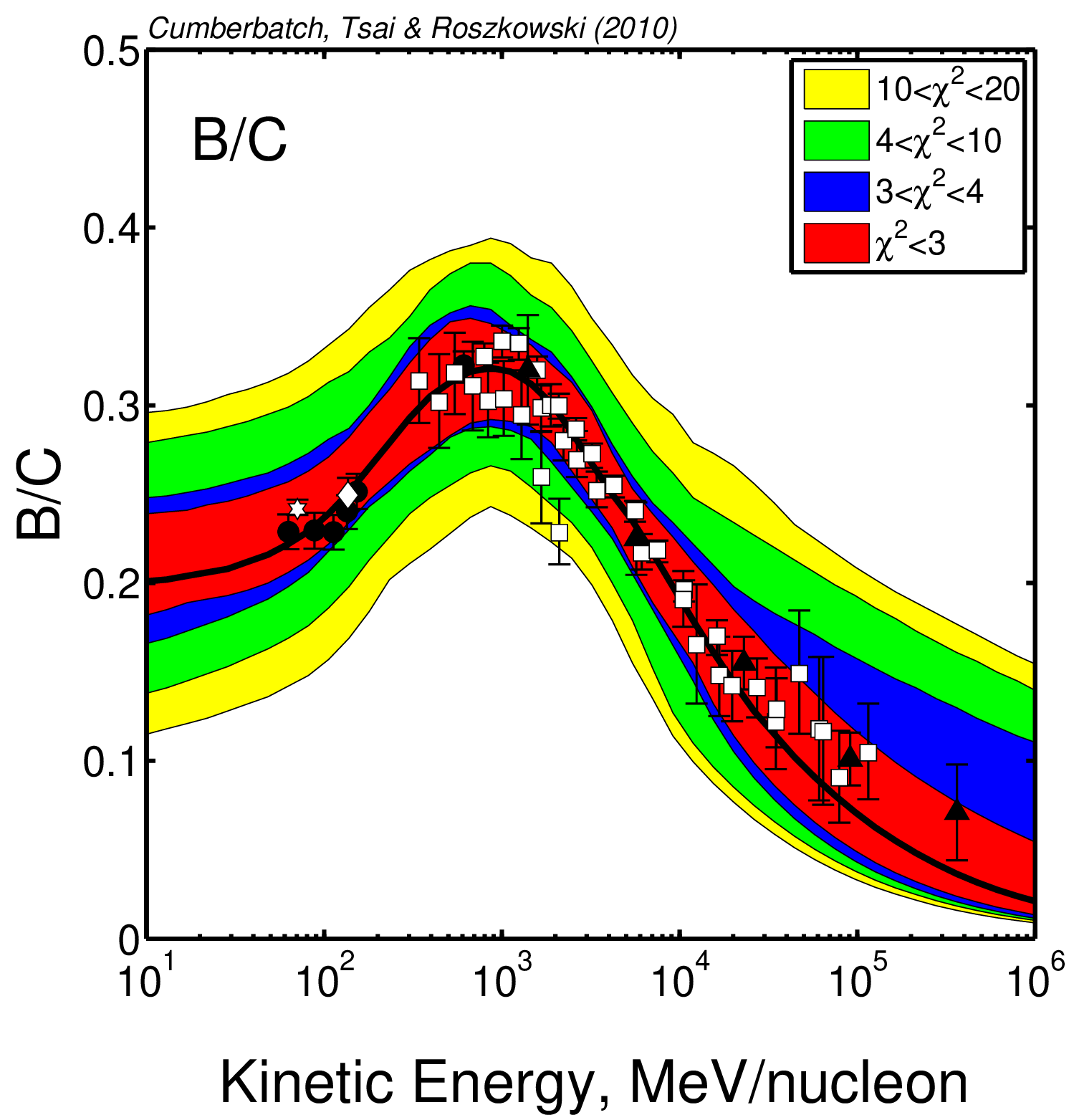}
	\includegraphics[width=6.6cm, keepaspectratio]{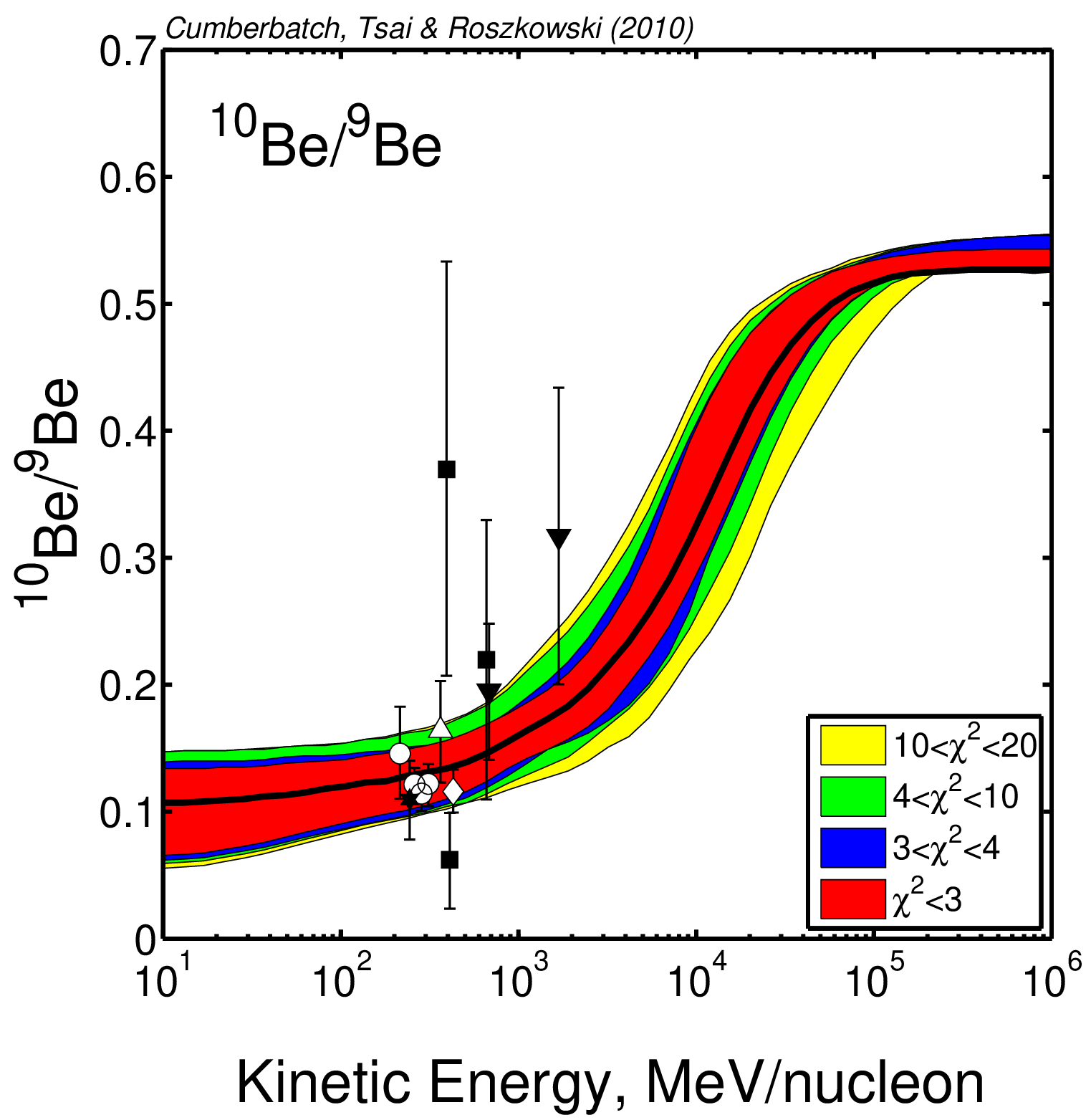}
	\caption{B/C (upper panel) and \bebe~ (lower panel) in the solar neighbourhood, as a function of kinetic energy per nucleon, calculated for various PPCs. Each configuration is grouped according to its $\chi_{\rm LAR}^2$ value per data point (innermost to outermost filled regions): $\chi_{\rm LAR}^2<3$ (red), $3<\chi_{\rm LAR}^2<4$ (blue), $4<\chi_{\rm LAR}^2<10$ (green) and $10<\chi_{\rm LAR}^2<20$ (yellow), calculated using all displayed data sets: ACE (B/C: black circles, \bebe: white circles), HEAO-3 (B/C: white squares), Ulysses (white diamonds), Voyager (B/C: white stars, \bebe: black stars), CREAM (B/C: black triangles), ISEE-3 (\bebe: white triangles), ISOMAX (\bebe: black triangles) and other balloon experiments (Balloon) (\bebe: black squares). We also display the spectra associated with the propagation parameter configuration generating the smallest value of $\chi_{\rm LAR}^2$ resulting from our scan (thick black curve).}
		\label{fig:b/c}
         \end{center}
\end{figure}

In Figure\,\ref{fig:b/c} we display the results of our high resolution scan. We display the resulting range of B/C (upper panel) and \bebe~ ratios (lower panel) in the solar neighbourhood, as a function of kinetic energy per nucleon, and grouped according to their associated values of $\chi_{\rm LAR}^2$ (innermost to outermost filled regions): $\chi_{\rm LAR}^2<3$ (red), $3<\chi_{\rm LAR}^2<4$ (blue), $4<\chi_{\rm LAR}^2<10$ (green) and $10<\chi_{\rm LAR}^2<20$ (yellow). In our calculations of $\chi_{\rm LAR}^2$ we utilised experimental data from ACE \cite{davis2000} (B/C: black circles, \bebe: white circles), HEAO-3 \cite{engelmann1990} (B/C: white squares), Ulysses \cite{duvernois1996} (white diamonds), Voyager \cite{lukasiak1999} (B/C: white stars, \bebe: black stars), the recent data from the balloon-borne experiment CREAM \cite{Ahn:2008my} (B/C: black triangles), ISEE-3 \cite{Wiedenbeck:1983} (\bebe: white triangles), ISOMAX \cite{Hams:2004} (\bebe: black triangles) and other balloon experiments (Balloon) \cite{Hagen:1977, Buffington:1978, Webber&Kish:1979} (\bebe: black squares). Also, in each respective plot we display the corresponding spectra associated with the propagation parameter configuration (PPC) generating the smallest value of $\chi_{\rm LAR}^2$ resulting from our scan (black curves).

The relatively larger spread in the values of B/C compared to corresponding results for \bebe~indicate the sensitive nature of the B/C with respect to the selected PPC. 
Despite the purpose of this paper being to highlight the overall uncertainties owing to propagation parameters rather than the effects of individual propagation parameters, we do mention that, whilst the changes in the high energy (i.e., $>$1\,GeV/n) B/C ratio is dominated by the effects of the diffusion parameters, the spread in the low energy B/C ratio owes to the different values of the solar modulation potential, $\phi$, utilised. This owes primarily to the nature of the {\it force field} approximation, discussed in Sec.\,\ref{sec:CRprop}, invoked here to estimate the effects of the solar wind, where the shift $\Delta E = -|Z|\phi/A$ in energy per nucleon, given by Eq.\,(\ref{eq:sma}), is equal to 225\,MeV/n for both $^{10}$B and $^{12}$C for $\phi=450\,$MV. Hence, this explains the aforementioned importance of $\phi$ at low energies, since we expect the effects of solar modulation to become significant for energies (per nucleon) near and below $E\sim|\Delta E|$.

Compared to the local B/C ratio, the sensitivity of \bebe~with respect to PPCs is much less significant, indicated in the lower plot of Figure\,\ref{fig:b/c} by the much narrower spread of each $\chi_{\rm LAR}^2$ group. Further, we can infer, from the closely related spread in \bebe~of each $\chi_{\rm LAR}^2$ group at low energies, that this data is much less significant in determining the overall $\chi_{\rm LAR}^2$ than the B/C data. This clearly reconciles with the poor nature of the \bebe~data relative to the local B/C data. Unfortunately, at higher energies (i.e., $>2$\,GeV), where the spread in the \bebe~spectra from PPCs yielding smaller $\chi_{\rm LAR}^2$ becomes increasingly narrow, and hence would be most significant in constraining propagation parameters, there is a complete lack of experimental data.

\section{\gray flux from the mid-latitude region}
\label{sec:results}

\noindent In this section, we present our results for the \gray flux predicted to emerge from the MLR
from astrophysical sources and compare them with the corresponding measurements by Fermi LAT.
We utilise a modified version of \texttt{GALPROP} to calculate the \gray flux for
PPCs with $\chi_{\rm LAR}^2 <20$, following the
high resolution scan of propagation parameter space discussed in Sec.\,\ref{subsec:b/c}.
We consider two different scenarios:
\begin{itemize}
\item Firstly, in Sec.\,\ref{subsec:bkgflux}, we consider the \gray flux from the SAB interactions
as well as accounting for contributions from those point sources identified by Fermi LAT
in the MLR.
\item Secondly, in Sec.\,\ref{subsec:bkgegbflux}, we simultaneously fit a power-law \gray
component to the Fermi LAT data, intended to represent the EGB flux, in addition to the SAB
and point source contributions.
\end{itemize}


\subsection{SAB and Point Source Contributions}
\label{subsec:bkgflux}

\begin{figure}[t]
	\includegraphics[width=\linewidth, keepaspectratio]{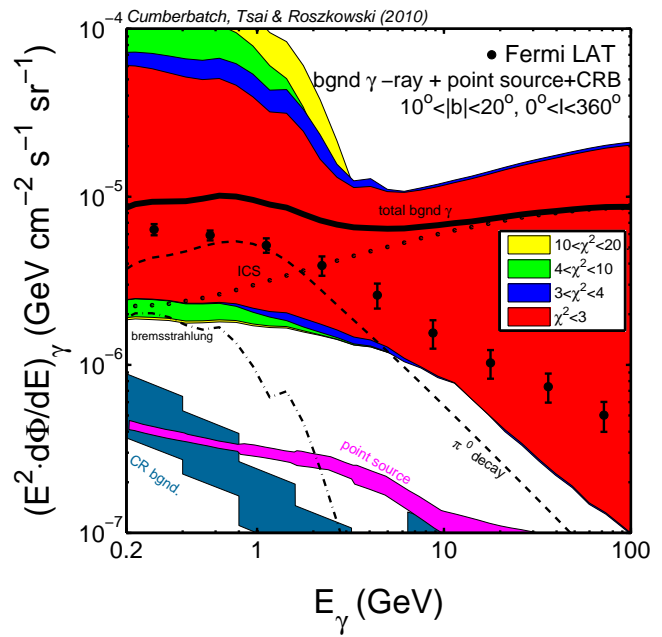}\\
	\caption{Predicted \gray flux from the MLR resulting from various astrophysical sources compared with the corresponding data from Fermi LAT (black data points, with 1$\sigma$ error bars). Predictions of the total \gray flux (grouped according to their $\chi_{\rm LAR}^2$ values as in Figure\,\ref{fig:b/c}) from the combined contributions from the SAB, point sources identified by Fermi LAT (magenta filled region, corresponding to 1\,$\sigma$ error range), and the Fermi estimated CR background contamination (teal filled region). We also display the total \gray spectra associated with the PPC generating the lowest $\chi_{\rm LAR}^2$ value (thick black curve), together with its associated spectral components.}
	\label{fig:BGND_E_chisq}
\end{figure}

\noindent In Figure\,\ref{fig:BGND_E_chisq} we display predictions for the total \gray flux (upper filled regions) from the MLR, consisting of \gray contributions from: (i) the SAB, (ii) point sources identified by Fermi LAT, and (iii) a residual particle background consisting of CRs misclassified as \grays (referred to hereafter as the {\it CR background}), as determined by the Fermi Collaboration. We compare these predictions with the corresponding measurements by Fermi LAT (black data points, with 1\,$\sigma$ error bars).

To account for point sources, we utilised the Fermi LAT point source data presented in \cite{collaboration:2010nz}, displayed in Figure\,\ref{fig:BGND_E_chisq} as the magenta filled region (corresponding to the 1\,$\sigma$ error range \cite{footnote:Fermi_systematics}). Secondly, we utilised estimates for the CR background determined by Monte Carlo simulations conducted by the Fermi Collaboration \cite{collaboration:2010nz}, displayed in Figure\,\ref{fig:BGND_E_chisq} as the teal filled region.

As mentioned above, to calculate the SAB contribution we utilised PPCs with $\chi_{\rm LAR}^2<20$ following our high resolution scan of propagation parameter space. For convenience, the spectra associated with these configurations are grouped according to their $\chi_{\rm LAR}^2$ values using an identical colour scheme to that displayed in Figure\,\ref{fig:b/c}.

In addition, following the earlier comments made in Sec.\,\ref{subsec:b/c}, for each PPC we varied the 4 propagation parameters: $B_0, z_0, e_{g_0}$\,and\,$e_{g_1}$, that were found to have no discernible influence on the value of $\chi_{\rm LAR}^2$, and hence were kept constant, but significantly affect the \gray flux from the MLR. The values of these parameters utilised to calculate the \gray flux are displayed in Table\,\ref{tab:grayparam}. 

Since the effects of solar modulation are obviously local in nature, leading to the strong $\phi$-dependence of the low energy B/C ratio observed in Figure\,\ref{fig:b/c}, they play no role in the spectrum of \grays generated by processes occurring almost entirely beyond the solar neighbourhood.
Therefore, for simplicity, the grouping of the \gray spectra displayed in Figure\,\ref{fig:BGND_E_chisq} corresponds to the values of $\chi_{\rm LAR}^2$ associated with a fixed value of $\phi=450$\,MV. To justify this, we investigated how the range of \gray fluxes changed for the PPCs yielding $\chi_{\rm LAR}^2<3$ and $3<\chi_{\rm LAR}^2<4$, for values of $\phi=100, 250$ and 450\,MV. We deduced that these changes were minimal in the Fermi LAT energy range and have no effect on our conclusions. 

\begin{table}[t]
\begin{tabular}{|c|c|c|c|}
\hline\hline
$B_0$ & $z_0$ & $e_{g_{0}}$ & $ e_{g_{1}}$\\
\hline
2.5, 5.0, 10.0 &  1.0, 2.0, 3.0 & 0.65, 1.60, 1.90 & 1.20, 2.50, 3.20 \\
\hline\hline
\end{tabular}
\caption{Ranges of the 4 propagation parameters, expressed as:
$B_{0}\,(\mu G)$, $z_{0}$\,(kpc), $e_{g_{0}}$, $e_{g_{1}}$, 
that were varied in our scan of propagation parameter space when 
calculating the \gray flux from the MLR, in addition to those 
displayed in Table\,\ref{tab:param}.}
\label{tab:grayparam}
\end{table}

From Figure\,\ref{fig:BGND_E_chisq} we can clearly see that the uncertainties in the \gray flux relating to propagation parameters, i.e., approximately 1-2 orders of magnitude in the Fermi energy range for our selected PPCs with $\chi_{\rm LAR}^2<30$, dominate those associated with other astrophysical sources, where in many cases the resulting \gray spectra far exceeds the Fermi LAT data. 
This excess can be alleviated by adjusting the normalisation of the various spectral components of the SAB, which correspond to related adjustments in the normalisation of the $e^{\pm}$ injection spectra and initial nuclear abundances.

For the spectra displayed in Figure\,\ref{fig:BGND_E_chisq} the normalisations of the $e^{\pm}$ spectra and nuclear abundances are all set equal to the default 
values specified in the {\it conventional} \texttt{galdef} parameter file.
This is because such normalisations are irrelevant when fitting the abundance ratios B/C and ~\bebe, since the predictions for these ratios associated with any given PPC are independent of absolute fluxes.
However, when making predictions for the \gray flux from the MLR, one must take into account that the normalisation of the injection spectra of electrons/positrons  as well as the abundances of light nuclei can vary.
Such variations translate into variations in the abundance of supernovae, pulsars, etc. that give rise to the initial abundance of electrons/positrons and light elements within our Galaxy.
In this study, we account for such uncertainties by allowing for a free variation of the three spectral components of the SAB: bremsstrahlung, inverse Compton and 
$\pi^0$-decay emission. 
As described in Sec.\,\ref{subsec:signal_components}, \grays resulting from bremsstrahlung and inverse Compton scattering primarily originate from the scattering of electrons/positrons with background nuclei or ISRFs. However, $\pi^0$-decay emission primarily originates from pions produced in nuclear interactions.

Consequently, here we represent changes in normalisation of the $e^{\pm}$ injection spectra by independently varying the combined bremsstrahlung+ICS flux by a factor $N$, relative to the default \texttt{GALPROP} normalisation, 
and represent the changes in the initial nuclear abundances by varying the normalisation of the $\pi^0$-decay flux by a factor $M$, relative to that corresponding to the results displayed in Figure\,\ref{fig:BGND_E_chisq}, hence, giving rise to the re-normalised SAB flux:
\begin{equation}
E_{\gamma}^2\frac{{\rm d}\Phi_{\rm SAB}}{{\rm d}E_{\gamma}}= 
N \left( E_{\gamma}^2\frac{{\rm d}\Phi_{\rm Brem.}}{{\rm d}E_{\gamma}} + E_{\gamma}^2\frac{{\rm d}\Phi_{\rm ICS}}{{\rm d}E_{\gamma}} \right) +
M E_{\gamma}^2\frac{{\rm d}\Phi_{\pi^0}}{{\rm d}E_{\gamma}}.
\label{eq:renormSAB}
\end{equation}
\begin{figure}[t]
	\includegraphics[width=\linewidth, keepaspectratio]{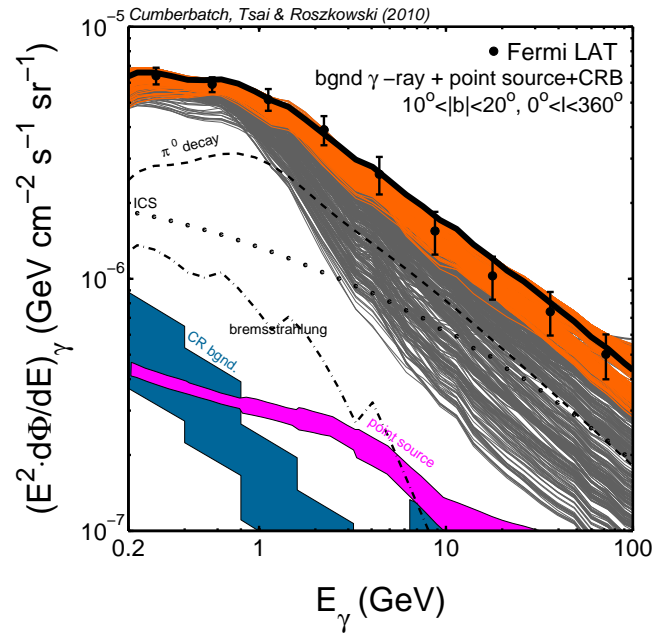}\\
	\caption{Best-fit spectra of the total \gray flux from the MLR for our selected PPCs possessing $\chi_{\rm LAR}\le4$, following 
	the fits described in the text, and grouped according to their associated $\chi_{\rm bgnd.}^2$ value: $\chi_{\rm bgnd.}^2\le4$ (orange), 
	$\chi_{\rm bgnd.}^2>4$ (grey).
	Each spectrum was fitted to the Fermi LAT data by varying the SAB components of each spectra as described in the text,
	using normalisation factors $N$ and $M$ in the range: $\frac{1}{X}\le N,M \le X$, for $X=10$. The corresponding results when using $X=5$ were found to be extremely similar (see main text). 
	We also display the total \gray flux associated with the PPC generating the smallest $\chi_{\rm bgnd.}^2$ value (thick black curve),
	together with its corresponding SAB spectral components.}
	\label{fig:Bkg_fit}
\end{figure}

In Figure\,\ref{fig:Bkg_fit}, we display the total \gray flux associated with PPCs possessing 
$\chi_{\rm LAR}^2\le4$ after renormalising their respective spectral components, as described by
Eq.\,(\ref{eq:renormSAB}), in order to obtain their best-fits to the Fermi LAT data when including the 
aforementioned contributions from point sources and CR background.
Best-fit spectra that exceeded the $1\sigma$ upper limit of the Fermi LAT data where necessarily adjusted to conform with the Fermi LAT
data whilst maintaining the best-fit possible.
The orange and grey spectra correspond to PPCs with $\chi_{\rm LAR}^2\le4$ that generate best-fit spectra possessing values of the reduced $\chi^2$ statistic $\chi_{\rm bgnd.}^2\le4$ and $\chi_{\rm bgnd.}^2>4$ respectively, where $\chi_{\rm bgnd.}^2$ is defined as
\begin{equation}
\chi_{\rm bgnd.}^2 =\frac{\sum_i^N\frac{\left(D_{i}-T_{i}\right)^2}{\sigma_{i}^2}}{n - f }.
\label{eq:chisqbkg}
\end{equation}
Here $\{D_i, \sigma_i\}$ refers to the $i$th data point of the Fermi LAT data, $T_i$ refers to 
our corresponding theoretical prediction,
$n=9$ is the number of Fermi LAT data points utilised, and $f$ is the number of degrees of freedom (see below).

Here we utilise the reduced $\chi^2$ statistic here since in what follows we are going to be directly comparing 
the fits of spectra described by different models of varying complexity. 
Consequently, we must invoke a penalty for the additional degrees of freedom possessed by each model, 
which we account for here by setting $f$ equal to the number of fitting parameters used {\it in addition} to those utilised to generate the
results displayed in Figure\,\ref{fig:BGND_E_chisq}, where no renormalisation took place.
Hence, for the current scenario, $f$ is to be set equal to the number of normalisation factors
used (i.e., $N$ and $M$), and hence, is equal to 2.

We considered two different allowed ranges for the normalisation
factors $N$ and $M$. In both cases these factors were allowed to vary within the range 
$\frac{1}{X}\le N, M \le X$. One set of results was generated using $X=10$, which
correspond to those displayed in Figure\,\ref{fig:Bkg_fit}, whilst the other set of results
was generated using $X=5$.

As can be observed from Figure\,\ref{fig:Bkg_fit}, we found that many of our selected PPCs provide excellent fits to the 
Fermi LAT data, for both $X=5$ and 10, at all relevant energies, with $\chi_{\rm bgnd.}^2$ values as low as 0.653. 
(For reference, Table\,\ref{tab:bkgonly} in Sec.\,\ref{sec:appendix} lists the values of the 
best-fit fitting parameters, together with their associated $\chi^2$ values, for the best-fitting PPCs resulting from our described fits.)

The justification of the exact choice of $X$ is a complicated question 
far beyond the scope of this paper, depending somewhat on the uncertainties relating to 
the abundance and distribution of various astrophysical sources throughout the Galaxy 
with respect to that modelled by \texttt{GALPROP}. 

Here we attempt to demonstrate the independence of our conclusions between the 
extreme choice of $X=10$ and the less extreme choice of $X=5$, provided that, 
in this scenario, we concern ourselves only with PPCs that give rise to a fitted spectra 
that survive a stringent cut of $\chi_{\rm bgnd.}^2\le4$.
Of course, as the value of $X$ increases we expect that the number of spectra that survive
such cuts will increase, since we allow for those spectra with a similar shape
to the Fermi LAT data, but with unsuitable normalisations of its SAB components, to be re-normalised to
provide the flux necessary to generate a  small $\chi_{\rm bgnd.}^2$ value.
However, one can observe that because all such spectra possess a similar shape
(owing primarily to the power law shape of the Fermi LAT data) the spread in the SAB 
contributions after renormalisation is of order 2$\sigma$ (as you would expect for
a cut of $\chi_{\rm bgnd.}^2\le4$) and hence is not very much different
when using either $X=10$ or $X=5$.

However, as a precaution we will continue to generate corresponding results for both $X=10$ and $X=5$
throughout, and attempt to explain any significant differences between them and discuss
their impact (if any) on our conclusions.


\subsection{SAB + Point Sources + EGB }
\label{subsec:bkgegbflux}

\noindent As mentioned in Sec.\,\ref{subsec:signal_components}, if one adopts an extragalactic
background that is similar in magnitude to that estimated by the Fermi Collaboration 
\cite{collaboration:2010nz}, it is proper to incorporate an EGB component in one's attempts 
to fit the Fermi LAT mid-latitude \gray data.

In this section, we present the results of our attempts to
simultaneously fit the Fermi LAT data with an EGB described by the power-law (\ref{eq:egbflux})
\begin{equation}
E_{\gamma}^2\frac{{\rm d}\Phi}{{\rm d}E_{\gamma}}=A\left(\frac{E_{\gamma}}{E_1}\right)^{\gamma},\nonumber
\end{equation}
where, for convenience, $E_1\simeq281\,$MeV is the energy of the lowest energy 
Fermi LAT data point displayed in Figure\,\ref{fig:Bkg_fit}, in addition to the SAB and point source components described in Sec.\,\ref{subsec:bkgflux}.
Once again, we utilised those PPCs with $\chi_{\rm LAR}^2\le4$ whose spectra were
used to perform the fits in Sec.\,\ref{subsec:bkgflux} (henceforth referred to as our ``bgnd.\,\,only'' fits).

We simultaneously determined the best-fit values of the parameters $A$, $\gamma$, $N$ and $M$ 
for each PPC that generated spectra not exceeding the $1\sigma$ upper limit of the Fermi LAT data,
where, once again, we restricted the values of $N$ and $M$ to the range $\frac{1}{X}\le N,M\le X$,
for both $X=5$ and 10. 
We constrained our scan to values of the EGB normalisation, $A$, and slope, $\gamma$,
to the respective ranges $10^{-8}$\,GeV\,cm$^{-2}$\,s$^{-1}$\,sr$^{-1}\le A \le \Phi_1$
and $-4\le\gamma\le0$, where $\Phi_1\simeq7\times10^{-6}$\,GeV\,cm$^{-2}$\,s$^{-1}$\,sr$^{-1}$ 
is the 1$\sigma$ upper limit of the Fermi LAT measurements of the total \gray flux from the MLR at $E_{\gamma}=E_1$. 
We restricted $A$ to values greater than $10^{-8}$\,GeV\,cm$^{-2}$\,s$^{-1}$\,sr$^{-1}$
since we regard smaller best-fit values to indicate a preferential absence of an EGB, and
restricted $A$ to values smaller than $\Phi_1$ since larger values would exceed the Fermi LAT data.
We restricted $\gamma$ to negative values since, irrespective of the various astrophysical motivations, 
it is clear from the spectral shape of the Fermi LAT data that only negative sloped EGB will benefit the 
fit to the data, and choose a maximum slope of -4 since we believe steeper slopes would be 
difficult to generate from the astrophysical sources mentioned in Sec.\,\ref{subsec:signal_components} 
that are likely to dominate the EGB.
\begin{figure}[t]
	\includegraphics[width=\linewidth, keepaspectratio]{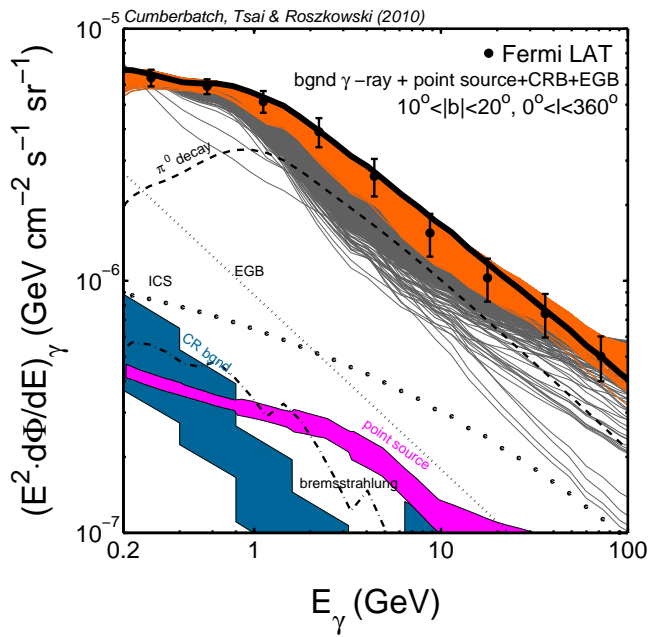}
	\caption{Same as in Figure\,\ref{fig:Bkg_fit} but now for our bgnd.+EGB fits, as described in the text. We also explicitly display the corresponding EGB flux component associated with the best-fit spectra possessing the lowest $\chi_{\rm DM}^2$ value (thin dotted curve).}
	\label{fig:EGB_fit}
\end{figure}

In Figure\,\ref{fig:EGB_fit} we display the results of our described fits (henceforth referred to as the
``Bkg\,+\,EGB'' fits), for $X=10$.
(We note that there were no significant differences between the trends displayed and the corresponding $X=5$
results that would effect our subsequent discussion, with the exception that, for $X=5$, owing to a smaller range 
for $N$ and $M$, there are fewer PPCs that do not exceed the Fermi LAT data, mostly with $\chi_{\rm EGB}^2>4$. 
This reaffirms our earlier remarks regarding the independence of our conclusions with respect to the choice of $X$, 
and hence, for clarity, we henceforth omit displaying results corresponding to $X=5$.)

Analogous to the bgnd.\,\,only results, the orange and grey spectra in Figure\,\ref{fig:EGB_fit} correspond to PPCs 
with $\chi_{\rm LAR}^2\le4$ that generate best-fit spectra possessing values of the reduced $\chi^2$ statistic $\chi_{\rm EGB}^2\le4$ and $\chi_{\rm EGB}^2>4$ respectively, 
where $\chi_{\rm EGB}^2$ is identical in form to $\chi_{\rm bgnd.}^2$ except now $T_i$ includes the EGB flux (\ref{eq:egbflux}) and $f$ has increased to 4 to account for the two additional fitting parameters $A$ and $\gamma$.
(For reference, in Table\,\ref{tab:bkgegb} of Sec.\,\ref{sec:appendix} we list the values of the 
best-fit fitting parameters, together with their associated $\chi^2$ values, 
for the best-fitting PPCs resulting from our described bgnd.+EGB fits.)

One can clearly observe from Figure\,\ref{fig:EGB_fit} that the range of fluxes displayed by the orange spectra is much less than that displayed in Figure\,\ref{fig:Bkg_fit}.
This is a consequence of the fact that $f$ has increased, increasing the value of $\chi_{\rm EGB}^2$ for
those PPCs that fail to improve their fits to the data when the additional degrees of freedom associated with the EGB are introduced.
Hence, those PPCs that possess $\chi_{\rm EGB}^2\le4$ still provide reasonable fits to the Fermi LAT
data when an EGB has been included either by: (i) virtue of the EGB (if the EGB component is
significant) or (ii) because of the excellent fit provided by the SAB component irrespective of an EGB.
Those PPCs that belong to category (i) should be associated with a significant improvement
in their fits to the Fermi LAT data when the EGB is included, compared to their corresponding bgnd.\,\,only fit. 
Such PPCs should possess negative (or very small) values of $\delta\chi_{\rm EGB}^2$ defined by
\begin{equation}
\delta\chi_{\rm EGB}^2 = \chi_{\rm EGB}^2 - \chi_{\rm bgnd.}^2.
\end{equation}
However, those PPCs which reside in category (ii) should have positive values of
$\delta\chi_{\rm EGB}^2$ owing to the corresponding increase of $f$ in $\chi_{\rm EGB}^2$
without any significant change in the predicted spectrum $T_i$ \cite{footnote:fits}.
\begin{figure}[t]
	\includegraphics[width=7.05cm, keepaspectratio]{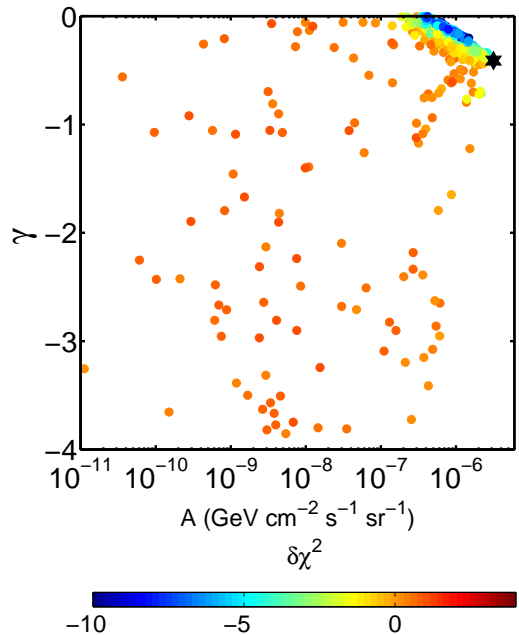}
	\caption{Best-fit values (in addition to several nearby points, as described in the text) 
	of the EGB normalisation $A$ and slope $\gamma$, together with their corresponding 
	value of $\delta\chi_{\rm EGB}^2$, for each of our selected PPCs possessing $\chi_{\rm EGB}^2\le4$
	following our bgnd.+EGB fits. We also indicate the best-fit EGB estimate by the Fermi Collaboration (black star).}
	\label{fig:deltschisqegb}
\end{figure}

In Figure\,\ref{fig:deltschisqegb}, we display the best-fit values of $A$ and $\gamma$, together
with their corresponding value of $\delta\chi_{\rm EGB}^2$, for each of our selected PPCs with 
$\chi_{\rm EGB}^2\le4$.
In order to better understand the trends displayed between $A, \gamma$ and $\delta\chi_{\rm EGB}^2$, for
each of the aforementioned best-fit points, in Figure\,\ref{fig:deltschisqegb} we have also plotted 
several points with values of $A$ and $\gamma$ close to their corresponding best-fit points, again, 
with their corresponding values of $\delta\chi_{\rm EGB}^2$.

We observe that the majority of best-fit values of $A<10^{-7}$\,GeV\,cm$^{-2}$\,s$^{-1}$\,sr$^{-1}$ 
fail to produce an improvement in the fit to the Fermi LAT data when an additional EGB component is
included. 
This owes primarily to the size of the error, $\sigma_{\rm res.}$, associated with the ``residual flux'', 
(which we set equal to $D_i$ in Eq.\,(\ref{eq:chisqbkg})\,) equal to the difference 
between the Fermi LAT measurement of the total \gray from the MLR 
and the sum of the corresponding fluxes from point sources and 
the CR background. Hence, we obtain $\sigma_{\rm res.}$ by adding the 
errors of the total, point source and CR background measurements 
($\sigma_{\rm Tot.}$, $\sigma_{\rm PS}$ and $\sigma_{\rm CRb}$ respectively) in quadrature:
\begin{eqnarray}
\sigma_{\rm res.}&=&\left(\sigma_{\rm CRB}^2+\sigma_{\rm PS}^2+\sigma_{\rm Tot.}^2\right)^{\frac{1}{2}}\nonumber\\
&\simeq&10^{-(6-7)}\,{\rm GeV\,cm}^{-2}\,{\rm s}^{-1}{\rm sr}^{-1}.\nonumber\\
\label{eq:sigmares}
\end{eqnarray}
Hence, when one attempts to simultaneously fit the Fermi LAT data using the SAB and 
an associated EGB with $A\ll\sigma_{\rm res.}$ one expects that the improvement in the fit, relative to that
when just fitting the SAB component (i.e., its ``bgnd.\,\,only'' fit), to be marginal. 
Also, since $f$ increases to 4 with our bgnd.+EGB fits we expect that PPCs that prefer
the EGB with $A\ll\sigma_{\rm res.}$ to predominantly have a positive value of $\delta\chi_{\rm EGB}^2$,
which is what we observe in Figure\,\ref{fig:deltschisqegb}.
Since such EGB components have minimal influence in the bgnd.+EGB fits, this also explains the lack of any strong correlation between the values of $A$ and $\gamma$ for $A<10^{-7}$\,GeV\,cm$^{-2}$\,s$^{-1}$\,sr$^{-1}$, indicating that the quality of the Fermi LAT data is insufficient to place constraints on components with fluxes $E_{\gamma}^2\frac{{\rm d}\Phi}{{\rm d}E_{\gamma}}\ll\sigma_{\rm res.}$.

However, correlations between the best-fit values of $A$ and $\gamma$ 
for $A\gtrsim\sigma_{\rm res.}$ allow us to place weak constraints on the 
parameters of the EGB. In Figure\,\ref{fig:deltschisqegb} we observe 
a preference for slopes with $\gamma\ge-1$, which
includes the best-fit EGB estimate by the
Fermi Collaboration: $A\simeq6\times10^{-6}$\,GeV\,cm$^{-2}$\,s$^{-1}$\,sr$^{-1}$ and
$\gamma = -0.41$ (black star) \cite{collaboration:2010nz}.
Such constraints are not surprising since they ultimately result from the slope 
of the Fermi LAT total flux measurements, which one can observe varies 
approximately within this range. 
Interestingly, the points in this region demonstrate 
a wide range of both positive and negative $\delta\chi_{\rm EGB}^2$,
indicating that there are PPCs that still provide excellent fits to the Fermi LAT
data when a significant EGB component is present. 

\section{Neutralino Dark Matter} 
\label{sec:DM} 

\noindent In this section we discuss the results obtained from our attempts to improve our 
previous fits to the Fermi LAT data by simultaneously fitting a \gray flux component from SUSY
dark matter in addition to those from the SAB and EGB.

\subsection{The Minimal Supersymmetric Standard Model} 

\noindent Supersymmetry is arguably the current most favoured
theory of particle physics beyond the Standard Model
\cite{Haber:1984rc}.  Not only can SUSY solve the hierarchy
problem \cite{Martin:1997ns}, in models where R-parity is
conserved, the lightest neutralino, assumed to be the lightest
supersymmetric particle (LSP), is a WIMP, whose total annihilation
cross section often results in a thermal relic density
$\Omega\,h^2\sim0.1$, similar to that expected for DM by
astrophysical observations \cite{Jarosik:2010iu}.  Hence, the LSP
is, in this respect, an excellent DM candidate
\cite{Jungman:1995df, Bertone:2004pz}.

Because of its large degree of freedom, owing to its 124
Lagrangian parameters \cite{Berger:2008cq}, and complicated
properties, it is a common practice to impose within the MSSM
some well-motivated boundary assumptions, usually in the form of
grand-unfication conditions on the MSSM masses and couplings. The
most economical, and popular, scenario of this type is the
Constrained MSSM (CMSSM) \cite{Kane:1993td}.

On the other hand, even in the general MSSM, only some parameters
play a real role in determining DM properties, including
detection rates and fluxes, which makes an MSSM-based analysis
managable. Furthermore, the LSP neutralino in the CMSSM (and
often in other simple unified models) is predominantly
gaugino-like and, with LEP constraints imposed, heavier than
roughly 100\,GeV, in which case, resulting
\gray fluxes will be uninterestingly small. For these reasons, we
conduct our analysis in the framework of the general MSSM. On the
other hand, it is not our intention here to investigate the MSSM
in detail, but rather to present typical fluxes resulting in the
model, along with the uncertainties associated with the
ambiguities in the propagation parameters describing the
diffusion of CRs. 

Therefore, in this paper we select three representative
cases of the neutralino in the way described below. Firstly, we narrow down the list of MSSM parameters 
to:
\begin{eqnarray}\label{eq:MSSMp}
    &&M_{2},~~\mu,~~m_{A},~~{\rm tan}\beta,\nonumber\\
    &&m_{\tilde{L}},m_{\tilde{E}},~~ m_{\tilde{Q}},~~m_{\tilde{U}},~~m_{\tilde{D}},\nonumber\\
    &&A_{\tau}, ~~A_b,~~A_t,\nonumber\\ 
\end{eqnarray}  
where $M_{2}$ denotes the soft wino mass, $\mu$ - the Higgs/higgsino
mass parameter, $m_{A}$ - the pseudoscalar Higgs mass, ${\rm
tan}\beta$ - the ratio of the vacuum expectation values of the
two Higgs fields , $m_{\tilde{I}}$ ($I=L,E,Q,U,D$) - the soft
masses of the slepton and squark doublet and singlet fields, respectively, and
finally $A_{i}$ ($i=\tau,b,t$) - the corresponding trilinear
couplings. (The remaining parameters not specified here are dependent on the values of those mentioned above, 
and related according to well-known equations, see, e.g.,\,\cite{Martin:1997ns}.)

In our choice of the above parameters we were guided by only varying
those which produce the largest effect on significant indirect
detection parameters, such as the thermally-averaged product of the DM
annihilation cross section and relative velocity,
$\langle\sigma_{\rm ann.}\upsilon\rangle$, the yield of \gray/$e^{\pm}$ per
annihilation, $\frac{{\rm d}N_{\gamma/e^{+}}}{{\rm d}E}$, and the mass
of the lightest neutralino, $\chi_1^0$, $m_{\chi}$, while
simultaneously obeying the constraints on the DM relic abundance.

We performed a scan of the above MSSM parameters over the ranges as specified in Table\,\ref{tab:MSSMp}.
We generated a MSSM chain, consisting of 90,000 points, by taking a flat prior and 
peforming a nested-sampling scan using a modified version of the
publicly available \texttt{SuperBayeS} package
\cite{superbayes}. 
(An even larger scan over 25 input parameters in the general MSSM was
performed in \cite{AbdusSalam:2009qd}, although not in the context of
indirect detection.) We took into consideration all current collider limits (see Tables\,1
and 2 in \cite{Roszkowski:2009sm} for details), as well as invoking
constraints published by the WMAP Collaboration regarding  the
cosmological DM relic abundance \cite{Jarosik:2010iu}. 
\begin{table}[t]
\begin{tabular}{|c|c|c|}
\hline
\hline
$0<M_{2}<2000$ & $50<\mu<1000$ & $0<m_{A}<1000$  \\
\hline
$2<{\rm tan}\beta<65$ & $0<m_{\tilde{I}}<2000$& $-3000<A_{i}<2000$ \\
\hline
\hline
\end{tabular}
\caption{Values, in GeV, of various masses and couplings, defined in the main text, used in our scan of MSSM parameter space.  We take flat prior and likelihood ranges 
were shown on above.}
\label{tab:MSSMp}
\end{table}


\subsection{Point selection}

\noindent The \gray flux originating from the particle cascade immediately following
each annihilation (which is almost entirely due to $\pi^{0}$-decay), could 
be considerable when observing the MLR. 
Since \grays are not subject to diffusion processes within the ISM, 
the differential \gray flux arriving from a line-of-sight (l.o.s.) 
at inclination $\psi$ relative to the direction of the GC is simply given by
\begin{equation}
E_{\gamma}^2\frac{{\rm d}\Phi_{\gamma}}{{\rm d} E_{\gamma}} (E_{\gamma}, \psi) = 
\frac{\langle\sigma_{\rm ann.}\upsilon\rangle}{8\pi m_{\chi}^{2}}E_{\gamma}^2\frac{{\rm d} N_{\gamma}}{{\rm d}
  E_{\gamma}}\int_{\text{l.o.s.}} \rho_{\chi}^2 {\rm d}l. 
\label{eq:diffgammaflux} 
\end{equation}
We see that all the particle physics information regarding DM is conveniently contained
within the factor $\langle\sigma_{\rm ann.}\upsilon\rangle m_{\chi}^{-2}\frac{{\rm d}N_{\gamma}}{{\rm d}E_{\gamma}}$.
Hence, for convenience, we define here the quantity
\begin{eqnarray}
f_{{\rm SUSY},i}&=&\left(\frac{\langle\sigma_{\rm ann.}\upsilon\rangle}{3\times10^{-26}\,{\rm cm}^{3}\,{\rm s}^{-1}}\right)
\left(\frac{100\,{\rm GeV}}{m_{\chi}}\right)^2\nonumber\\
&&\times\left(\frac{F_i(E,m_{\chi})}{1\,{\rm GeV}}\right), 
\label{eq:susyf}
\end{eqnarray} 
where $i=\gamma, e^{\pm}$, $F_{\gamma}(E,m_{\chi})$ is defined as the maximum value of $E_{\gamma}^{2}\frac{{\rm d}N_{\gamma}}{{\rm d}E_{\gamma}}$
within the energy range 0.1\,GeV$<E_{\gamma}<m_{\chi}$, 
and $F_{e^+}(E,m_{\chi})$ is the value of 
$m_{\chi}\int_{0.1\,{\rm GeV}}^{m_{\chi}} \frac{{\rm d}N_{e^{+}}}{{\rm d}E_{e^{+}}} {\rm d}E_{e^{+}}$,
where the lower energy limit of 0.1\,GeV associated with both quantities corresponds to the 
approximate minimum energy of the Fermi LAT sensitivity range.  

From the above, it is clear that we should expect DM candidate points possessing larger values of $f_{{\rm SUSY}, \gamma}$ to
generate larger \gray fluxes within the Fermi sensitivity range 
that originate from the particle cascade following each annihilation. 
However, in addition, since a large proportion of \grays are expected to arise from
interactions involving $e^{\pm}$ generated by DM annihilations,
we expect that the most optimistic \gray fluxes will be produced by
candidate points that generate larger yields of $e^{\pm}$
involved in producing \grays with energies in the Fermi LAT sensitivity range,
i.e.,~candidate points that possess larger values of   
$\int_{0.1{\rm GeV}}^{m_{\chi}}\frac{{\rm d}N_{e^{+}}}{{\rm d}E_{e^{+}}} {\rm d}E_{e^{+}}$.
However, with regards to the $e^{\pm}$ yields generated, rather than invoking a selection
criterion for candidate points involving, for example, 
$\langle\sigma_{\rm ann.}\upsilon\rangle m_{\chi}^{-2}\int_{0.1{\rm GeV}}^{m_{\chi}} \frac{{\rm d}N_{e^{+}}}{{\rm d}E_{e^{+}}} {\rm d}E_{e^{+}}$ alone,
which preferentially selects candidate points with a high yield of $e^{\pm}$ within the specified energy range, using
$f_{{\rm SUSY}, e^{\pm}}=\langle\sigma_{\rm ann.}\upsilon\rangle m_{\chi}^{-1}\int_{0.1{\rm GeV}}^{m_{\chi}} \frac{{\rm d}N_{e^{+}}}{{\rm d}E_{e^{+}}} {\rm d}E_{e^{+}}$ 
preferentially selects points with a high yield of $e^{\pm}$ {\it and} higher
mass neutralinos, which generally produce a greater number of (higher energy)
$e^{\pm}$. These $e^{\pm}$ contribute significantly more to the rate of ICS and bremsstrahlung
in the Fermi LAT sensitvity range, which, as we shall see, dominates the total DM \gray
flux from the MLR. 


Since we desire our benchmark models to adequately reflect the variation of $f_{{\rm SUSY},\gamma}$ and $f_{{\rm SUSY},e^{\pm}}$,
as well as providing an observable \gray flux in the direction of the MLR, 
given the above, we selected the two candidate points from our MSSM chain with masses 
$m_{\chi}\simeq100$\,GeV and 50\,GeV corresponding to the points with approximately the largest and smallest
values of $f_{{\rm SUSY}, \gamma}$ and $f_{{\rm SUSY}, e^{\pm}}$ respectively.
For our third benchmark candidate point, we selected the point corresponding
to the largest posterior probability with respect to all experimental constraints utilised in
generating our MSSM chain (denoted hereafter as our {\it best-fit} candidate point).

\begin{table}[t]
\begin{center}
\begin{tabular}{|c|c|c|c|}
\hline\hline
Parameter& best-fit & gaugino & mixed \\
\hline\hline
$M_{2}$ (GeV)& 283.6 & 104.8 & 224.5 \\
\hline
$\mu$ (GeV)& 201.5 & 305.8 & 165.0 \\
\hline
$m_{A}$ (GeV)& 982.9 & 471.8 & 856.2 \\
\hline
tan$\beta$ & 25.6 & 24.3 & 39.8 \\
\hline
$m_{\tilde{L}}$ (GeV)& 408.7 & 510.0 & 692.0 \\
\hline
$m_{\tilde{E}}$ (GeV)& 1486.5 & 292.2 & 627.5 \\
\hline
$m_{\tilde{Q}}$ (GeV)& 1548.1 & 699.4 & 1964.4 \\
\hline
$m_{\tilde{U}}$ (GeV)& 1452.7 & 1393.5 & 1150.3 \\
\hline
$m_{\tilde{D}}$ (GeV)& 1222.0 & 1153.6 & 1162.0 \\
\hline
$A_{t}$ (GeV)& 1534.0 & 328.4 & 1355.5 \\
\hline
$A_{b}$ (GeV)& 1823.3 & 95.5 & -316.2\\
\hline
$A_{\tau}$ (GeV)& -46.0 & 1289.0 & -790.6 \\
\hline\hline
$m_{\chi}$ (GeV)& 132.6 & 52.2 & 100.6 \\
\hline
$g_{f}$ & 0.733 & 0.973 & 0.686 \\
\hline
$\Omega_{\chi}h^{2}$ & 0.112 & 0.116 & 0.087 \\
\hline
$\langle\sigma\upsilon\rangle$ (cm$^{3}$s$^{-1}$)& $1.75\times 10^{-26} $ & $3.24\times 10^{-28} $ & $2.06\times 10^{-26} $\\
\hline
$f_{{\rm SUSY},{\gamma}}$ & $5.90 $ & $0.29 $ & $9.27 $\\
\hline
$f_{{\rm SUSY},{e^+}}$& 0.29  & 0.03 & 0.51\\
\hline\hline
\end{tabular}
\end{center} 
\caption{Table displaying the properties of our three benchmark MSSM DM candidate points. 
All three points are consistent with our experimental constraints within the 95\% C.L.}
\label{tab:susyparam}
\end{table}

The properties of our three benchmark candidate points are displayed in Table\,\ref{tab:susyparam}. All three points are 
consistent with our experimental constraints to within the 95\% C.L.

\subsection{Dark matter halo profiles}
\label{subsec:halo}
\noindent To calculate our results we adopt two very different, spherically symmetrical  
halo profiles to describe the DM density within the Galactic halo: the Einasto 
profile \cite{einasto89} and the Burkert profile \cite{burkert:1995}. 
Owing to indications that cuspy profiles are inconsistent with observations, 
specifically regarding the rotation curves of small-scale galaxies 
\cite{flores1994,moore1994,weldrake2003,donato2004,gentile2007, salucci:2000}, 
which are more likely to be consistent with density profiles possessing flattened cores,
we also consider the Burkert density profile \cite{burkert:1995}
\begin{equation}
\rho_{\rm Bur.}(r)=\frac{\rho_s}{ \left[1+(r/r_s)\right]\left[ 1+(r/r_s)^{2}\right]}.
\label{eq:burkert}
\end{equation}
Here we use $r_s=11.68$\,kpc and $\rho_s=0.79$\,GeV\,cm$^{-3}$, where once again, the latter is 
determined by normalising to the local DM density. The values of these parameters are 
consistent with a whole set of dynamical informations (see, e.g.,~\cite{edsjo:2004} and references therein), including, constraints from the trajectories of stars in the solar neighbourhood, estimates of the total mass of the Galaxy from the motion of satellites in its outer regions, and the Galactic rotation curve. The Einasto profile can be described by 
\begin{equation}
\rho_{\rm Ein.}(r)=\rho_s\,{\rm exp}\left[-\frac{2}{\alpha}\left\{\left(\frac{r}{r_s}\right)^{\alpha}-1\right\}\right], 
\label{eq:einasto}
\end{equation} 
where $r$ is the distance from the GC. 
We use best-fit values for the scale radius $r_{s}=21.5\,{\rm kpc}$ and the slope 
$\alpha=0.17$, determined from recent numerical simulations of the Galactic halo 
\cite{Diemand:2008in}. 
Here, we normalise Eq.\,(\ref{eq:einasto}) to a local density of 0.3\,GeV\,cm$^{-3}$ at $r=r_{\odot}\simeq8.5\,{\rm kpc}$, 
giving rise to a normalisation density factor $\rho_{s} = 0.0538$\,GeV\,cm$^{-3}$.   

\subsection{Dark matter component}
\label{subsec:dmflux}

\noindent In this section we present the results of our attempts to simultaneously fit a \gray flux component from each of our DM candidate points, in addition to the SAB and EGB, to the Fermi LAT data.
\begin{figure}
	\includegraphics[width=6.2cm, keepaspectratio]{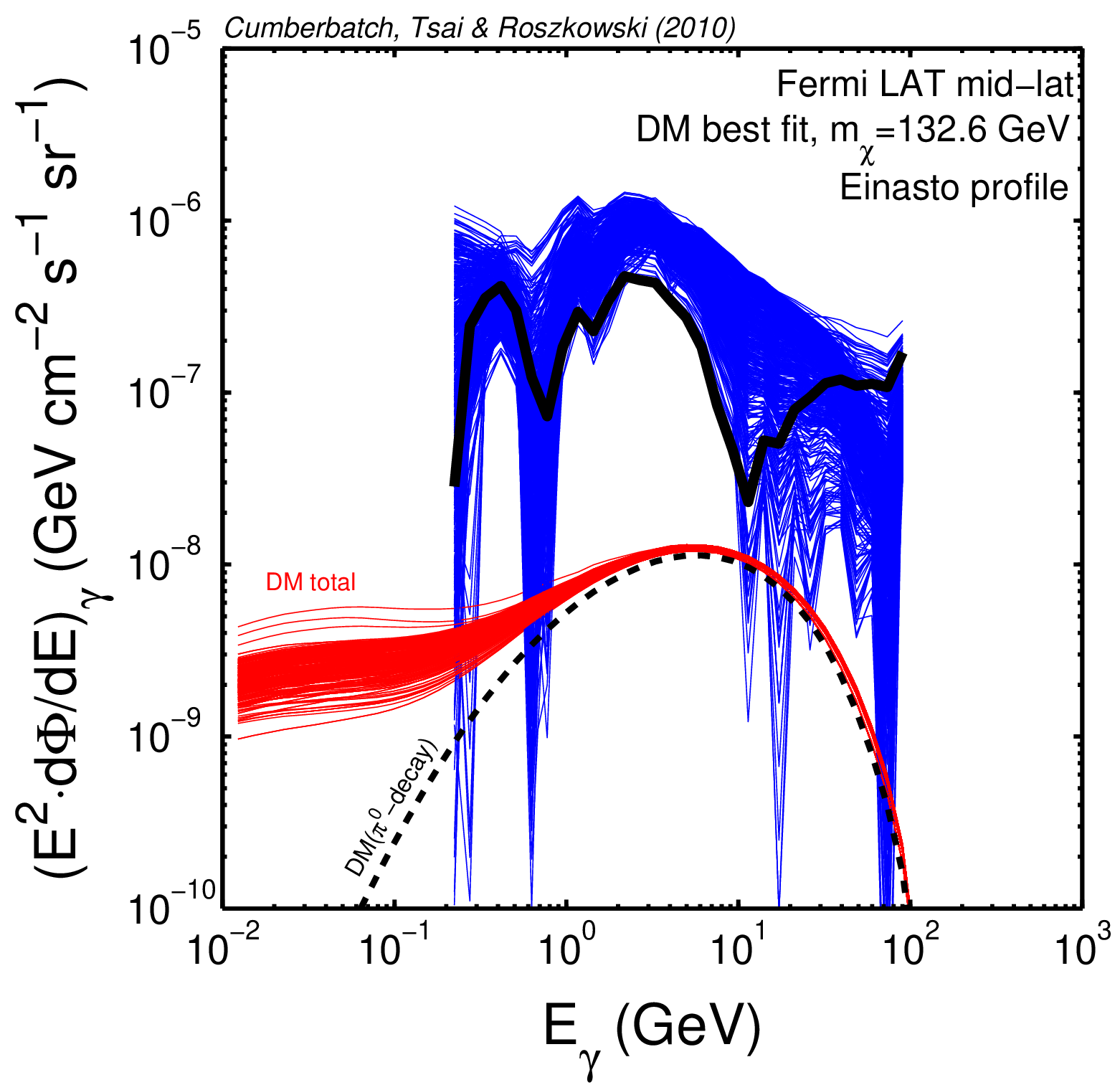}\\
	\includegraphics[width=6.2cm, keepaspectratio]{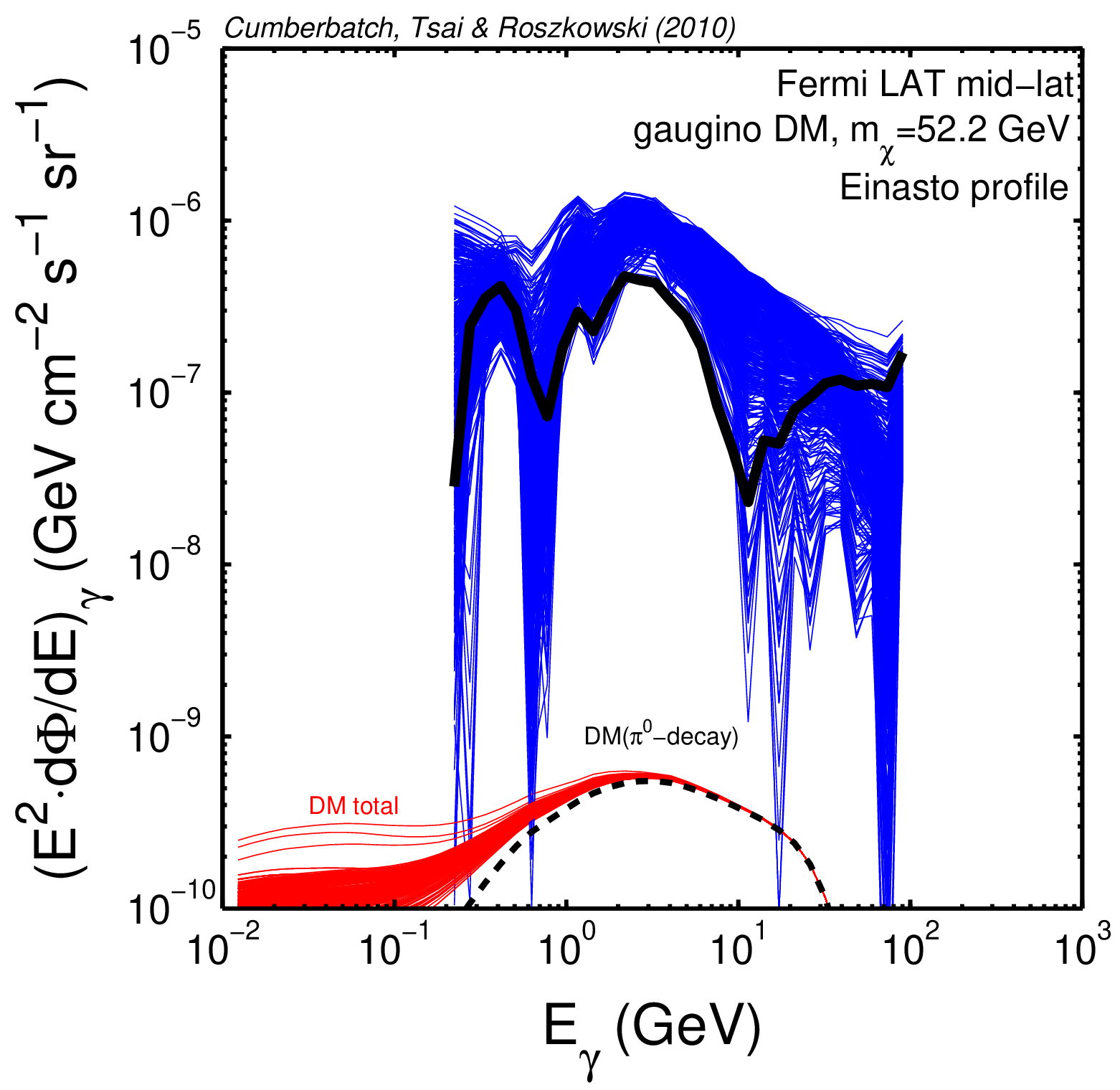}\\
	\includegraphics[width=6.2cm, keepaspectratio]{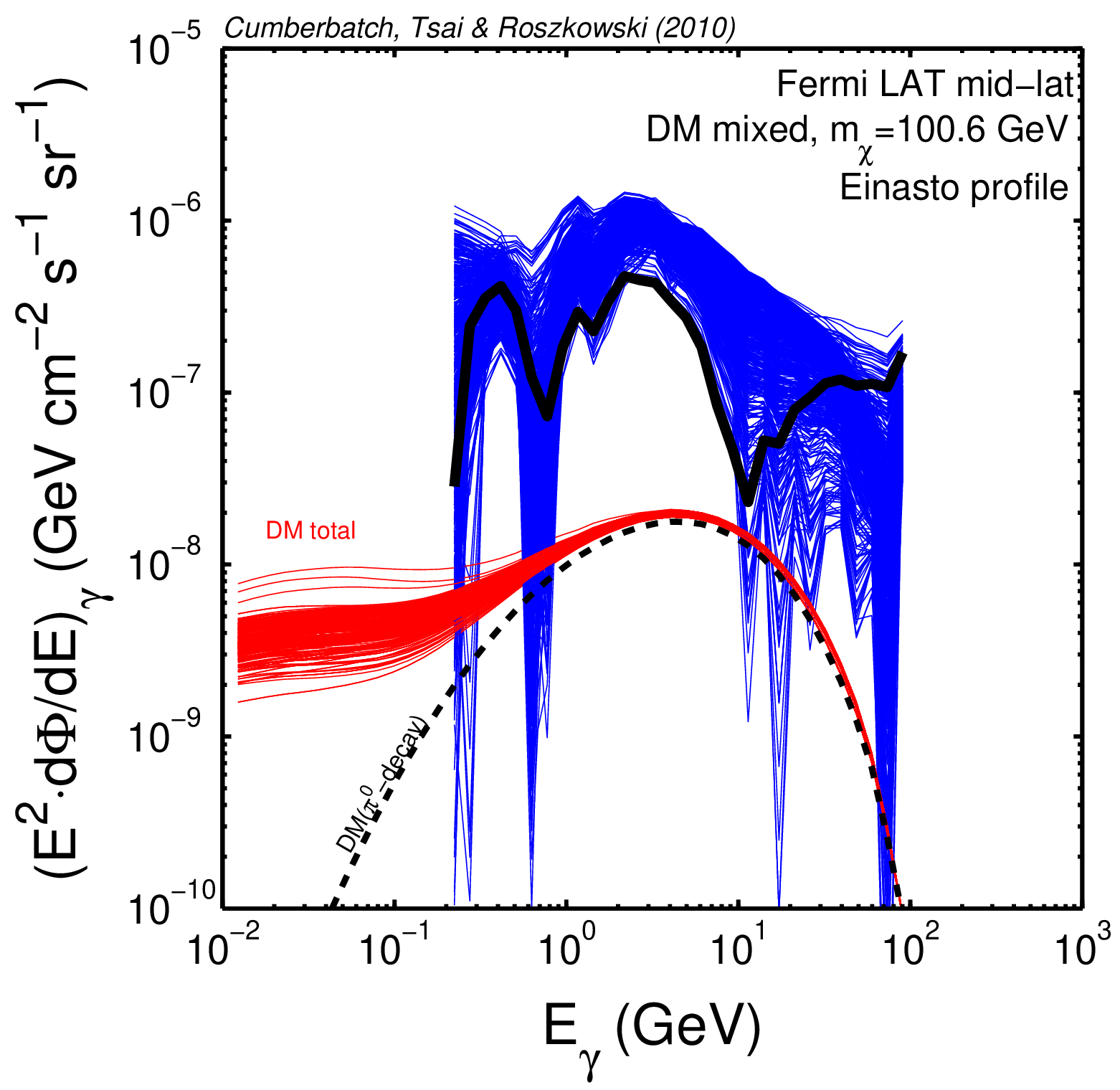}
	\caption{Results for the unaccounted for, residual flux (blue curves) following the bgnd.+EGB fits described in Sec.\,\ref{subsec:bkgegbflux} for
	PPCs with best-fit spectra possessing $\chi_{\rm EGB}^2\le4$. For comparison, we superimpose 
	the corresponding (un-boosted) DM \gray fluxes (red curves) obtained when using our best-fit (upper), gaugino (centre) and mixed (lower) 
	DM candidate points, together with an Einasto DM density profile. 
	In each plot we also indicate the $\pi^0$-decay spectral
	component associated with the displayed flux from each respective DM candidate point (dashed black curve).
	We also display the residual flux associated with the PPC whose
	best-fit spectrum possessed the smallest $\chi_{\rm EGB}^2$ value following our fits (solid black curve).}
	\label{fig:residual}
\end{figure}

In Figure\,\ref{fig:residual}, we display the residual (i.e. unaccounted for) flux 
following the bgnd.+EGB fits described in Sec.\,\ref{subsec:bkgegbflux} for PPCs with 
best-fit spectra possessing $\chi_{\rm EGB}^2\le4$ (blue curves).
For comparison, we superimpose the corresponding un-boosted DM \gray fluxes (red curves)
obtained when using our best-fit (upper), gaugino (centre) and mixed (lower) 
DM candidate points, and a Galactic DM density distribution described by our Einasto profile 
(see below for a discussion regarding the corresponding results for our Burkert profile).

We can observe that for energies $E_{\gamma}\lesssim1\,{\rm GeV}$, it is inaccurate to approximate the DM \gray flux to its $\pi^0$-decay component, given approximately by Eq.(\ref{eq:diffgammaflux}), which is almost entirely independent of propagation parameters.
In each panel of Figure\,\ref{fig:residual}, we explicitly illustrate the \gray flux from $\pi^0$-decays (black dashed curves). We see that for energies $E_{\gamma}\gtrsim1\,$GeV, the $\pi^0$-decay flux dominates the total \gray flux, and since this contribution is generated local to its source, the associated errors relating to uncertainties in CR propagation are negligible.
However, for energies $E_{\gamma}\lesssim1\,{\rm GeV}$, the $\pi^0$-decay component has virtually no impact on the size or shape of the total \gray flux from DM, which is now dominated by bremsstrahlung and inverse Compton radiation, each with uncertainties relating to CR propagation.

Clearly, one can see that the DM component, whilst being consistent with the residual flux at all but a few narrow ranges of energies (that can be compensated for by appropriately adjusting one or more of our fitting parameters) is at least a factor of 10-100 times smaller than its corresponding residual flux.
Hence, if the \gray flux from our DM candidate points are to have any significant effect when attempting to fit the Fermi LAT data, it is clear that such components require substantial enhancement.
To do so, we re-normalised the un-boosted DM flux by the commonly adopted {\it boost factor}, BF $\ge1$.

Therefore, for each of the six combinations of our three DM candidate points and two DM density profiles, we simultaneously determined the best-fit values of the parameters $N$, $M$, $A$, $\gamma$ and BF, for BF$\ge1$, and using the same ranges of values for the other fitting parameters as described in Sec.\,\ref{subsec:bkgegbflux}. 
\begin{figure}[t]
	\includegraphics[width=\linewidth, keepaspectratio]{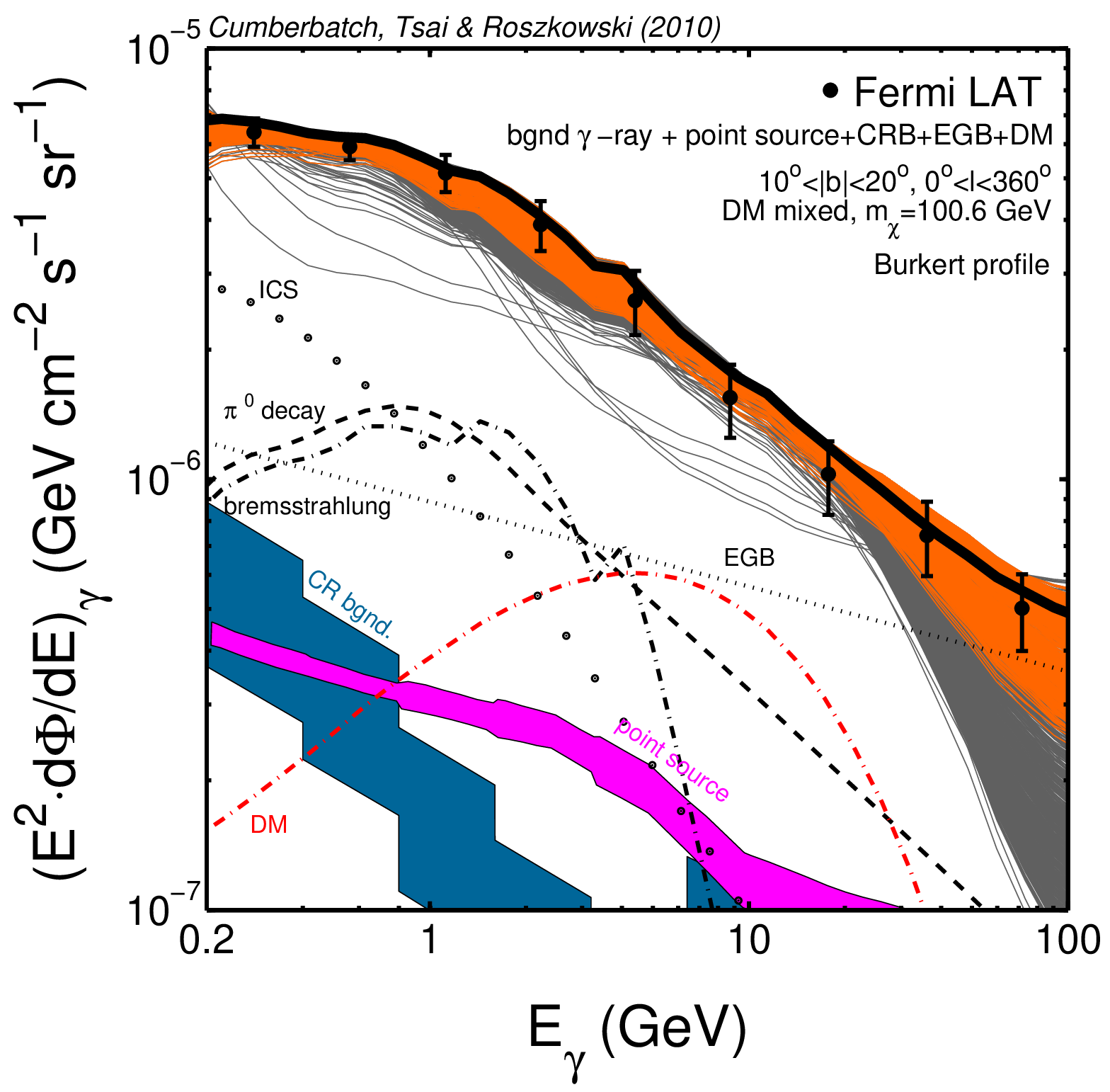}\\
	\includegraphics[width=\linewidth, keepaspectratio]{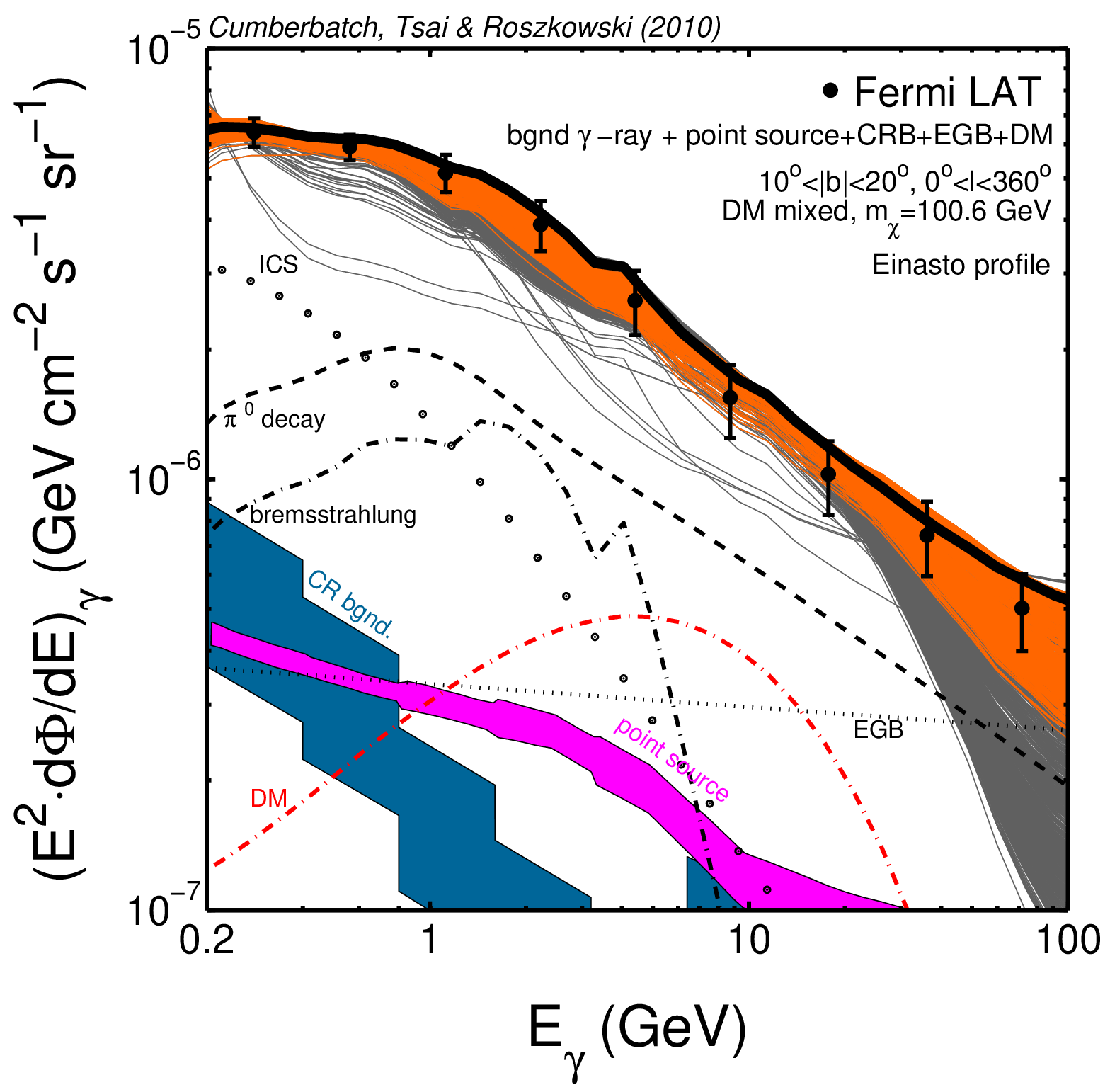}
	\caption{Same as in Figure\,\ref{fig:EGB_fit} but now for our bgnd.+EGB+DM fits, as described in the text, when using our mixed DM candidate point, together with a Burkert (upper panel) or Einasto (lower panel) DM density profile. In each plot we also explicitly display the corresponding DM flux component associated with the best-fit spectra possessing the lowest $\chi_{\rm DM}^2$ value (orange dot-dashed curves).}
	\label{fig:bkgegbdm}
\end{figure}

In Figure \ref{fig:bkgegbdm}, we display the results of our fits (henceforth referred to as the ``bgnd.+EGB+DM'' fits), for $X=10$, for our mixed DM candidate point when using our Burkert (upper panel), and Einasto (lower panel) DM density profiles. 
(Once again, for the reasons discussed in Sec.\,\ref{subsec:bkgegbflux}, we omit the corresponding results for $X=5$.)
Owing to the flexibility of our fitting procedure and the ultimately subdominant effects of the DM component the plots corresponding to the other two DM models are similar to that displayed, and hence, for clarity, we omit them here.
Further, we observe that because of these reasons, the results displayed, corresponding to the mixed DM candidate point for the Burkert and Einasto profiles are also extremely similar.

Analogously with the bgnd.\,\,only and bgnd.+EGB fits, the orange and grey spectra in Figure\,\ref{fig:bkgegbdm} correspond to PPCs with $\chi_{\rm LAR}^2\le4$ and best-fit spectra possessing values of the reduced $\chi^2$ statistic $\chi_{\rm DM}^2\le4$ and $\chi_{\rm DM}^2>4$ respectively, where $\chi_{\rm DM}^2$ is identical in form to $\chi_{\rm EGB}^2$ except now $T_i$ includes the DM \gray flux and $f$ has increased to 5 to account for the boost factor BF. 
(For reference, in Tables\,\ref{tab:bkgegbdmbestfitburkert}-\ref{tab:bkgegbdmmixedeinasto} of Sec.\,\ref{sec:appendix} we list the values of the best-fit fitting parameters, together with their associated $\chi^2$ values, 
for the best-fitting PPCs resulting from our described bgnd.+EGB+DM fits.)

Consequently, as described in Sec.\,\ref{subsec:bkgegbflux} with regards to the bgnd.\,\,only and bgnd.+EGB results, because of this further increase in $f$ we once again observe a noticeable reduction in the spread of the best-fit spectra with $\chi_{\rm DM}^2\le4$ compared to that corresponding to the bgnd.+EGB results.
\begin{figure}[t]
	\includegraphics[width=4.2cm, keepaspectratio]{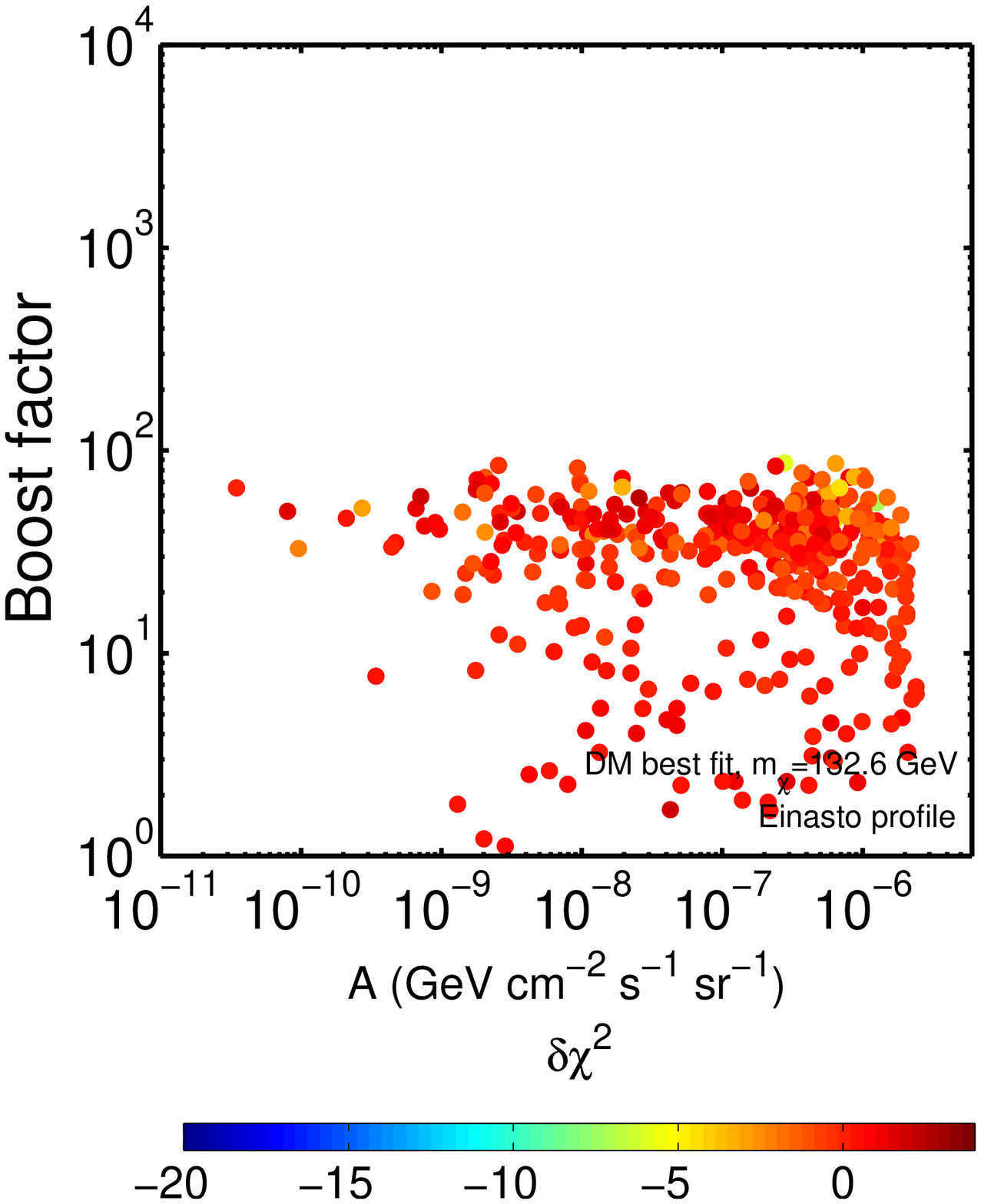}
	\includegraphics[width=4.2cm, keepaspectratio]{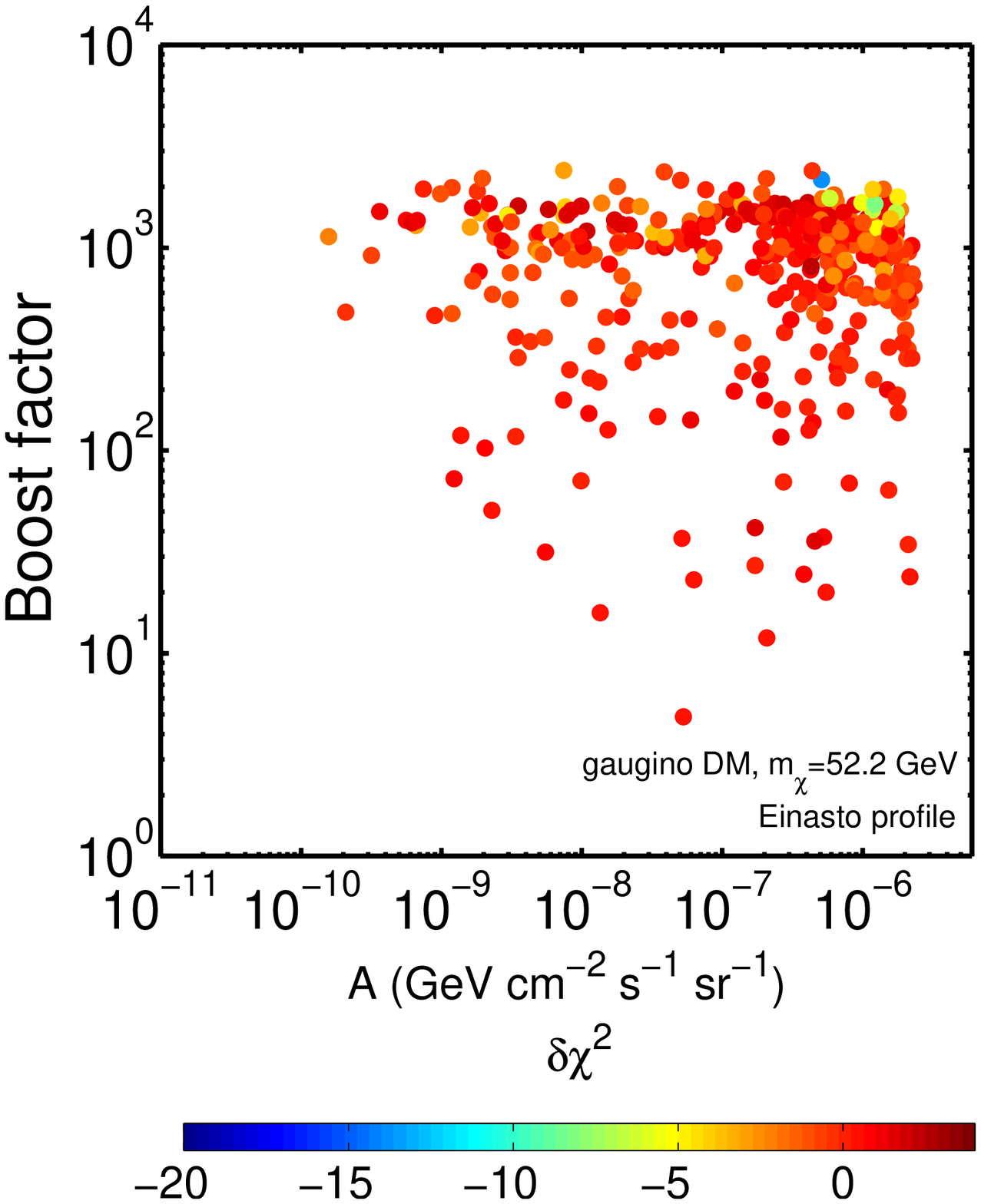}\\
	\includegraphics[width=4.2cm, keepaspectratio]{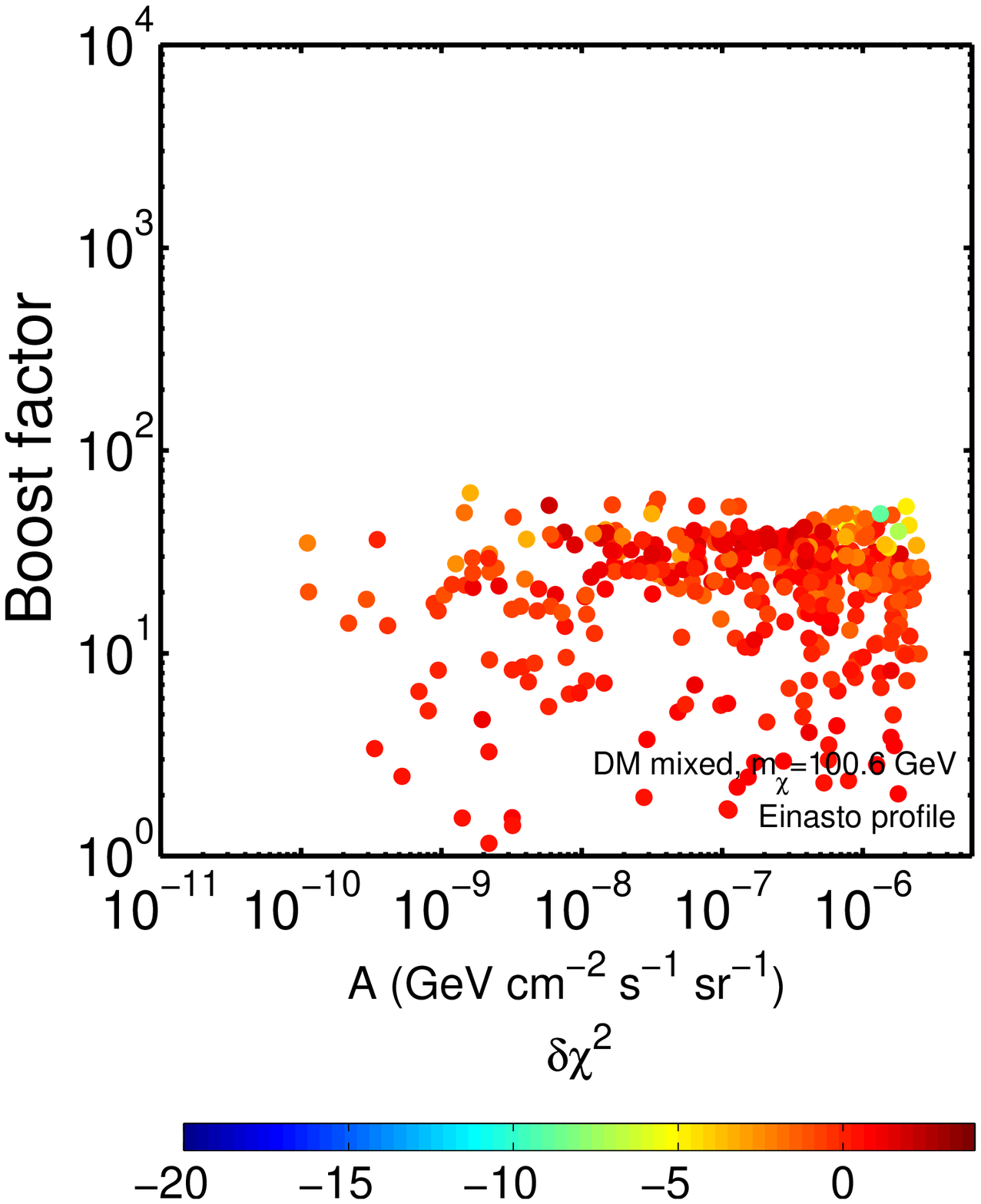}
	\caption{Results for the values of the boost factor BF, EGB normalisation $A$, and corresponding value of $\delta\chi_{\rm DM}^2$ associated with the best-fit points resulting from our bgnd.+EGB+DM fits, when using an Einasto DM density profile, together with our best-fit (upper left panel), gaugino (upper right panel) and mixed (lower panel) DM candidate points.}
	\label{fig:bfvsavsdeltachisq}
\end{figure}

To understand the impact of the DM flux in our bgnd.+EGB+DM fits, and to determine whether the presence of a DM component has actually improved the fit, one needs to look at the correlation between the preferred value of the boost factor BF and the associated improvement in the fit. 
Analogously to the bgnd.+EGB fits, we assess this level of improvement through the quantity $\delta\chi_{\rm DM}^2$, defined by 
\begin{equation}
\delta\chi_{\rm DM}^2 = \chi_{\rm DM}^2 - \chi_{\rm EGB}^2,
\end{equation}
where negative values of $\delta\chi_{\rm DM}^2$ represent an improvement in the reduced $\chi^2$ value when the DM contribution is included.

Hence, in Figure \ref{fig:bfvsavsdeltachisq}, we display the best-fit points from our bgnd.+EGB+DM fits, with the associated values of BF and $\delta\chi_{\rm DM}^2$ indicated, when using our Einasto profile. In order to compare the relative importance of the DM and EGB components in the best-fit spectra, we also plot corresponding values of $A$ on a third (colour scale) axis.

We can clearly observe that the DM component has a less significant effect in improving the bgnd.+EGB fits than the EGB did for the bgnd.\,\,only fits, with the majority of points possessing positive $\delta\chi_{\rm DM}^2$, with minimum $\delta\chi_{\rm DM}^2$ values of -8.95, -6.9 and -8.0 for the mixed, best-fit and gaugino models respectively in Einasto Profile.

We also observe that the distribution of points is very similar for each DM model, where in each case there is a strong preference for a small range of boost factors approximately equal to 40, 1500 and 30 for the best-fit, gaugino and mixed DM models respectively.
Such correlations are expected since, analogous to our discussion in Sec.\,\ref{subsec:bkgegbflux} regarding the best-fit values of $A$ following our bgnd.+EGB fits, these are the approximate boost factors necessary to generate respective DM fluxes of order $\sigma_{\rm res.}$, given by Eq.\,(\ref{eq:sigmares}). 
Hence, we do not expect there to be a strong preference for boost factors much less than these preferred ranges, with upper limits on BF determined by the fact that the total theoretical flux not violate the Fermi LAT data.

The distribution of points corresponding to the use of a Burkert profile are extremely similar to those displayed in Figure\,\ref{fig:bfvsavsdeltachisq}. 
Their relationship can be approximately obtained by rescaling the values of BF displayed in Figure\,\ref{fig:bfvsavsdeltachisq} by the ratio of the integrals of $\rho_{\chi}^2$ along line of sights, averaged over the MLR, when using the Burkert and Einasto profiles.
This ratio is given by
\begin{equation}
\frac{\int_{\rm MLR} {\rm d}\Omega\int_{\text{l.o.s.}} \rho_{\rm Bur.}^2(r,\psi){\rm d}l}{\int_{\rm MLR} {\rm d}\Omega\int_{\text{l.o.s.}} \rho_{\rm Ein.}^2(r,\psi){\rm d}l}\simeq0.30,
\label{eq:dmgammaratio}  
\end{equation}
and is identical to the ratio of the respective DM $\pi^0$-decay components when using these two profiles.
This relationship follows because, as can be observed from Figure\,\ref{fig:residual}, the energies at which the $\pi^0$-decay component dominates the DM flux is very similar to those at which the displayed residual flux is calculated.
Consequently, since the shape of the $\pi^0$-decay spectrum is independent of the DM density profile, for a given PPC, the ratio of the BF's corresponding to the Einasto and Burkert profiles should be approximately equal to the ratio (\ref{eq:dmgammaratio}).

Since the DM annihilation rate is proportional to the square of the DM density, a possible source of the above enhancements could partially originate from DM substructures.
Recent semi-analytical studies indicate that boost factors of up to 20 may arise from substructures present in the Galactic halo \cite{Lavalle:2006vb}. 
From Figure\,\ref{fig:bfvsavsdeltachisq} we observe that the mixed and best-fit candidates together with an Einasto profile are the only two of our scenarios that generate sufficient flux to significantly effect the bgnd.+EGB+DM fits when using BF$\,\simeq20$.
One should bear in mind however that since it is likely that the distribution of substructures in the Galactic halo is not congruent with the smooth DM density profile \cite{Diemand:2008in}, the resulting boost factor of the DM \gray flux will not be spatially independent and hence will alter the shape of the DM spectrum in a non-trivial way.


In Figure\,\ref{fig:bfvsavsdeltachisq}, as we might expect, we observe that points with the most negative values of $\delta\chi_{\rm DM}^2$ have a slight preference for larger values of the EGB normalisation where the EGB has an increasingly significant effect on the bgnd.+EGB+DM fits.
\begin{figure}[t]
	\includegraphics[width=4.2cm, keepaspectratio]{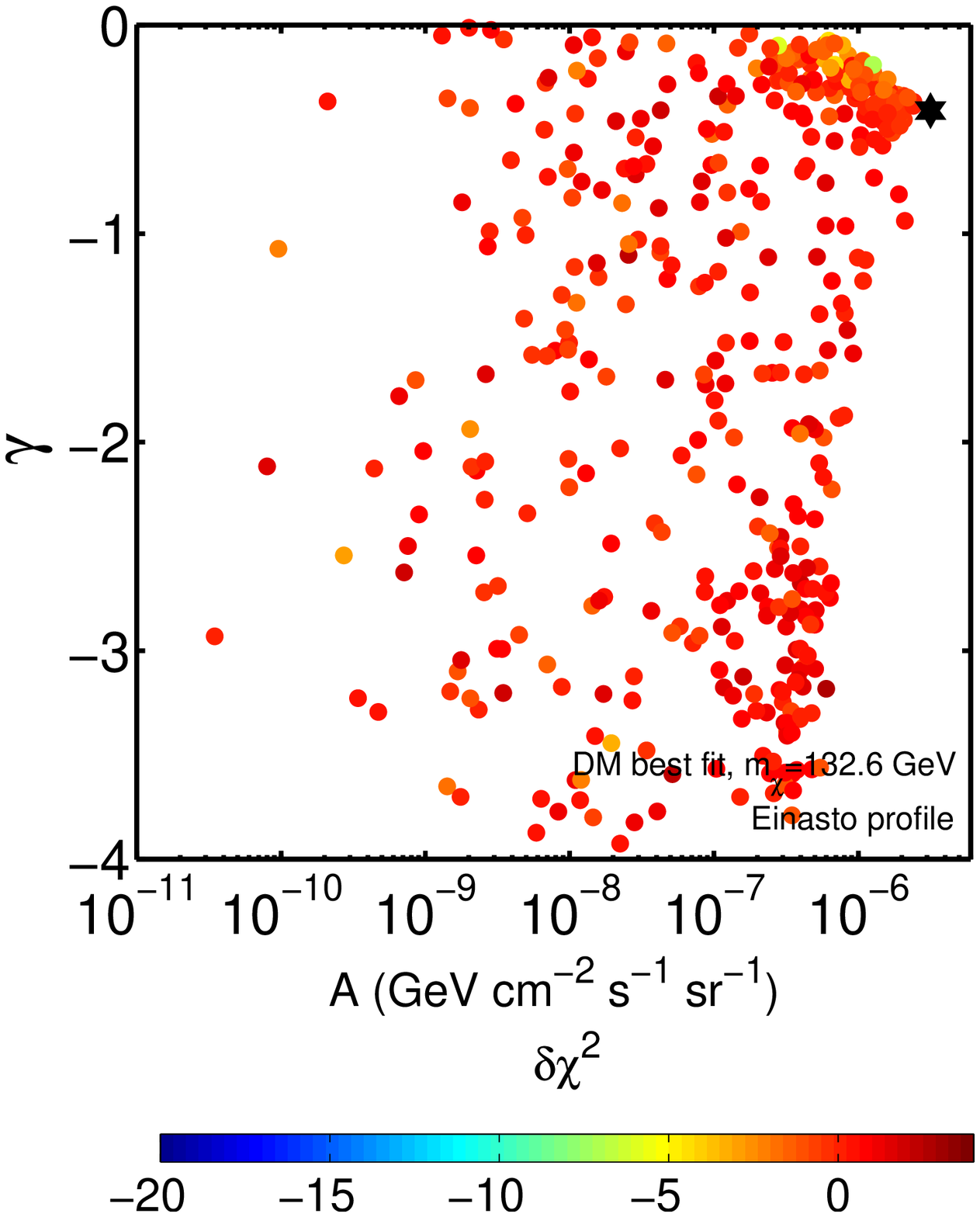}
	\includegraphics[width=4.2cm, keepaspectratio]{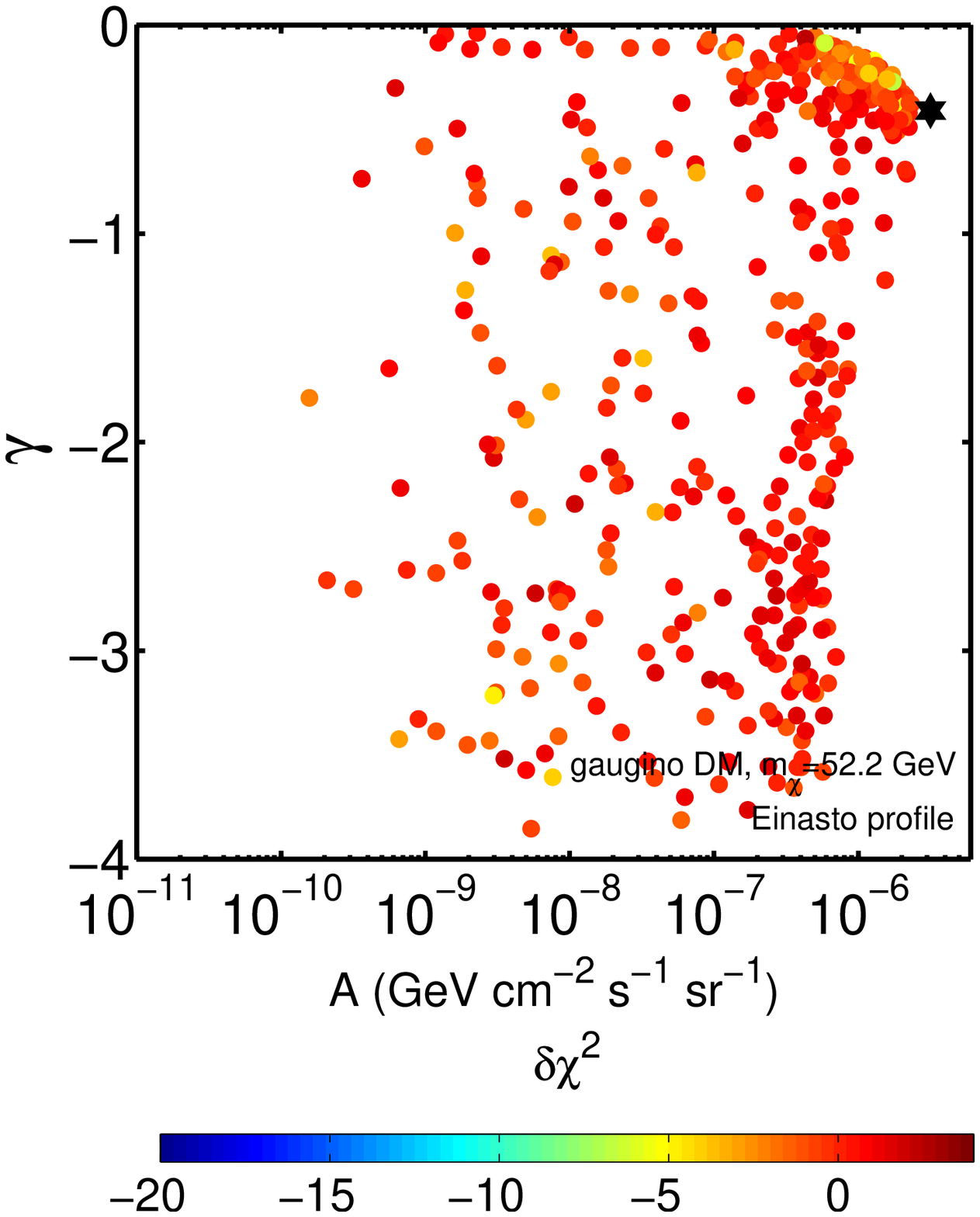}\\
	\includegraphics[width=4.2cm, keepaspectratio]{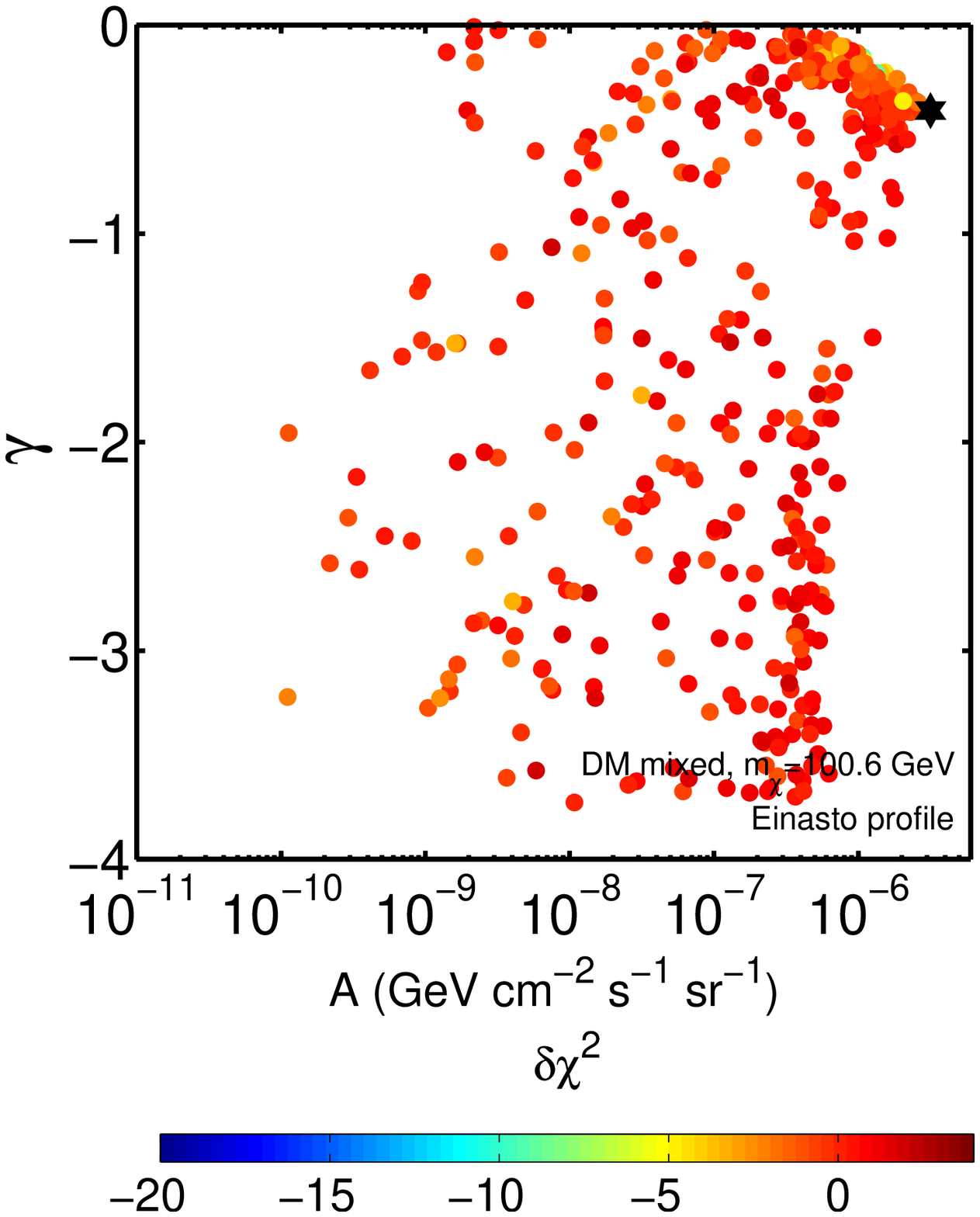}
	\caption{Results for the values of the boost factor BF, EGB normalisation $A$, and corresponding value of $\delta\chi_{\rm DM}^2$ associated with the best-fit points displayed in Figure\,\ref{fig:bfvsavsdeltachisq}. Once again, we highlight the location of the Fermi LAT best-fit EGB estimate (black stars).}
	\label{fig:gvsavsdeltadmchisq}
\end{figure}
In order to further explore such correlations, in Figure\,\ref{fig:gvsavsdeltadmchisq}, we plot the best-fit EGB parameters corresponding to those points displayed in Figure\,\ref{fig:bfvsavsdeltachisq}.

We can clearly observe that the strong correlation displayed in Figure\,\ref{fig:deltschisqegb} between large values of $A$ and $\gamma$ is much diminished, being most distinctive for the results corresponding to the gaugino dominated DM point.
However, interestingly, not only do the best-fit points possessing the most negative values of $\delta\chi_{\rm DM}^2$ possess significant EGB {\it and} DM components in order to achieve their fit, their associated EGB component resides in the bgnd.+EGB favoured region (which we recall includes the Fermi LAT best-fit EGB estimate - indicated by the black star in Figure\,\ref{fig:deltschisqegb}), despite the weak correlation displayed by the bgnd.+EGB+DM fits.
(For reference, the values [$A/10^{-6}$,\,$\gamma$,\,BF,\,$\delta\chi_{\rm DM}^2$] associated with these best-fit points are 
[$1.27,-0.19,55.3,-6.9$], 
[$1.23,-0.22,1644.5,-8.0$] and 
[$1.34,-0.23,48.8,-8.95$] for the best-fit, gaugino and mixed DM candidate points respectively in Einasto profile.)


\section{Summary}
\label{sec:summary}

\noindent In this paper we investigated the extent to which the uncertainties in the propagation of Galactic CRs impact upon our estimates of the \gray flux from the MLR. 

For each PPC considered, we calculated the corresponding $\chi_{\rm LAR}^2$ value resulting from fits to current experimental data on the B/C and $^{10}$Be/$^9$Be local abundance ratios, and found that a significant number of them were consistent with the data, producing $\chi^2$ values as low as 1.5 per data point.

We then calculated the corresponding \gray spectra resulting from standard astrophysical background processes involving energetic CRs and interstellar nuclei or ISRFs.
We deduced that when the normalisation of the various components of the SAB remained fixed, the uncertainties in the \gray flux generated exceeded several orders of magnitude. 
However, when these normalisations were freely adjusted to fit the Fermi LAT data the uncertainties in the \gray spectra were much reduced, with many PPCs producing acceptable fits.

We then investigated how these fits may be improved by simultaneously fitting an EGB component, described by a free-form power law, with the SAB to the Fermi LAT data.
We found substantial improvement in the fits of many PPCs, which themselves provided good fits to the data, with the largest improvements (i.e. as large as $\delta\chi_{\rm EGB}^2\simeq-10$) being associated with EGB's of order $A\simeq10^{-7} - 10^{-6}$\,GeV\,cm$^{-2}$\,s$^{-1}$\,sr$^{-1}$ at $E_1=200\,$MeV and slopes $0>\gamma\gtrsim-1$, which also includes the EGB estimate determined by the Fermi Collaboration.

We then investigated how these fits may be further improved by simultaneously fitting a \gray flux component from dark matter described by three different candidate points within the MSSM, together with the SAB and the EGB to the Fermi LAT data.
We found that the un-boosted flux, generated from motivated versions of the Burkert and Einasto profiles, was insufficient to impact the fits to the data.

However, when we artificially re-normalised the DM flux using boost factors
of approximately 40, 1500 and 30 with our best-fit, gaugino-dominated and mixed candidate points together with our Einasto profile (and BF's approximately 3.3 times larger when using our Burkert profile), we found that there exist many PPCs that provide further improvement in their fits to the Fermi data.
Moreover, there exist PPCs that not only provide good fits to the Fermi data when including the SAB, EGB and DM components, that are significantly better than the corresponding fits when omitting the DM flux, but also require both substantial EGB and DM fluxes which are crucial in determining the quality of the fit with the parameters of the EGB component similar to those estimated previously by the Fermi Collaboration.  
A possible source of these enhancements could arise from local DM substructures within the Galactic halo, thought to be able to give rise to boost factors as large as 20.\\

\noindent {\bf Acknowledgments:} 
We would like to thank Gudlaugur Johannesson, Tim Linden, Igor
Moskalenko, Stefano Profumo and Andrew Strong for their helpful
comments and technical assistance. 
DTC is funded by the Science and Technology Facilities Council.


\section{Appendix: Best-fit PPCs and associated parameter values}
\label{sec:appendix}
\begin{table*}[h]
\begin{center}
\begin{tabular}{|c|c|c|c|c|c|c|c|c|c|c|c|c|c|c|c|c|}
\hline
$D_0$ & $B_0$ & $\alpha$ & $z_0$ & $e_{g_0}$& $e_{g_1}$ & $n_{g_1}$ &
$ n_{g_2}$ & $v_{\rm A}$ & $L$ & $\frac{{\rm d}V}{{\rm d}z}$ & $\phi_{\odot}$ & $\chi_{\rm LAR}^2$  &
$N$ & $M$ & $\chi_{\rm bgnd.}^2$\\
\hline
\hline
5& 5& 0.4& 2& 1.6& 2.5& 2.15& 2.3& 20& 7& 6& 250& 1.71& 2.2& 1.66& 0.653\\
6.5& 5& 0.4& 2& 1.6& 2.5& 1.7& 2.3& 35& 11& 10& 450& 3.35& 2.95& 0.954& 0.682\\
6.5& 5& 0.4& 2& 1.6& 2.5& 2.15& 2.3& 20& 11& 2& 250& 3.03& 2.38& 1.3& 0.84\\
5& 2.5& 0.4& 3& 0.65& 3.6& 1.7& 2& 35& 4& 2& 450& 1.98& 0.637& 2.27& 0.841\\
5.75& 5& 0.34& 2& 1.6& 2.5& 1.82& 2.18& 36& 4& 3& 450& 3.69& 3.14& 1.04& 0.901\\
8& 5& 0.45& 2& 1.6& 2.5& 1.8& 2.18& 40& 14& 8& 450& 3.34& 1.57& 1.77& 0.902\\
6.5& 5& 0.4& 2& 1.6& 2.5& 2.15& 2.3& 20& 7& 2& 250& 2.15& 2.19& 1.43& 0.905\\
6.5& 5& 0.6& 2& 1.6& 2.5& 2.15& 2& 20& 11& 2& 250& 3.49& 1.22& 2.15& 0.97\\
6.5& 2.5& 0.4& 1& 0.65& 3.6& 1.7& 2& 35& 11& 6& 450& 3.94& 0.603& 2.2& 1\\
5& 5& 0.6& 2& 1.6& 2.5& 1.7& 2& 20& 7& 2& 100& 2.82& 2.37& 1.34& 1.02\\
5& 5& 0.6& 2& 1.6& 2.5& 1.7& 2& 35& 11& 10& 250& 1.62& 1.11& 2.26& 1.03\\
5& 5& 0.4& 2& 1.6& 2.5& 2.15& 2.3& 20& 11& 6& 250& 3.8& 2.04& 1.74& 1.06\\
5& 5& 0.4& 2& 1.6& 2.5& 1.7& 2.1& 35& 4& 2& 450& 1.92& 1.88& 1.78& 1.06\\
6.5& 10& 0.4& 1& 0.65& 3.6& 1.7& 2& 35& 11& 6& 450& 3.94& 0.499& 2.24& 1.08\\
5& 10& 0.4& 3& 0.65& 3.6& 1.7& 2& 35& 7& 10& 450& 3.21& 0.592& 2.11& 1.16\\
5& 5& 0.4& 2& 1.6& 2.5& 1.7& 2.6& 20& 11& 6& 250& 3.64& 3.15& 0.698& 1.3\\
5& 2.5& 0.4& 1& 0.65& 3.6& 1.7& 2& 35& 7& 10& 450& 3.21& 0.683& 2.04& 1.35\\
5& 5& 0.4& 2& 1.6& 2.5& 1.7& 2& 35& 7& 10& 450& 3.21& 3.14& 0.726& 1.35\\
8.5& 2.5& 0.4& 1& 0.65& 3.6& 1.7& 2& 35& 11& 2& 450& 1.79& 0.503& 2.31& 1.37\\
6.5& 5& 0.4& 2& 1.6& 2.5& 1.7& 2& 35& 7& 2& 450& 3.53& 2.43& 1.05& 1.39\\
8.5& 2.5& 0.4& 3& 0.65& 3.6& 2.15& 2& 35& 11& 6& 450& 1.97& 0.514& 2.17& 1.4\\
5& 5& 0.38& 2& 1.6& 2.5& 1.8& 2.15& 35& 4& 2.5& 450& 1.63& 1.36& 2.03& 1.41\\
5& 5& 0.6& 2& 1.6& 2.5& 1.7& 2& 35& 7& 10& 250& 3.23& 1.01& 2.19& 1.43\\
6.5& 5& 0.6& 2& 1.6& 2.5& 1.7& 2& 35& 11& 6& 250& 2.7& 0.785& 2.4& 1.44\\
5& 5& 0.4& 2& 1.6& 2.5& 1.7& 2& 35& 4& 2& 450& 1.98& 2.84& 0.99& 1.45\\
8.5& 2.5& 0.4& 1& 0.65& 3.6& 2.15& 2& 35& 11& 6& 450& 1.97& 0.489& 2.21& 1.47\\
6.5& 5& 0.4& 2& 1.6& 2.5& 1.7& 2& 35& 11& 6& 450& 3.94& 2.95& 0.711& 1.5\\
5.75& 2.5& 0.34& 1& 1.9& 2.95& 1.82& 2.18& 36& 4& 3& 450& 3.69& 0.415& 2.7& 1.58\\
8& 5& 0.45& 2& 1.6& 2.5& 1.8& 2.18& 35& 14& 3& 450& 2.68& 1.04& 1.99& 1.62\\
6.5& 5& 0.4& 2& 1.6& 2.5& 1.7& 2& 35& 7& 6& 450& 1.93& 3.18& 0.605& 1.65\\
6.5& 2.5& 0.4& 3& 0.65& 3.6& 1.7& 2& 35& 11& 10& 450& 3.31& 0.793& 1.86& 1.65\\
8.5& 5& 0.4& 2& 1.6& 2.5& 1.7& 2& 35& 11& 2& 450& 1.79& 2.88& 0.639& 1.68\\
5& 5& 0.6& 2& 1.6& 2.5& 2.15& 2& 20& 11& 6& 250& 2.04& 2.61& 1.1& 1.7\\
5& 5& 0.6& 2& 1.6& 2.5& 1.7& 2.3& 20& 11& 6& 250& 2.79& 2.19& 1.11& 1.74\\
5.5& 5& 0.38& 2& 1.6& 2.5& 1.8& 2.25& 35& 4& 2.5& 450& 3.24& 1.46& 1.78& 1.75\\
6.5& 5& 0.4& 2& 1.6& 2.5& 1.7& 2.3& 35& 7& 6& 450& 1.92& 1.69& 1.5& 1.76\\
5& 5& 0.4& 2& 1.6& 2.5& 1.7& 2.6& 20& 7& 6& 250& 2.27& 2.68& 0.868& 1.76\\
8.5& 5& 0.4& 2& 1.6& 2.5& 1.7& 2.3& 35& 11& 2& 450& 2.08& 1.52& 1.48& 1.76\\
5& 5& 0.4& 2& 1.6& 2.5& 1.7& 2.3& 30& 4& 2& 250& 3.36& 1.6& 1.71& 1.79\\
8.5& 10& 0.4& 1& 0.65& 3.6& 1.7& 2& 35& 11& 2& 450& 1.79& 0.532& 1.93& 1.79\\
5& 5& 0.4& 2& 1.6& 2.5& 1.7& 2.3& 35& 7& 10& 450& 3.78& 1.63& 1.58& 1.81\\
6.5& 2.5& 0.4& 3& 0.65& 3.6& 1.7& 2& 35& 7& 6& 450& 1.93& 0.71& 1.85& 1.84\\
8.5& 2.5& 0.2& 3& 1.9& 2.95& 2.6& 2.3& 20& 7& 2& 450& 3.69& 0.557& 2.44& 1.85\\
5.75& 10& 0.34& 3& 0.65& 3.6& 1.82& 2.18& 36& 4& 3& 450& 3.69& 0.215& 2.79& 1.87\\
6.5& 10& 0.4& 3& 0.65& 3.6& 2.15& 2& 35& 11& 10& 450& 1.9& 0.516& 2& 1.87\\
5.75& 10& 0.34& 1& 0.65& 3.6& 1.82& 2.18& 36& 4& 3& 450& 3.69& 0.215& 2.79& 1.88\\
5& 5& 0.4& 2& 1.6& 2.5& 1.7& 2& 35& 4& 6& 450& 3.97& 3.41& 0.671& 1.9\\
6.5& 5& 0.4& 2& 1.6& 2.5& 1.7& 2.6& 20& 11& 2& 250& 2.62& 2.54& 0.767& 1.91\\
5& 2.5& 0.2& 1& 1.9& 2.95& 2.6& 2.3& 20& 4& 6& 450& 3.85& 0.558& 2.47& 1.93\\
6.5& 5& 0.4& 2& 1.6& 2.5& 1.7& 2.6& 20& 7& 2& 250& 3.2& 2.51& 0.791& 1.94\\

 \hline
\end{tabular}
\caption{Tabulated values of those
PPCs utilised in our bgnd.\,\,only fits (i.e. $\chi_{\rm LAR}^2$) 
that generated the smallest values of $\chi_{\rm bgnd.}^2$ for $X=10$. 
We also provide the best-fit values of the
normalisation parameters $N$ and $M$ of each PPC
as well as the corresponding value of $\chi_{\rm bgnd.}^2$.
The units of the propagation parameters listed are as follows: 
$D_0/10^{28}$\,(cm$^2$\,s$^{-1})$, $\alpha$, $L\,$(kpc), $B_0\,(\mu\,G)$, 
$z_0\,$(kpc), d$V/$d$z$\,(km\,s$^{-1}$\,kpc$^{-1})$, $v_{\rm A}$\,(km\,s$^{-1})$,
$n_{g_1}$, $n_{g_2}$, $e_{g_0}$, $e_{g_1}$ and $\phi_{\odot}$ (MV).}
\label{tab:bkgonly}
\end{center}
\end{table*}

\vspace{1.0cm}

\begin{table*}[h]
\begin{center}
\begin{tabular}{|c|c|c|c|c|c|c|c|c|c|c|c|c|c|c|c|c|c|c|}
\hline
$D_0$ & $B_0$ & $\alpha$ & $z_0$ & $e_{g_0}$& $e_{g_1}$ & $n_{g_1}$ &
$ n_{g_2}$ & $v_{\rm A}$ & $L$ & $\frac{{\rm d}V}{{\rm d}z}$ & $\phi_{\odot}$ & $\chi_{\rm LAR}^2$  &
$N$ & $M$ & \texttt{log}$_{10}(A)$ & $\gamma$ & $\chi_{\rm EGB}^2$ & $\delta\chi_{\rm EGB}^2$\\
\hline
\hline
6.5& 5& 0.4& 2& 1.6& 2.5& 1.7& 2.3& 20& 7& 2& 250& 3.28& 0.994& 2.03& -5.68& -0.692& 0.826&-1.5\\
6.5& 5& 0.4& 2& 1.6& 2.5& 1.7& 2.3& 35& 11& 10& 450& 3.35& 0.922& 2.35& -5.82& -1.22& 0.887&0.191\\
5& 5& 0.4& 2& 1.6& 2.5& 2.15& 2.3& 20& 7& 6& 250& 1.71& 2.2& 1.66& -7.33& -2.71& 0.893&0.24\\
5& 5& 0.6& 2& 1.6& 2.5& 1.7& 2.3& 20& 11& 6& 250& 2.79& 1.35& 1.3& -5.88& -0.376& 0.897&-0.841\\
6.5& 5& 0.4& 2& 1.6& 2.5& 1.7& 2.3& 35& 7& 6& 450& 1.92& 1.26& 1.32& -5.84& -0.355& 0.93&-0.988\\
6.5& 5& 0.4& 2& 1.6& 2.5& 2.15& 2.3& 20& 7& 2& 250& 2.15& 1.22& 1.68& -6.11& -0.296& 0.948&0.0436\\
5& 5& 0.4& 2& 1.6& 2.5& 2.15& 2.3& 20& 11& 6& 250& 3.8& 2.04& 1.74& -6.4& -1.04& 0.953&-0.103\\
6.5& 5& 0.4& 2& 1.6& 2.5& 1.7& 2.3& 20& 11& 2& 250& 2.38& 1.82& 1.63& -5.85& -0.766& 0.99&-1.55\\
5& 2.5& 0.4& 3& 0.65& 3.6& 1.7& 2& 35& 4& 2& 450& 1.98& 0.637& 2.27& -6.67& -3.2& 1.03&0.19\\
5& 5& 0.4& 2& 1.6& 2.5& 1.7& 2.6& 20& 11& 6& 250& 3.64& 2.3& 0.979& -6.51& -0.123& 1.05&-0.242\\
5& 5& 0.4& 2& 1.6& 2.5& 1.7& 2.3& 20& 7& 6& 250& 2.21& 1.45& 1.86& -5.69& -0.721& 1.06&-1.43\\
5& 5& 0.4& 2& 1.6& 2.5& 1.7& 2.3& 35& 7& 10& 450& 3.78& 0.995& 1.4& -5.71& -0.42& 1.06&-0.744\\
8& 5& 0.4& 2& 1.6& 2.5& 1.8& 2.36& 40& 14& 8& 450& 3.24& 1.2& 1.15& -5.89& -0.301& 1.08&-1.2\\
8& 5& 0.45& 2& 1.6& 2.5& 1.8& 2.18& 40& 14& 8& 450& 3.34& 1.57& 1.77& -6.59& -3.72& 1.08&0.18\\
8& 5& 0.45& 2& 1.6& 2.5& 1.8& 2.36& 35& 14& 5.5& 450& 2.14& 0.843& 1.09& -5.69& -0.348& 1.08&-1.8\\
8& 5& 0.45& 2& 1.6& 2.5& 1.8& 2.18& 35& 14& 3& 450& 2.68& 1& 1.47& -5.82& -0.408& 1.12&-0.494\\
6.5& 5& 0.4& 2& 1.6& 2.5& 1.7& 2.6& 20& 7& 2& 250& 3.2& 1.94& 0.84& -6.19& -0.206& 1.14&-0.795\\
5& 5& 0.4& 2& 1.6& 2.5& 1.7& 2.3& 35& 4& 6& 450& 3.76& 0.948& 1.46& -5.73& -0.412& 1.15&-0.931\\
5.75& 5& 0.35& 2& 1.6& 2.5& 1.82& 2.36& 36& 4& 3& 450& 3.77& 1.08& 1.29& -5.84& -0.31& 1.17&-1.48\\
6.5& 5& 0.4& 2& 1.6& 2.5& 2.15& 2.3& 20& 11& 2& 250& 3.03& 2.38& 1.3& -8.35& -1.82& 1.18&0.335\\
5& 5& 0.4& 2& 1.6& 2.5& 1.6& 2.3& 35& 4& 2& 450& 2.14& 0.839& 1.33& -5.68& -0.397& 1.19&-1.65\\
6.5& 5& 0.4& 2& 1.6& 2.5& 1.7& 2.6& 20& 11& 2& 250& 2.62& 1.37& 0.842& -5.75& -0.361& 1.2&-0.708\\
5& 5& 0.36& 2& 1.6& 2.5& 1.8& 2.25& 35& 4& 2.5& 450& 1.76& 0.82& 1.41& -5.68& -0.403& 1.21&-1.31\\
5.75& 5& 0.34& 2& 1.6& 2.5& 1.82& 2.36& 36& 4& 2.5& 450& 3.17& 0.828& 1.21& -5.65& -0.377& 1.23&-1.37\\
5.75& 5& 0.34& 2& 1.6& 2.5& 1.82& 2.36& 36& 4& 3& 450& 3.52& 1.39& 1.31& -6.04& -0.251& 1.24&-1.13\\
5.75& 5& 0.34& 2& 1.6& 2.5& 1.82& 2.18& 36& 4& 3& 450& 3.69& 3.14& 1.04& -7.52& -2.1& 1.25&0.349\\
5& 5& 0.4& 2& 1.6& 2.5& 1.7& 2.3& 35& 4& 2.5& 450& 1.96& 0.62& 1.29& -5.66& -0.354& 1.25&-2.04\\
5.75& 5& 0.34& 2& 1.6& 2.5& 1.82& 2.36& 38& 4& 3& 450& 2.92& 0.797& 1.15& -5.64& -0.374& 1.26&-1.85\\
8& 5& 0.45& 2& 1.6& 2.5& 1.8& 2.36& 35& 14& 5& 450& 1.83& 0.601& 1.09& -5.62& -0.367& 1.26&-1.88\\
5& 5& 0.4& 2& 1.6& 2.5& 1.7& 2.1& 35& 4& 2& 450& 1.92& 1.13& 1.93& -5.79& -0.619& 1.28&0.1\\
8& 5& 0.45& 2& 1.6& 2.5& 1.8& 2.36& 40& 14& 8.5& 450& 3.97& 0.862& 1.31& -5.95& -0.261& 1.34&-1.68\\
5& 5& 0.6& 2& 1.6& 2.5& 1.7& 2& 35& 11& 10& 250& 1.62& 1.05& 2.24& -6.23& -1.79& 1.35&-0.0777\\
6.5& 5& 0.6& 2& 1.6& 2.5& 2.15& 2& 20& 11& 2& 250& 3.49& 1.22& 2.15& -8.53& -3.31& 1.36&0.387\\
5.75& 2.5& 0.34& 3& 0.65& 3.6& 1.82& 2.18& 36& 4& 3& 450& 3.69& 0.756& 1.44& -6.42& -0.0669& 1.37&-1.24\\
8.5& 5& 0.45& 2& 1.6& 2.5& 1.8& 2.36& 35& 14& 3& 450& 1.8& 0.71& 1.08& -5.75& -0.313& 1.37&-2.59\\
6.5& 5& 0.6& 2& 1.6& 2.5& 1.7& 2.3& 20& 11& 2& 100& 3.91& 0.719& 1.15& -5.7& -0.356& 1.38&-0.633\\
7.5& 5& 0.45& 2& 1.6& 2.5& 1.8& 2.36& 40& 14& 8& 450& 2.18& 0.832& 1.07& -5.78& -0.313& 1.38&-2.47\\
5& 2.5& 0.4& 1& 0.65& 3.6& 1.7& 2& 35& 4& 6& 450& 3.97& 0.894& 1.6& -6.66& -0.0973& 1.39&-1.27\\
6.5& 2.5& 0.4& 1& 0.65& 3.6& 1.7& 2& 35& 11& 6& 450& 3.94& 0.603& 2.2& -8.92& -3.39& 1.4&0.399\\
5& 5& 0.38& 2& 1.6& 2.5& 1.79& 2.25& 35& 4& 2.5& 450& 1.64& 1.02& 1.25& -5.75& -0.354& 1.42&-1.57\\
5& 5& 0.6& 2& 1.6& 2.5& 1.7& 2& 20& 7& 2& 100& 2.82& 2.37& 1.34& -9.82& -3.65& 1.42&0.407\\
5.2& 5& 0.38& 2& 1.6& 2.5& 1.8& 2.3& 35& 4& 2& 450& 1.72& 0.792& 1.24& -5.72& -0.343& 1.44&-2.12\\
5& 5& 0.38& 2& 1.6& 2.5& 1.8& 2.15& 35& 4& 2.5& 450& 1.63& 1.02& 2.25& -6.06& -1.65& 1.44&-0.234\\
6.5& 2.5& 0.4& 3& 0.65& 3.6& 1.7& 2.3& 35& 11& 10& 450& 3.35& 0.641& 1.36& -6.39& -0.0441& 1.44&-1.8\\
5& 5& 0.4& 2& 1.6& 2.5& 1.7& 2.3& 20& 11& 6& 250& 3.46& 1.44& 1.8& -5.66& -0.713& 1.45&-2.64\\
5& 2.5& 0.4& 1& 0.65& 3.6& 1.7& 2.3& 35& 4& 6& 450& 3.76& 0.562& 1.17& -6.04& -0.163& 1.46&-5.76\\
5& 5& 0.38& 2& 1.6& 2.5& 1.8& 2.27& 35& 4& 2.5& 450& 1.63& 0.599& 1.28& -5.62& -0.387& 1.46&-1.96\\
5& 5& 0.34& 2& 1.6& 2.5& 1.82& 2.36& 36& 4& 3& 450& 2.6& 0.95& 1.02& -5.73& -0.326& 1.46&-2.82\\
8& 5& 0.45& 2& 1.6& 2.5& 1.8& 2.36& 42& 14& 8& 450& 2.79& 0.903& 1.07& -5.91& -0.255& 1.49&-2.67\\
8& 5& 0.45& 2& 1.6& 2.5& 1.8& 2.36& 35& 14& 4& 450& 1.9& 0.868& 0.914& -5.69& -0.337& 1.5&-2.3\\

\hline
\end{tabular}
\caption{Same as for Table\,\ref{tab:bkgonly} but now for our bgnd.+EGB fits.
For each PPC listed we also provide the best-fit values of
the EGB normalisation $A$ (GeV\,cm$^{-2}$\,s$^{-1}$\,sr$^{-1}$)
and slope $\gamma$, as well as the corresponding values of $\chi_{\rm EGB}^2$
and $\delta\chi_{\rm EGB}^2$.}
\label{tab:bkgegb}
\end{center}
\end{table*}

\vspace{1.0cm}


\begin{table*}[h]
\begin{center}
\begin{tabular}{|c|c|c|c|c|c|c|c|c|c|c|c|c|c|c|c|c|c|c|c|c|c|c|c|c|c|c|}
\hline
\multicolumn{12}{|c|}{Propagation Parameter Combination} & & \multicolumn{7}{|c|}{Burkert Profile}\\
\hline
$D_0$ & $B_0$ & $\alpha$ & $z_0$ & $e_{g_0}$& $e_{g_1}$ & $n_{g_1}$ &
$ n_{g_2}$ & $v_{\rm A}$ & $L$ & $\frac{{\rm d}V}{{\rm d}z}$ & $\phi_{\odot}$ & $\chi_{\rm LAR}^2$  &
$N$ & $M$ & \texttt{log}$_{10}(A)$ & $\gamma$ &  \texttt{log}$_{10}({\rm BF})$ & $\chi_{\rm DM}^2$ & $\delta\chi_{\rm DM}^2$\\
\hline
5&2.5&0.4&1&0.65&3.6&1.7&2&35&4&2&450&1.98&0.672&1.94&-7.24&-0.0566&1.94&0.842&-1.36\\
5&2.5&0.2&3&0.65&3.6&2.6&2&20&4&6&450&3.67&1.28&1.3&-9.24&-2.59&2.15&0.979&-2.03\\
5.75&10&0.34&1&0.65&3.6&1.82&2.36&36&4&2.5&450&3.17&0.552&0.663&-5.96&-0.188&2.12&1.02&-1.37\\
5.75&10&0.34&3&0.65&3.6&1.82&2.36&38&4&3&450&2.92&0.556&0.596&-5.99&-0.171&2.18&1.07&-1.87\\
5&2.5&0.2&1&0.65&3.6&2.6&2&20&4&6&450&3.67&1.33&1.12&-7.92&-1.04&2.22&1.1&-1.9\\
5&5&0.4&2&1.6&2.5&2.15&2.3&20&11&6&250&3.8&1.42&2.04&-6.07&-0.954&1.07&1.11&0.156\\
5.75&2.5&0.35&3&0.65&3.6&1.82&2.36&36&4&3&450&3.77&0.571&0.814&-5.86&-0.256&2.01&1.11&-1.19\\
5&5&0.4&2&1.6&2.5&2.15&2.3&20&7&6&250&1.71&1.73&1.91&-6.61&-1.83&1.06&1.14&0.243\\
8&5&0.4&2&1.6&2.5&1.8&2.36&40&14&8&450&3.24&0.84&1.22&-5.69&-0.438&1.76&1.16&0.0817\\
5&10&0.2&3&0.65&3.6&2.6&2&20&4&6&450&3.67&1.04&1.12&-8.05&-3.55&2.36&1.17&-3.69\\
6.5&5&0.4&2&1.6&2.5&1.7&2.3&35&11&10&450&3.35&2.13&1.51&-6.22&-1.73&0.884&1.18&0.228\\
8&5&0.45&2&1.6&2.5&1.8&2.18&40&14&8&450&3.34&1.4&1.88&-6.4&-2.46&1.29&1.19&0.095\\
6.5&10&0.4&1&0.65&3.6&1.7&2&35&11&10&450&3.31&0.648&1.64&-7.65&-0.869&2.21&1.23&-2.43\\
6.5&5&0.4&2&1.6&2.5&2.15&2.3&20&11&2&250&3.03&2.25&1.36&-7.29&-2.02&1.3&1.28&0.107\\
5&5&0.36&2&1.6&2.5&1.8&2.3&35&4&2.5&450&1.8&0.9&1.23&-5.74&-0.359&1.59&1.34&-0.303\\
6.5&5&0.4&2&1.6&2.5&2.15&2.3&20&7&2&250&2.15&1.31&1.72&-6.21&-0.426&1.39&1.35&0.398\\
5&2.5&0.4&3&0.65&3.6&2.15&2&35&4&10&450&3.8&0.735&1.53&-6.98&-0.994&2.25&1.38&-1.14\\
4.8&5&0.4&2&1.6&2.5&1.7&2.3&35&4&2&450&1.87&0.708&1.08&-5.62&-0.396&1.75&1.38&-0.587\\
6.5&5&0.4&2&1.6&2.5&1.7&2.3&20&7&2&250&3.28&1.71&1.82&-5.9&-1.21&0.892&1.4&0.575\\
5&2.5&0.4&1&0.65&3.6&2.15&2&35&4&10&450&3.8&0.752&1.55&-7.55&-1.27&2.21&1.42&-1.75\\
7&5&0.45&2&1.6&2.5&1.8&2.36&40&14&8&450&2.41&0.74&0.933&-5.72&-0.349&1.85&1.43&-0.402\\
8&5&0.45&2&1.6&2.5&1.8&2.36&34&14&3&450&2.59&0.572&1.09&-5.67&-0.385&1.74&1.43&-0.755\\
5&10&0.2&1&0.65&3.6&2.6&2&20&4&6&450&3.67&1.05&1.04&-9.41&-3.54&2.35&1.43&-2.58\\
5&5&0.6&2&1.6&2.5&1.7&2.3&20&11&6&250&2.79&2.25&1.12&-8.57&-1.83&1.68&1.45&0.549\\
6.5&5&0.6&2&1.6&2.5&1.7&2.3&20&11&2&100&3.91&0.719&1.15&-5.7&-0.356&1.16&1.45&0.065\\
5&5&0.4&2&1.6&2.5&2.15&2.6&20&7&6&250&1.78&0.545&0.935&-5.76&-0.322&1.99&1.45&-0.509\\
5.75&5&0.35&2&1.6&2.5&1.82&2.36&36&4&3&450&3.77&0.904&1.31&-5.79&-0.38&1.74&1.46&-0.614\\
8&5&0.45&2&1.6&2.5&1.8&2.36&42&14&8&450&2.79&0.952&0.858&-5.76&-0.341&1.8&1.47&-0.278\\
5&2.5&0.4&3&0.65&3.6&1.7&2&35&4&6&450&3.97&0.916&1.3&-6.59&-0.0576&1.66&1.47&-0.106\\
8&10&0.45&1&0.65&3.6&1.8&2.36&40&14&5.5&450&2.13&0.386&0.64&-6.22&-0.13&2.33&1.48&-5.67\\
5.1&5&0.38&2&1.6&2.5&1.8&2.25&35&4&2.5&450&1.66&1.04&1.36&-5.84&-0.344&1.48&1.49&-0.128\\
8&5&0.45&2&1.6&2.5&1.8&2.18&35&14&3&450&2.68&1.25&1.62&-6.44&-0.232&1.26&1.5&0.371\\
5&5&0.38&2&1.6&2.5&1.8&2.15&35&4&2.5&450&1.63&1.24&2.12&-6.39&-3.48&1.45&1.5&0.0606\\
5.2&5&0.38&2&1.6&2.5&1.8&2.3&35&4&2.5&450&1.83&0.798&1.19&-5.68&-0.386&1.59&1.54&-0.299\\
6.5&5&0.6&2&1.6&2.5&1.7&2&35&11&6&250&2.7&0.909&2.19&-6.28&-2.39&1.5&1.55&-0.011\\
5&5&0.4&2&1.6&2.5&1.7&2.3&35&7&10&450&3.78&1.69&1.45&-6.36&-2.22&1.9&1.58&-0.412\\
8.5&5&0.4&2&1.6&2.5&1.7&2.6&35&11&2&450&2.58&1.07&0.64&-6.07&-0.215&2.06&1.59&-0.716\\
5&5&0.4&2&1.6&2.5&1.7&2.3&40&4&2&450&2.01&0.653&1.05&-5.79&-0.341&2.01&1.6&-0.608\\
4.8&5&0.38&2&1.6&2.5&1.8&2.3&35&4&2.5&450&2.19&0.63&1.09&-5.62&-0.404&1.8&1.61&-0.478\\
7.5&10&0.45&3&0.65&3.6&1.8&2.36&40&14&8&450&2.18&0.451&0.398&-5.97&-0.162&2.28&1.63&-1.61\\
5&5&0.38&2&1.6&2.5&1.8&2.35&35&4&2.5&450&1.65&0.705&0.981&-5.64&-0.391&1.88&1.63&-0.183\\
6.5&5&0.4&2&1.6&2.5&1.7&2.3&35&7&6&450&1.92&1.83&1.33&-6.55&-3.39&1.86&1.65&0.704\\
5&5&0.4&2&1.6&2.5&1.8&2.3&35&4&3&450&1.82&0.712&1.21&-5.76&-0.385&1.94&1.66&-1.35\\
5.75&5&0.34&2&1.6&2.5&1.82&2.18&36&4&3&450&3.69&3.15&1&-7.98&-1.42&0.598&1.67&0.416\\
8&10&0.45&3&0.65&3.6&1.8&2.36&35&10&3&450&1.76&0.402&0.618&-6.2&-0.109&2.31&1.67&-5.25\\
5&5&0.4&2&1.6&2.5&1.7&2.6&20&7&6&250&2.27&2.76&0.798&-8.7&-0.862&1.87&1.68&-0.747\\
5&5&0.6&2&1.6&2.5&1.7&2&35&11&10&250&1.62&1.2&2.06&-6.63&-2.98&1.31&1.68&0.248\\
5&5&0.4&2&1.6&2.5&1.7&2.1&35&4&2&450&1.92&1.95&1.66&-6.57&-3.13&1.29&1.68&0.228\\
8.5&5&0.4&2&1.6&2.5&1.7&2.3&35&11&2&450&2.08&1.34&1.55&-6.39&-3.41&1.89&1.69&0.17\\
5&5&0.38&2&1.6&2.5&1.8&2.25&34.5&4&2.5&450&1.67&0.816&1.52&-5.8&-0.471&1.84&1.72&-0.258\\

 \hline
\end{tabular}
\caption{Same as for Table\,\ref{tab:bkgegb} but now for our bgnd.+EGB+DM fits
when using our best-fit DM candidate point with our Burkert DM density
profile. For each PPC listed we also provide the best-fit values of the best-fit boost factors 
BF as well as the corresponding values of $\chi_{\rm DM}^2$ and $\delta\chi_{\rm DM}^2$.}
\label{tab:bkgegbdmbestfitburkert}
\end{center}
\end{table*}


\begin{table*}[h]
\begin{center}
\begin{tabular}{|c|c|c|c|c|c|c|c|c|c|c|c|c|c|c|c|c|c|c|c|c|c|c|c|c|c|c|}
\hline
\multicolumn{12}{|c|}{Propagation Parameter Combination} & & \multicolumn{7}{|c|}{Einasto Profile}\\
\hline
$D_0$ & $B_0$ & $\alpha$ & $z_0$ & $e_{g_0}$& $e_{g_1}$ & $n_{g_1}$ &
$ n_{g_2}$ & $v_{\rm A}$ & $L$ & $\frac{{\rm d}V}{{\rm d}z}$ & $\phi_{\odot}$ & $\chi_{\rm LAR}^2$  &
$N$ & $M$ & \texttt{log}$_{10}(A)$ & $\gamma$ &  \texttt{log}$_{10}({\rm BF})$ & $\chi_{\rm DM}^2$ & $\delta\chi_{\rm DM}^2$\\
\hline
6.5&2.5&0.4&3&0.65&3.6&1.7&2&35&7&6&450&1.93&0.71&1.85&-10&-1.07&1.52&0.492&-2.08\\
5.75&2.5&0.34&3&0.65&3.6&1.82&2.36&36&4&3&450&3.52&0.598&0.893&-5.97&-0.226&1.55&0.86&-0.749\\
6.5&5&0.4&2&1.6&2.5&1.7&2.6&20&11&2&250&2.62&1.37&0.842&-5.75&-0.361&0.931&0.953&-0.246\\
5&5&0.4&2&1.6&2.5&1.7&2.6&20&11&6&250&3.64&2.11&0.899&-6.02&-0.351&0.998&0.965&-0.0893\\
5&2.5&0.34&3&0.65&3.6&1.82&2.36&36&4&3&450&2.6&0.545&0.804&-6.03&-0.211&1.66&0.989&-1.5\\
5&2.5&0.2&3&0.65&3.6&2.6&2&20&4&6&450&3.67&1.27&1.3&-7.64&-0.853&1.6&1.03&-1.98\\
8&2.5&0.45&3&0.65&3.6&1.8&2.36&35&14&2.5&450&3.81&0.403&0.687&-6.07&-0.144&1.65&1.04&-2.58\\
5.75&5&0.34&2&1.6&2.5&1.82&2.36&36&4&3&450&3.52&0.899&1.29&-5.71&-0.379&0.983&1.13&-0.232\\
6.5&5&0.6&2&1.6&2.5&1.7&2.3&20&11&2&100&3.91&1.38&1.14&-6.13&-0.299&1.14&1.14&-0.246\\
6.5&5&0.4&2&1.6&2.5&2.15&2.3&20&7&2&250&2.15&1.2&1.76&-6.2&-0.302&0.47&1.14&0.195\\
5&5&0.4&2&1.6&2.5&2.15&2.3&20&11&6&250&3.8&1.06&2.28&-6&-1.12&0.663&1.14&0.191\\
5&5&0.6&2&1.6&2.5&1.7&2.6&20&11&6&250&2.82&0.848&0.738&-5.84&-0.298&1.46&1.16&-1.01\\
8&5&0.45&2&1.6&2.5&1.8&2.36&40&14&8.5&450&3.97&0.785&1.15&-5.73&-0.354&0.97&1.16&-0.252\\
5&5&0.4&2&1.6&2.5&2.15&2.3&20&7&6&250&1.71&1.61&2&-6.66&-3.5&0.225&1.17&0.276\\
5&10&0.4&1&0.65&3.6&1.7&2.3&35&7&10&450&3.78&0.569&0.583&-6.21&-0.0958&1.7&1.18&-1.52\\
8&5&0.45&2&1.6&2.5&1.8&2.36&35&14&5&450&1.83&0.601&1.09&-5.62&-0.367&0.799&1.19&-0.0714\\
8&5&0.45&2&1.6&2.5&1.8&2.18&40&14&8&450&3.34&1.2&1.96&-6.09&-1.38&0.931&1.19&0.0976\\
5.75&5&0.34&2&1.6&2.5&1.82&2.36&36&4&2.5&450&3.17&0.828&1.21&-5.65&-0.377&0.773&1.2&-0.0317\\
5&2.5&0.2&1&0.65&3.6&2.6&2&20&4&6&450&3.67&1.35&1.03&-8.85&-3.65&1.7&1.23&-1.78\\
8&5&0.45&2&1.6&2.5&1.8&2.18&35&14&3&450&2.68&0.929&1.5&-5.8&-0.421&0.652&1.25&0.122\\
6.5&10&0.4&3&0.65&3.6&2.15&2&35&11&10&450&1.9&0.528&1.77&-7.33&-0.0875&1.55&1.26&-1.36\\
6.5&5&0.4&2&1.6&2.5&2.15&2.3&20&11&2&250&3.03&1.74&1.69&-6.82&-3.7&0.871&1.29&0.119\\
6.5&5&0.4&2&1.6&2.5&1.7&2.3&35&11&10&450&3.35&2.7&1.1&-6.86&-2.95&0.276&1.33&0.443\\
5&2.5&0.4&1&0.65&3.6&2.15&2&35&4&10&450&3.8&0.764&1.5&-7.95&-1.33&1.69&1.36&-1.81\\
5.2&5&0.38&2&1.6&2.5&1.8&2.3&35&4&2.5&450&1.83&0.668&1.37&-5.74&-0.358&0.963&1.38&-0.185\\
5&2.5&0.4&3&0.65&3.6&2.15&2&35&4&10&450&3.8&0.676&1.73&-7.12&-2.15&1.58&1.38&-1.14\\
5&5&0.38&2&1.6&2.5&1.8&2.25&35&4&2.4&450&1.65&0.783&1.4&-5.74&-0.39&1.1&1.39&-0.418\\
5&5&0.38&2&1.6&2.5&1.8&2.27&35&4&2.5&450&1.63&0.599&1.28&-5.62&-0.387&0.835&1.39&-0.066\\
5&5&0.4&2&1.6&2.5&1.7&2.3&35&7&10&450&3.78&1.06&1.46&-5.77&-0.516&1.15&1.4&-0.59\\
5&2.5&0.4&1&0.65&3.6&1.7&2.3&35&4&2&450&1.86&0.697&0.545&-6.17&-0.088&1.55&1.41&-0.899\\
5&5&0.38&2&1.6&2.5&1.8&2.3&34&4&2&450&1.71&0.816&1.28&-5.78&-0.318&0.867&1.45&-0.0798\\
5&10&0.2&3&0.65&3.6&2.6&2&20&4&6&450&3.67&1.05&1.06&-7.71&-3.44&1.82&1.46&-3.39\\
5&5&0.38&2&1.6&2.5&1.8&2.25&35&4&3.5&450&1.69&0.872&1.57&-5.88&-0.376&1.1&1.47&-0.4\\
5&5&0.6&2&1.6&2.5&1.7&2.3&20&11&6&250&2.79&2.16&1.17&-7.62&-0.688&1.14&1.48&0.579\\
8&5&0.45&2&1.6&2.5&1.8&2.36&35&14&5.5&450&2.14&1.57&0.934&-6.15&-0.27&1.2&1.48&0.391\\
5&5&0.38&2&1.6&2.5&1.8&2.25&35&4&3&450&1.65&0.992&1.33&-5.78&-0.383&1.03&1.48&-0.302\\
8&5&0.45&2&1.6&2.5&1.7&2.36&35&14&3&450&2.41&1.26&0.992&-6.13&-0.247&1.27&1.49&-0.333\\
5&2.5&0.4&1&0.65&3.6&1.7&2&35&7&10&450&3.21&0.683&2.04&-6.69&-2.4&0.841&1.5&-0.228\\
8&5&0.45&2&1.6&2.5&1.8&2.36&40&14&5.5&450&2.13&0.619&0.993&-5.75&-0.371&1.43&1.53&-0.867\\
5&5&0.4&2&1.6&2.5&1.7&2.3&35&4&2&450&1.86&0.766&1.19&-5.69&-0.407&1.28&1.53&-0.175\\
8&2.5&0.45&3&0.65&3.6&1.8&2.36&40&14&5.5&450&2.13&0.398&0.571&-5.9&-0.225&1.6&1.53&-1.48\\
5&10&0.4&1&0.65&3.6&1.7&2.3&35&4&6&450&3.76&0.598&0.748&-6.13&-0.215&1.76&1.56&-0.766\\
5&5&0.38&2&1.6&2.5&1.8&2.15&35&4&2.5&450&1.63&1.08&2.02&-5.95&-1.13&1.12&1.58&0.135\\
8&2.5&0.45&1&0.65&3.6&1.8&2.36&35&14&3&450&2.93&0.404&0.597&-5.96&-0.183&1.62&1.59&-1.3\\
5&5&0.38&2&1.6&2.5&1.8&2.35&35&4&2&450&1.71&0.715&0.99&-5.68&-0.344&1.2&1.59&-0.349\\
5&10&0.2&1&0.65&3.6&2.6&2&20&4&6&450&3.67&1.04&0.715&-6.42&-0.167&1.73&1.61&-2.41\\
6.5&2.5&0.4&1&1.9&2.95&1.7&2.6&35&11&10&450&3.49&0.54&0.732&-5.84&-0.267&1.56&1.61&-2.64\\
6.5&5&0.4&2&1.6&2.5&1.7&2.6&35&7&6&450&2.02&0.985&0.739&-6.01&-0.231&1.47&1.61&-0.962\\
5&10&0.6&1&0.65&3.6&1.7&2.6&35&11&10&250&1.5&0.293&0.325&-5.97&-0.169&1.75&1.62&-5.71\\
6.5&2.5&0.4&3&0.65&3.6&1.7&2&35&11&10&450&3.31&0.793&1.86&-7.84&-3.8&1.08&1.63&-0.673\\

 \hline
\end{tabular}
\caption{Same as for Table\,\ref{tab:bkgegbdmbestfitburkert} but now using our Einasto DM density profile.}
\label{tab:bkgegbdmbestfiteinasto}
\end{center}
\end{table*}


\begin{table*}[h]
\begin{center}
\begin{tabular}{|c|c|c|c|c|c|c|c|c|c|c|c|c|c|c|c|c|c|c|c|c|c|c|c|c|c|c|}
\hline
\multicolumn{12}{|c|}{Propagation Parameter Combination} & & \multicolumn{7}{|c|}{Burkert Profile}\\
\hline
$D_0$ & $B_0$ & $\alpha$ & $z_0$ & $e_{g_0}$& $e_{g_1}$ & $n_{g_1}$ &
$ n_{g_2}$ & $v_{\rm A}$ & $L$ & $\frac{{\rm d}V}{{\rm d}z}$ & $\phi_{\odot}$ & $\chi_{\rm LAR}^2$  &
$N$ & $M$ & \texttt{log}$_{10}(A)$ & $\gamma$ &  \texttt{log}$_{10}({\rm BF})$ & $\chi_{\rm DM}^2$ & $\delta\chi_{\rm DM}^2$\\
\hline
5&2.5&0.2&3&0.65&3.6&2.6&2&20&4&6&450&3.67&1.24&1.3&-7.88&-0.826&3.61&0.743&-2.27\\
5.5&10&0.34&3&0.65&3.6&1.82&2.36&36&4&3&450&2.25&0.532&0.621&-6.03&-0.169&3.71&0.795&-1.62\\
6.5&10&0.4&3&0.65&3.6&1.7&2.6&35&11&10&450&3.49&0.497&0.414&-6&-0.165&3.72&0.917&-2.64\\
5&2.5&0.2&1&0.65&3.6&2.6&2&20&4&6&450&3.67&1.26&1.31&-7.33&-1.08&3.52&0.957&-2.05\\
8&2.5&0.45&1&0.65&3.6&1.8&2.36&40&14&5.5&450&2.13&0.403&0.53&-5.95&-0.166&3.6&0.969&-2.32\\
5&10&0.2&1&0.65&3.6&2.6&2&20&4&6&450&3.67&1&1.27&-9.34&-3.03&3.65&0.973&-3.04\\
8&5&0.45&2&1.6&2.5&1.8&2.36&40&14&8.5&450&3.97&0.697&1.12&-5.67&-0.365&2.94&1.02&-0.322\\
5&2.5&0.4&1&0.65&3.6&1.7&2&35&4&6&450&3.97&0.805&1.95&-8.23&-2.17&3.23&1.04&-0.347\\
5&5&0.4&2&1.6&2.5&2.15&2.3&20&11&6&250&3.8&1.17&2.26&-6.07&-1.32&2.55&1.07&0.119\\
6.5&5&0.4&2&1.6&2.5&1.7&2.3&35&7&6&450&1.92&0.949&1.38&-5.7&-0.455&2.98&1.08&0.142\\
5&5&0.4&2&1.6&2.5&2.15&2.3&20&7&6&250&1.71&2.2&1.66&-7.33&-2.71&1.32&1.1&0.21\\
5&2.5&0.4&3&0.65&3.6&1.7&2&35&4&6&450&3.97&0.847&1.74&-6.85&-0.0873&2.96&1.11&-0.473\\
5&10&0.4&1&0.65&3.6&1.7&2.3&35&7&10&450&3.78&0.56&0.313&-6.06&-0.127&3.71&1.12&-1.58\\
8.5&5&0.4&2&1.6&2.5&1.7&2.6&35&11&2&450&2.58&0.489&0.627&-5.57&-0.352&3.2&1.14&-1.06\\
6.5&5&0.4&2&1.6&2.5&2.15&2.3&20&7&2&250&2.15&1.22&1.68&-6.11&-0.296&1.8&1.14&0.194\\
5.75&10&0.34&1&0.65&3.6&1.82&2.36&36&4&2.5&450&3.17&0.533&0.444&-5.85&-0.241&3.7&1.17&-1.22\\
6.5&5&0.4&2&1.6&2.5&1.7&2.3&35&11&10&450&3.35&2.79&1.07&-7.08&-2.8&1.76&1.17&0.287\\
8&5&0.45&2&1.6&2.5&1.8&2.18&40&14&8&450&3.34&1.23&2.03&-6.3&-3.07&2.64&1.18&0.0997\\
5&2.5&0.34&1&0.65&3.6&1.82&2.36&36&4&3&450&2.6&0.584&0.504&-5.85&-0.264&3.68&1.19&-1.68\\
6.5&5&0.4&2&1.6&2.5&1.7&2.6&20&7&2&250&3.2&0.646&0.97&-5.68&-0.36&3.3&1.2&0.0573\\
8&2.5&0.45&3&0.65&3.6&2&2.36&40&14&8&450&2.07&0.395&0.647&-6.02&-0.137&3.53&1.22&-2.67\\
6.5&5&0.4&2&1.6&2.5&2.15&2.3&20&11&2&250&3.03&1.91&1.55&-6.86&-1.94&2.95&1.27&0.0904\\
8&5&0.45&2&1.6&2.5&1.8&2.36&35&14&5.5&450&2.14&1.48&1.02&-6.31&-0.187&3.23&1.28&-1.09\\
8.5&5&0.45&2&1.6&2.5&1.8&2.36&35&14&3&450&1.8&0.653&0.955&-5.68&-0.372&3.35&1.29&-0.911\\
8&5&0.45&2&1.6&2.5&1.8&2.36&35&14&3&450&2.93&0.579&0.876&-5.6&-0.373&3.22&1.3&-0.374\\
5&5&0.4&2&1.6&2.5&1.7&2.3&20&7&6&250&2.21&1.62&1.83&-5.73&-0.793&1.81&1.3&0.246\\
5&5&0.34&2&1.6&2.5&1.82&2.36&36&4&3&450&2.6&0.804&1.01&-5.68&-0.349&3.12&1.31&-0.148\\
5&2.5&0.4&1&0.65&3.6&1.7&2&35&4&2&450&1.98&0.807&0.961&-6.31&-0.151&3.48&1.32&-0.876\\
8&10&0.45&3&0.65&3.6&1.8&2.36&35&14&5&450&1.83&0.441&0.511&-6.03&-0.158&3.67&1.32&-2.58\\
8&5&0.45&2&1.6&2.5&1.8&2.36&35&10&3&450&1.76&0.609&1.04&-5.78&-0.312&3.25&1.32&-1.03\\
5&5&0.6&2&1.6&2.5&2.15&2.3&20&11&6&250&2.04&0.923&1.13&-5.97&-0.261&3.27&1.34&-0.599\\
5&5&0.4&2&1.6&2.5&1.7&2.1&35&4&2&450&1.92&1.24&1.98&-5.86&-0.794&2.75&1.37&-0.0904\\
6.5&5&0.4&2&1.6&2.5&1.7&2.3&20&7&2&250&3.28&1.01&2.37&-5.83&-1.42&2.24&1.37&0.543\\
5.75&5&0.34&2&1.6&2.5&1.82&2.36&36&4&2.5&450&3.17&1.3&1.03&-5.75&-0.412&3.27&1.37&-0.0017\\
5&5&0.38&2&1.6&2.5&1.8&2.25&35&4&3.5&450&1.69&0.938&1.46&-5.84&-0.398&3.16&1.41&-0.814\\
8&10&0.45&1&0.65&3.6&1.8&2.36&35&14&5.5&450&2.14&0.481&0.326&-6&-0.13&3.65&1.45&-1.89\\
5&5&0.38&2&1.6&2.5&1.8&2.25&35&4&3&450&1.65&0.896&1.37&-5.77&-0.391&3.03&1.46&-0.325\\
5.75&5&0.34&2&1.6&2.5&1.82&2.36&36&4&3&450&3.52&1.06&1.2&-5.76&-0.449&3.37&1.46&0.0961\\
5&5&0.6&2&1.6&2.5&1.7&2.3&20&11&6&250&2.79&2.44&0.944&-9.65&-2.01&3.19&1.46&0.231\\
5&5&0.38&2&1.6&2.5&1.8&2.15&35&4&2.5&450&1.63&1.09&2.08&-6.12&-1.33&3.12&1.48&-0.0124\\
8&5&0.5&2&1.6&2.5&1.8&2.36&35&14&3&250&2.32&0.559&0.885&-5.77&-0.327&3.47&1.48&-0.904\\
6.5&10&0.4&1&0.65&3.6&1.7&2.3&35&11&10&450&3.35&0.507&0.828&-5.96&-0.207&3.41&1.49&-1.27\\
5&5&0.4&2&1.6&2.5&1.7&2.3&35&4&2&450&1.86&0.91&0.982&-5.65&-0.354&2.91&1.5&-0.288\\
5&5&0.6&2&1.6&2.5&1.7&2.3&20&7&2&100&2.75&1.14&1.13&-6.36&-0.173&3.46&1.5&-1.49\\
6.5&5&0.4&2&1.6&2.5&1.7&2.6&20&11&2&250&2.62&2.21&0.691&-6.22&-0.386&3.33&1.53&0.331\\
8&5&0.45&2&1.6&2.5&1.8&2.18&35&14&3&450&2.68&1.01&1.88&-6.2&-2.48&3.26&1.53&0.408\\
5&5&0.6&2&1.6&2.5&1.7&2.6&20&11&6&250&2.82&0.911&0.775&-6.09&-0.198&3.48&1.53&-0.628\\
8&2.5&0.45&1&0.65&3.6&1.8&2.54&35&14&3&450&3.23&0.326&0.437&-5.82&-0.248&3.6&1.54&-3.45\\
5&5&0.6&2&1.6&2.5&1.7&2&35&11&10&250&1.62&0.994&2.07&-6.09&-0.72&2.8&1.54&0.108\\
5&5&0.38&2&1.6&2.5&1.8&2.25&35&4&2.4&450&1.65&0.854&1.34&-5.76&-0.39&3.01&1.56&-0.243\\

 \hline
\end{tabular}
\caption{Same as for Table\,\ref{tab:bkgegbdmbestfitburkert} but now for our gaugino-dominated DM candidate point.}
\label{tab:bkgegbdmgauginoburkert}
\end{center}
\end{table*}


\begin{table*}[h]
\begin{center}
\begin{tabular}{|c|c|c|c|c|c|c|c|c|c|c|c|c|c|c|c|c|c|c|c|c|c|c|c|c|c|c|}
\hline
\multicolumn{12}{|c|}{Propagation Parameter Combination} & & \multicolumn{7}{|c|}{Einasto Profile}\\
\hline
$D_0$ & $B_0$ & $\alpha$ & $z_0$ & $e_{g_0}$& $e_{g_1}$ & $n_{g_1}$ &
$ n_{g_2}$ & $v_{\rm A}$ & $L$ & $\frac{{\rm d}V}{{\rm d}z}$ & $\phi_{\odot}$ & $\chi_{\rm LAR}^2$  &
$N$ & $M$ & \texttt{log}$_{10}(A)$ & $\gamma$ &  \texttt{log}$_{10}({\rm BF})$ & $\chi_{\rm DM}^2$ & $\delta\chi_{\rm DM}^2$\\
\hline
6.5&2.5&0.4&1&0.65&3.6&1.7&2&35&11&6&450&3.94&0.603&2.2&-8.92&-3.39&2.68&0.45&-0.949\\
5&2.5&0.34&1&0.65&3.6&1.82&2.36&36&4&3&450&2.6&0.608&0.565&-5.97&-0.157&3.03&0.503&-2.37\\
5&2.5&0.2&1&0.65&3.6&2.6&2&20&4&6&450&3.67&1.25&1.36&-8.08&-3.06&3.01&0.596&-2.41\\
5&2.5&0.2&3&0.65&3.6&2.6&2&20&4&6&450&3.67&1.23&1.32&-9.8&-1.79&3.06&0.761&-2.25\\
6.5&10&0.4&1&0.65&3.6&1.7&2&35&11&6&450&3.94&0.499&2.24&-8.27&-3.85&2.56&0.762&-0.746\\
5&2.5&0.4&1&0.65&3.6&1.7&2.3&35&4&2&450&1.86&0.563&0.548&-5.88&-0.194&3&0.775&-1.54\\
5&5&0.4&2&1.6&2.5&1.7&2.6&20&11&6&250&3.64&2.17&0.784&-5.92&-0.357&2.35&0.869&-0.185\\
5&10&0.2&3&0.65&3.6&2.6&2&20&4&6&450&3.67&0.972&1.25&-8.12&-3.61&3.2&0.877&-3.98\\
5&2.5&0.4&3&0.65&3.6&1.7&2.3&35&4&6&450&3.76&0.69&0.496&-5.98&-0.18&3.12&0.934&-0.987\\
5.75&2.5&0.34&1&0.65&3.6&1.82&2.54&36&4&3&450&3.34&0.539&0.544&-6.02&-0.141&3.09&0.939&-2.37\\
6.5&5&0.4&2&1.6&2.5&1.7&2.3&20&7&2&250&3.28&0.994&2.03&-5.68&-0.692&1.54&0.962&0.137\\
6.5&5&0.4&2&1.6&2.5&1.7&2.3&35&11&10&450&3.35&0.922&2.35&-5.82&-1.22&1.8&0.974&0.0869\\
8&10&0.45&3&0.65&3.6&1.8&2.18&35&14&3&450&2.68&0.33&1.2&-6.13&-0.136&2.94&1&-2.59\\
6.5&5&0.4&2&1.6&2.5&1.7&2.6&35&11&10&450&3.49&0.915&0.753&-5.66&-0.363&2.74&1.04&-0.505\\
8&2.5&0.45&1&0.65&3.6&1.8&2.36&40&14&8.5&450&3.97&0.521&0.636&-6.16&-0.131&3.14&1.07&-1.43\\
6.5&5&0.4&2&1.6&2.5&1.7&2.3&35&7&6&450&1.92&0.776&1.52&-5.71&-0.456&2.53&1.08&0.0529\\
8&2.5&0.45&1&0.65&3.6&1.8&2.36&34&14&3&450&2.59&0.495&0.607&-6.3&-0.0546&3.14&1.1&-1.75\\
5&5&0.4&2&1.6&2.5&2.15&2.3&20&7&6&250&1.71&2.07&1.72&-6.76&-3.36&1.43&1.14&0.247\\
8&5&0.45&2&1.6&2.5&1.8&2.36&35&14&5&450&1.83&0.863&1.09&-5.76&-0.325&2.26&1.15&-0.114\\
6.5&2.5&0.4&3&0.65&3.6&1.7&2.6&35&7&6&450&2.02&0.389&0.386&-5.76&-0.238&3.03&1.16&-2.52\\
5&5&0.4&2&1.6&2.5&2.15&2.6&20&7&6&250&1.78&0.965&0.815&-5.82&-0.28&2.74&1.17&-0.792\\
5.75&10&0.34&3&0.65&3.6&1.82&2.54&36&4&3&450&3.34&0.604&0.346&-6.21&-0.0893&3.26&1.17&-2.93\\
5&2.5&0.4&1&0.65&3.6&2.15&2.3&35&4&10&450&3.49&0.614&0.505&-5.96&-0.168&3.03&1.19&-2.28\\
8&2.5&0.45&1&0.65&3.6&1.8&2.18&40&14&8&450&3.34&0.609&0.376&-6.1&-0.114&3.16&1.22&-1.77\\
6.5&5&0.4&2&1.6&2.5&2.15&2.3&20&11&2&250&3.03&1.81&1.62&-6.85&-3.19&2.39&1.24&0.0627\\
8&5&0.45&2&1.6&2.5&1.8&2.18&40&14&8&450&3.34&1.33&1.85&-6.18&-1.86&2.36&1.28&0.194\\
5&5&0.4&2&1.6&2.5&2.15&2.3&20&11&6&250&3.8&1.12&2.37&-6.26&-2.61&1.3&1.28&0.325\\
8&5&0.4&2&1.6&2.5&1.8&2.36&40&14&8&450&3.24&1.55&1.14&-6.45&-0.204&2.72&1.28&-0.423\\
5&2.5&0.4&1&0.65&3.6&2.15&2&35&4&6&450&2.07&0.648&0.942&-6.16&-0.221&3.08&1.28&-1.83\\
5&2.5&0.4&1&0.65&3.6&1.7&2&35&4&6&450&3.97&0.793&1.93&-9.05&-3.33&2.67&1.29&-0.0918\\
6.5&5&0.4&2&1.6&2.5&1.7&2.6&35&7&6&450&2.02&0.635&0.699&-5.73&-0.297&2.84&1.3&-0.716\\
6.5&5&0.4&2&1.6&2.5&2.15&2.3&20&7&2&250&2.15&2.26&1.36&-8.13&-2.91&2.25&1.3&0.354\\
8.5&5&0.45&2&1.6&2.5&1.8&2.36&35&14&3&450&1.8&0.875&1.08&-5.98&-0.267&2.75&1.33&-0.998\\
5.2&5&0.38&2&1.6&2.5&1.8&2.3&35&4&2.5&450&1.83&0.804&1.24&-5.74&-0.387&2.7&1.35&-0.487\\
5.75&5&0.34&2&1.6&2.5&1.82&2.36&36&4&3&450&3.52&1.51&0.996&-5.81&-0.365&2.51&1.35&0.113\\
5.5&5&0.34&2&1.6&2.5&1.82&2.36&36&4&3&450&2.25&0.795&1.25&-5.73&-0.403&2.74&1.35&-0.167\\
5.2&5&0.4&2&1.6&2.5&1.7&2.3&35&4&2&450&2.46&0.972&1.18&-5.75&-0.344&2.27&1.38&-0.194\\
5&5&0.36&2&1.6&2.5&1.8&2.3&35&4&2&450&1.9&0.86&1.1&-5.68&-0.375&2.5&1.41&-0.575\\
5&5&0.38&2&1.6&2.5&1.8&2.15&35&4&2.5&450&1.63&1.46&1.83&-6.27&-2.25&2.61&1.42&-0.0187\\
5.75&5&0.34&2&1.6&2.5&1.82&2.36&36&4&2.5&450&3.17&1.05&1.15&-5.74&-0.42&2.77&1.42&0.195\\
6.5&2.5&0.4&1&0.65&3.6&2.15&2&35&11&10&450&1.9&0.636&1.65&-7.12&-0.707&2.96&1.42&-3.25\\
5&5&0.36&2&1.6&2.5&1.8&2.3&35&4&2.5&450&1.8&0.974&1.3&-5.92&-0.34&2.76&1.42&-0.214\\
5&5&0.38&2&1.6&2.5&1.8&2.25&35&4&2.4&450&1.65&0.951&1.13&-5.69&-0.392&2.6&1.44&-0.815\\
8&5&0.45&2&1.6&2.5&1.8&2.36&40&14&5.5&450&2.13&0.627&0.885&-5.69&-0.327&2.59&1.44&-0.513\\
5&5&0.6&2&1.6&2.5&1.7&2.3&20&11&6&250&2.79&2.3&1.03&-7.24&-2.22&2.65&1.47&0.57\\
4.8&5&0.4&2&1.6&2.5&1.7&2.3&35&4&2&450&1.87&0.895&0.975&-5.69&-0.344&2.45&1.48&-0.488\\
5&5&0.38&2&1.6&2.5&1.8&2.25&35&4&2&450&1.69&0.68&1.32&-5.69&-0.448&2.79&1.49&-1.58\\
5&5&0.4&2&1.6&2.5&1.7&2.1&35&4&2&450&1.92&1.55&1.9&-6.12&-1.09&2.19&1.49&0.0385\\
8&5&0.45&2&1.6&2.5&1.8&2.18&35&14&3&450&2.68&1.02&1.87&-6.24&-2.74&2.73&1.5&0.373\\
8.5&2.5&0.2&3&0.65&3.6&2.6&2&20&7&2&450&3.62&0.973&1.27&-7.49&-1.6&3.08&1.53&-3.54\\

 \hline
\end{tabular}
\caption{Same as for Table\,\ref{tab:bkgegbdmbestfiteinasto} but now for our gaugino-dominated DM candidate point.}
\label{tab:bkgegbdmgauginoeinasto}
\end{center}
\end{table*}


\begin{table*}[h]
\begin{center}
\begin{tabular}{|c|c|c|c|c|c|c|c|c|c|c|c|c|c|c|c|c|c|c|c|c|c|c|c|c|c|c|}
\hline
\multicolumn{12}{|c|}{Propagation Parameter Combination} & & \multicolumn{7}{|c|}{Burkert Profile}\\
\hline
$D_0$ & $B_0$ & $\alpha$ & $z_0$ & $e_{g_0}$& $e_{g_1}$ & $n_{g_1}$ &
$ n_{g_2}$ & $v_{\rm A}$ & $L$ & $\frac{{\rm d}V}{{\rm d}z}$ & $\phi_{\odot}$ & $\chi_{\rm LAR}^2$  &
$N$ & $M$ & \texttt{log}$_{10}(A)$ & $\gamma$ &  \texttt{log}$_{10}({\rm BF})$ & $\chi_{\rm DM}^2$ & $\delta\chi_{\rm DM}^2$\\
\hline
5&2.5&0.4&1&0.65&3.6&1.7&2.3&35&4&2&450&1.86&0.599&0.65&-5.94&-0.197&2.04&0.542&-1.77\\
5&2.5&0.4&3&0.65&3.6&1.7&2&35&4&6&450&3.97&0.85&1.63&-6.84&-0.106&1.87&0.587&-0.993\\
8&10&0.45&1&0.65&3.6&1.8&2.18&40&14&8&450&3.34&0.417&1.21&-6.13&-0.145&1.87&0.629&-1.88\\
5.75&2.5&0.34&3&0.65&3.6&1.82&2.36&38&4&3&450&2.92&0.575&0.743&-5.87&-0.236&1.95&0.691&-1.32\\
5&2.5&0.4&1&0.65&3.6&1.7&2.3&35&7&10&450&3.78&0.672&0.72&-6.26&-0.0915&2.07&0.833&-1.13\\
5.75&2.5&0.34&3&0.65&3.6&1.82&2.36&36&4&2.5&450&3.17&0.643&0.661&-5.96&-0.181&2&0.871&-1.41\\
5&2.5&0.4&1&0.65&3.6&1.7&2&35&4&6&450&3.97&0.893&0.838&-6.22&-0.121&1.93&0.871&-0.515\\
5&5&0.4&2&1.6&2.5&1.7&2.3&35&7&10&450&3.78&1.01&1.37&-5.7&-0.431&1.3&0.877&-0.186\\
6.5&5&0.4&2&1.6&2.5&1.7&2.6&20&11&2&250&2.62&1.74&0.88&-6.08&-0.251&1.41&0.879&-0.321\\
5&5&0.4&2&1.6&2.5&1.7&2.6&20&11&6&250&3.64&2.3&0.979&-6.51&-0.123&1.24&0.91&-0.144\\
6.5&5&0.4&2&1.6&2.5&1.7&2.6&20&7&2&250&3.2&1.94&0.84&-6.19&-0.206&1.31&0.913&-0.228\\
5&2.5&0.2&1&0.65&3.6&2.6&2&20&4&6&450&3.67&1.27&1.34&-7.82&-1.22&1.97&0.915&-2.09\\
5.5&10&0.34&3&0.65&3.6&1.82&2.36&36&4&3&450&2.25&0.532&0.665&-5.95&-0.212&2.1&0.942&-1.47\\
6.5&10&0.4&1&0.65&3.6&2.15&2&35&7&10&450&2.59&0.52&1.7&-6.55&-0.177&1.85&0.95&-1.91\\
5&5&0.6&2&1.6&2.5&1.7&2.3&20&11&6&250&2.79&1.35&1.3&-5.88&-0.376&0.869&0.957&0.0602\\
8&5&0.4&2&1.6&2.5&1.8&2.36&40&14&8&450&3.24&0.863&1.17&-5.69&-0.385&1.37&0.971&-0.107\\
5.75&2.5&0.34&3&0.65&3.6&1.82&2.36&36&4&3&450&3.52&0.63&0.764&-5.94&-0.183&1.8&0.979&-0.63\\
5&2.5&0.6&3&0.65&3.6&1.7&2&35&11&10&250&1.62&0.534&0.966&-6.24&-0.0951&1.96&0.982&-3.11\\
5&2.5&0.2&3&0.65&3.6&2.6&2&20&4&6&450&3.67&1.26&1.33&-8.87&-2.97&1.97&0.992&-2.02\\
5&2.5&0.4&1&0.65&3.6&2.15&2&35&4&6&450&2.07&0.709&0.886&-6.28&-0.107&2.06&0.997&-2.11\\
8&5&0.45&2&1.6&2.5&1.8&2.36&42&14&8&450&2.79&0.644&0.926&-5.63&-0.361&1.57&1.11&-0.646\\
6.5&5&0.6&2&1.6&2.5&1.7&2.6&20&11&2&100&3.81&0.793&0.564&-5.81&-0.271&1.85&1.11&-1.2\\
8.5&2.5&0.4&1&0.65&3.6&2.15&2&35&11&6&450&1.97&0.542&1.83&-6.82&-0.226&1.83&1.11&-0.921\\
8&5&0.45&2&1.6&2.5&1.8&2.18&40&14&8&450&3.34&1.35&1.89&-6.23&-2.04&1.22&1.15&0.0709\\
8&10&0.45&3&0.65&3.6&1.8&2.36&34&14&3&450&2.59&0.418&0.495&-6.05&-0.118&2.1&1.16&-2.63\\
5&5&0.4&2&1.6&2.5&2.15&2.3&20&7&6&250&1.71&1.57&2.01&-6.8&-3.18&0.939&1.2&0.303\\
6.5&5&0.4&2&1.6&2.5&1.7&2.6&35&11&10&450&3.49&1.44&0.894&-6.29&-0.119&1.69&1.23&-0.323\\
5&10&0.2&3&0.65&3.6&2.6&2&20&4&6&450&3.67&1.03&1.11&-7.73&-1.96&2.23&1.24&-3.61\\
6.5&5&0.4&2&1.6&2.5&1.7&2.3&20&7&2&250&3.28&1.12&1.92&-5.7&-0.707&0.86&1.25&0.422\\
6.5&5&0.4&2&1.6&2.5&1.7&2.3&20&11&2&250&2.38&1.32&1.92&-5.75&-0.812&0.812&1.25&0.259\\
8&5&0.45&2&1.6&2.5&1.8&2.36&35&14&3&450&2.93&0.743&0.974&-5.76&-0.323&1.61&1.25&-0.836\\
6.5&5&0.4&2&1.6&2.5&1.7&2.3&35&11&10&450&3.35&2.86&0.979&-6.92&-2.26&0.754&1.27&0.387\\
6.5&5&0.4&2&1.6&2.5&2.15&2.3&20&11&2&250&3.03&2.04&1.5&-7.19&-2.56&1.18&1.29&0.113\\
5.5&5&0.34&2&1.6&2.5&1.82&2.36&36&4&3&450&2.25&0.835&1.19&-5.67&-0.41&1.61&1.29&-0.268\\
5&5&0.38&2&1.6&2.5&1.8&2.3&35&4&3.5&450&1.68&0.751&1.33&-5.77&-0.344&1.54&1.31&-0.519\\
6.5&5&0.4&2&1.6&2.5&2.15&2.3&20&7&2&250&2.15&1.06&1.81&-6.12&-0.443&1.46&1.31&0.364\\
8&5&0.45&2&1.6&2.5&1.8&2.36&35&14&5.5&450&2.14&1.46&1.07&-6.33&-0.164&1.58&1.32&0.232\\
5&5&0.38&2&1.6&2.5&1.8&2.25&34.5&4&2.5&450&1.67&0.69&1.34&-5.63&-0.438&1.58&1.33&-0.65\\
5.2&5&0.4&2&1.6&2.5&1.7&2.3&35&4&2&450&2.46&0.974&1.14&-5.73&-0.366&1.58&1.33&-0.244\\
5&5&0.38&2&1.6&2.5&1.8&2.3&35&4&3&450&1.64&0.713&1.16&-5.65&-0.365&1.36&1.34&-0.231\\
8&5&0.45&2&1.6&2.5&1.8&2.36&35&14&4&450&1.9&1.07&0.879&-5.8&-0.319&1.52&1.34&-0.561\\
5&5&0.4&2&1.6&2.5&1.7&2.3&35&4&6&450&3.76&1.4&1.44&-6.11&-0.319&1.5&1.36&-0.422\\
5&2.5&0.4&1&0.65&3.6&1.7&2&35&4&2&450&1.98&0.86&0.708&-6.41&-0.0764&2.05&1.37&-0.828\\
5&5&0.4&2&1.6&2.5&1.7&2.3&35&4&2.5&450&1.96&1.03&1.05&-5.73&-0.34&1.46&1.39&-0.378\\
5.2&5&0.38&2&1.6&2.5&1.8&2.3&35&4&2.5&450&1.83&0.757&1.31&-5.75&-0.36&1.55&1.4&-0.442\\
5.75&10&0.34&1&0.65&3.6&1.82&2.18&36&4&3&450&3.69&0.721&0.777&-6.3&-0.147&2.17&1.4&-0.526\\
5&2.5&0.4&3&0.65&3.6&2.15&2&35&4&10&450&3.8&0.677&1.66&-7.89&-1.54&2.06&1.42&-1.1\\
5&5&0.38&2&1.6&2.5&1.8&2.15&35&4&2.5&450&1.63&1.4&1.94&-6.37&-3.47&1.52&1.44&-0.053\\
5&5&0.38&2&1.6&2.5&1.8&2.23&35&4&2.5&450&1.63&0.936&1.41&-5.81&-0.405&1.63&1.45&-0.28\\
5&5&0.4&2&1.6&2.5&1.7&2.3&35&4&3&450&2.1&0.875&1.28&-5.79&-0.358&1.58&1.45&-0.153\\

 \hline
\end{tabular}
\caption{Same as for Table\,\ref{tab:bkgegbdmbestfitburkert} but now for our mixed DM candidate point.}
\label{tab:bkgegbdmmixedburkert}
\end{center}
\end{table*}


\begin{table*}[h]
\begin{center}
\begin{tabular}{|c|c|c|c|c|c|c|c|c|c|c|c|c|c|c|c|c|c|c|c|c|c|c|c|c|c|c|}
\hline
\multicolumn{12}{|c|}{Propagation Parameter Combination} & & \multicolumn{7}{|c|}{Einasto Profile}\\
\hline
$D_0$ & $B_0$ & $\alpha$ & $z_0$ & $e_{g_0}$& $e_{g_1}$ & $n_{g_1}$ &
$ n_{g_2}$ & $v_{\rm A}$ & $L$ & $\frac{{\rm d}V}{{\rm d}z}$ & $\phi_{\odot}$ & $\chi_{\rm LAR}^2$  &
$N$ & $M$ & \texttt{log}$_{10}(A)$ & $\gamma$ &  \texttt{log}$_{10}({\rm BF})$ & $\chi_{\rm DM}^2$ & $\delta\chi_{\rm DM}^2$\\
\hline

5.75&2.5&0.34&3&0.65&3.6&1.82&2.18&36&4&3&450&3.69&0.752&1.3&-6.45&-0.0536&1.38&0.421&-0.946\\
8&2.5&0.4&1&0.65&3.6&1.8&2.36&40&14&8&450&3.24&0.536&0.664&-6.06&-0.138&1.49&0.747&-1.41\\
5&2.5&0.2&1&0.65&3.6&2.6&2&20&4&6&450&3.67&1.26&1.35&-7.47&-0.381&1.39&0.943&-2.06\\
5&2.5&0.2&3&0.65&3.6&2.6&2&20&4&6&450&3.67&1.28&1.22&-7.73&-0.517&1.5&0.944&-2.06\\
5.75&10&0.34&1&0.65&3.6&1.82&2.18&36&4&3&450&3.69&0.691&0.927&-6.26&-0.107&1.39&0.969&-0.955\\
6.5&5&0.4&2&1.6&2.5&1.7&2.6&20&11&2&250&2.62&1.75&0.914&-6.22&-0.17&0.871&0.98&-0.22\\
5&10&0.2&1&0.65&3.6&2.6&2&20&4&6&450&3.67&1.01&1.2&-7.83&-0.659&1.61&0.996&-3.02\\
5.75&2.5&0.34&1&0.65&3.6&1.82&2.54&36&4&3&450&3.34&0.507&0.535&-5.81&-0.235&1.39&1.03&-2.28\\
6.5&5&0.4&2&1.6&2.5&1.7&2.6&35&11&10&450&3.49&0.745&0.792&-5.6&-0.367&0.997&1.03&-0.518\\
8&10&0.45&3&0.65&3.6&1.8&2.54&35&14&3&450&3.23&0.349&0.391&-5.96&-0.173&1.62&1.08&-5.56\\
6.5&2.5&0.4&3&0.65&3.6&2.15&2&35&7&10&450&2.59&0.723&0.968&-6.38&-0.106&1.53&1.13&-2.66\\
8&10&0.45&3&0.65&3.6&1.8&2.36&35&14&5&450&1.83&0.45&0.599&-6.13&-0.138&1.65&1.14&-2.76\\
8&5&0.45&2&1.6&2.5&1.8&2.18&40&14&8&450&3.34&1.02&1.66&-5.78&-0.537&0.696&1.16&0.0664\\
5.75&5&0.34&2&1.6&2.5&1.82&2.36&36&4&3&450&3.52&0.782&1.3&-5.68&-0.381&0.865&1.17&-0.196\\
5.5&2.5&0.34&1&0.65&3.6&1.82&2.36&36&4&3&450&2.25&0.621&0.657&-5.87&-0.251&1.41&1.19&-1.21\\
5.75&2.5&0.35&1&0.65&3.6&1.82&2.36&36&4&3&450&3.77&0.546&0.67&-5.76&-0.293&1.41&1.21&-1.14\\
5&10&0.4&1&0.65&3.6&2.15&2&35&4&10&450&3.8&0.703&0.735&-6.28&-0.132&1.62&1.21&-2.18\\
8.5&5&0.4&2&1.6&2.5&1.7&2.3&35&11&2&450&2.08&1.48&1.35&-6.52&-0.146&0.827&1.24&-0.285\\
8&5&0.45&2&1.6&2.5&1.8&2.36&35&14&5&450&1.83&0.844&1.01&-5.69&-0.371&0.991&1.25&-0.0153\\
5.5&5&0.34&2&1.6&2.5&1.82&2.36&36&4&3&450&2.25&1.07&1.18&-5.8&-0.358&1.05&1.26&-0.297\\
6.5&5&0.4&2&1.6&2.5&1.7&2.3&35&11&10&450&3.35&1.79&1.73&-6.1&-1.67&0.371&1.29&0.331\\
5&5&0.4&2&1.6&2.5&2.15&2.3&20&7&6&250&1.71&1.63&1.97&-7.56&-0.329&0.29&1.3&0.405\\
8&10&0.45&3&0.65&3.6&1.8&2.36&35&10&3&450&1.76&0.401&0.565&-6.11&-0.135&1.63&1.3&-5.62\\
8&5&0.45&2&1.6&2.5&1.8&2.36&35&14&4&450&1.9&0.865&1.09&-5.9&-0.28&1.05&1.31&-0.516\\
6.5&2.5&0.4&1&1.9&2.95&1.7&2.6&35&11&10&450&3.49&0.538&0.928&-6.22&-0.0806&1.4&1.32&-2.94\\
6.5&5&0.4&2&1.6&2.5&2.15&2.3&20&11&2&250&3.03&1.95&1.54&-7.26&-2.12&0.747&1.32&0.149\\
5&5&0.38&2&1.6&2.5&1.8&2.15&35&4&2.5&450&1.63&0.937&1.91&-5.87&-0.544&0.831&1.34&-0.104\\
8.5&2.5&0.4&1&0.65&3.6&2.15&2&35&11&6&450&1.97&0.555&1.76&-7.51&-0.198&1.4&1.37&-0.666\\
6.5&5&0.4&2&1.6&2.5&2.15&2.3&20&7&2&250&2.15&1.73&1.66&-7.84&-0.648&0.854&1.38&0.429\\
5&10&0.4&3&0.65&3.6&1.7&2&35&7&10&450&3.21&0.597&2.02&-8.11&-1.95&0.98&1.39&-0.174\\
8&10&0.45&1&0.65&3.6&1.8&2.36&35&14&4&450&1.9&0.421&0.521&-6.04&-0.128&1.47&1.46&-2.32\\
5&5&0.4&2&1.6&2.5&1.7&2.3&30&4&2&250&3.36&1.34&1.72&-6.42&-0.22&0.766&1.47&-0.146\\
5.75&5&0.34&2&1.6&2.5&1.82&2.54&36&4&3&450&3.34&0.745&0.784&-5.67&-0.353&1.27&1.47&-0.688\\
5&5&0.38&2&1.6&2.5&1.81&2.25&35&4&2.5&450&1.65&0.841&1.24&-5.71&-0.385&1&1.47&-0.607\\
5&5&0.6&2&1.6&2.5&1.7&2.3&20&11&6&250&2.79&2.07&1.23&-6.84&-3.26&1.03&1.48&0.582\\
5&5&0.4&2&1.6&2.5&1.7&2.3&35&7&10&450&3.78&1.75&1.34&-6.7&-0.244&1.11&1.5&0.441\\
5&5&0.4&2&1.6&2.5&1.7&2.3&40&4&2&450&2.01&0.58&0.968&-5.65&-0.356&1&1.51&-0.7\\
4.8&5&0.38&2&1.6&2.5&1.8&2.3&35&4&2.5&450&2.19&0.841&1.03&-5.74&-0.34&1.14&1.52&-0.569\\
5&5&0.38&2&1.6&2.5&1.8&2.25&35&4&2&450&1.69&0.78&1.37&-5.76&-0.419&1.15&1.52&-0.208\\
5&2.5&0.4&1&0.65&3.6&1.7&2&35&4&6&450&3.97&0.949&0.769&-6.43&-0.0506&1.37&1.52&0.137\\
5&10&0.4&1&0.65&3.6&2.15&2.3&35&4&6&450&1.88&0.5&0.444&-6.12&-0.1&1.58&1.52&-3.38\\
8&5&0.45&2&1.6&2.5&1.8&2.18&35&14&3&450&2.68&1.16&1.81&-6.38&-3.26&1.08&1.53&0.409\\
5&5&0.38&2&1.6&2.5&1.8&2.25&35&4&3&450&1.65&1.01&1.28&-5.79&-0.407&1.11&1.54&-0.938\\
5.75&2.5&0.34&3&0.65&3.6&1.82&2.36&36&4&2.5&450&3.17&0.499&1.26&-6.3&-0.0601&1.03&1.54&-0.744\\
5&5&0.34&2&1.6&2.5&1.82&2.36&36&4&3&450&2.6&0.932&0.928&-5.67&-0.406&1.26&1.55&-0.258\\
5.75&5&0.34&2&1.6&2.5&1.82&2.36&36&4&2.5&450&3.17&1.18&1.16&-5.78&-0.436&1.18&1.55&0.179\\
8&10&0.45&3&0.65&3.6&1.8&2.36&40&14&8.5&450&3.97&0.495&0.295&-6.01&-0.134&1.63&1.56&-1.63\\
5&5&0.6&2&1.6&2.5&1.7&2&35&11&10&250&1.62&0.965&2.11&-6.05&-0.942&0.881&1.58&0.0818\\
5&5&0.6&2&1.6&2.5&1.7&2.3&35&11&10&250&1.53&0.477&0.935&-5.66&-0.377&1.32&1.59&-0.686\\
8.5&10&0.4&3&0.65&3.6&2.15&2&35&11&6&450&1.97&0.484&1.68&-7.26&-1.91&1.54&1.6&-1.12\\

 \hline
\end{tabular}
\caption{Same as for Table\,\ref{tab:bkgegbdmbestfiteinasto} but now for our mixed DM candidate point.}
\label{tab:bkgegbdmmixedeinasto}
\end{center}
\end{table*}


\begin{thebibliography}{99}
\bibitem{footnote:rmax}
We note that increasing $r_{\rm max.}$ above this value has a negligible effect on our results.
\bibitem{footnote:reac}
\texttt{GALPROP} possesses several other diffusion models, namely Kolmogorov +  damping and Kraichnan + damping in addition to the conventional reacceleration model that we adopt in this study, however, a investigation of these models is not pursued here.
\bibitem{footnote:fits}
Of course, it is possible that a PPC that generates a small positive value of $\delta\chi_{\rm EGB}^2$
may still provide a better fit to the Fermi LAT data based solely upon its shape. However, following the 
discussion in Sec.\,\ref{subsec:bkgflux}, we believe that it is necessary to compare fits based upon their
respective values of the relevant reduced $\chi^2$ statistic of each fit, in order to properly account for the 
different levels of flexibility associated with each of the fitting procedures utilised.
\end{thebibliography}
\end{document}